\documentclass[fleqn,usenatbib,twocolumn]{mnras}

\usepackage{newtxtext,newtxmath}
\usepackage[T1]{fontenc}
\usepackage{cuted}
\DeclareRobustCommand{\VAN}[3]{#2}
\let\VANthebibliography\thebibliography
\def\thebibliography{\DeclareRobustCommand{\VAN}[3]{##3}\VANthebibliography}

\usepackage{graphicx}	
\usepackage{multirow}% Including figure files
\usepackage{amsmath}
\usepackage{ulem}% Advanced maths commands
\usepackage{xcolor}

\title[]{ Dynamical dark energy in the  Bianchi Type-V Universe with \textit{DESI DR2 BAO}, \textit{SNIa} compilation and \textit{RSD} measurements}

\author[S. Sahlu et al.]
{ S. Sahlu$^{1,2}$, \'{A}. de la Cruz-Dombriz$^{3,4}$,   A. H. A. Alfedeel$^{1,5}$, G. J. Olmo$^{6}$, A. Abebe$^{1,2}$, 
\\
$^{1}$Centre for Space Research, North-West University, Potchefstroom 2531, South Africa\\
$^{2}$National Institute for Theoretical and Computational Sciences (NITheCS), Potchefstroom 2520, South Africa\\
$^{3}$Departamento de F\'{i}sica Fundamental, Universidad de Salamanca, 37008 Salamanca, Spain\\
$^{4}$Cosmology and Gravity Group, Department of Mathematics and Applied Mathematics, University of Cape Town, 7700 Rondebosch, South Africa\\
$^{5}$Department of Mathematics and Statistics, Imam Mohammad Ibn Saud Islamic University (IMSIU), 13818, Riyadh, Saudi Arabia \\
$^{6}$Instituto de Física Corpuscular (IFIC), CSIC‐Universitat de València, Spain\\
{sahlushambel@gmail.com}
}

\date{Accepted XXX. Received YYY; in original form ZZZ}

\pubyear{\the\year{}}

\begin{document}
\label{firstpage}
\pagerange{\pageref{firstpage}--\pageref{lastpage}}
\maketitle

\begin{abstract}
We investigate the cosmological implications of dynamical dark energy (DDE) models 
within an anisotropic, spatially homogeneous Bianchi Type-V spacetime framework 
using a $1+3$ covariant thermodynamics approach. By implementing both constant ($w$) 
and time-varying ($w_0, w_a$) parameterized equations of state, we evaluate the 
background expansion history and track linear matter perturbations via the 
quasi-static approximation. We confront these scenarios with the latest 
cosmological datasets, including the Dark Energy Spectroscopic Instrument (DESI) 
DR2 Baryon Acoustic Oscillations (BAO), the Union3 and Dark Energy Survey 5-year 
(DESY5) Type Ia Supernovae compilations, Cosmic Chronometers (CC), and 
Redshift-Space Distortion (RSD) measurements. Our joint statistical analyses 
reveal that the introduction of spatial anisotropy coupled with DDE efficiently 
accommodates recent late-time measurements and provides a viable mechanism to 
mitigate the persistent $H_0$ and $S_8$ cosmological tensions. Model selection 
metrics show that while Akaike criteria strongly support the extended Bianchi 
Type-V scenarios across most joint data combinations, Bayesian criteria continue 
to favor the simpler standard $\Lambda$CDM baseline due to its 
lower dimensionality. Finally, we establish tight constraints on the current 
matter density parameter $\Omega_{m,0}$, the shear parameter $\Omega_{\sigma,0}$, 
and the dark energy evolution parameters, confirming that anisotropic extensions 
remain viable and testable frameworks for modern precision cosmology.
%\end{abstract}

%%%%%%%

%\url{https://arxiv.org/pdf/2507.00779}
% The recent work in \citep{rezaei2024evidence} broadly discussed the evidence of the evolving dark energy for a non-flat universe with different cosmological measurements.
 
\end{abstract}

%\begin{keywords}
%\keywords{cosmological parameters - dark energy – cosmology: observations – cosmology: theory.}
%\end{keywords}
%\url{https://arxiv.org/pdf/2010.02728}, \url{https://arxiv.org/pdf/2410.02386v2}
%%%%%%%%%%%%%%%%%%%%%%%%%%%%%%%%%%%%%%%%%%%%%%%%%%
%%%%%%%%%%%%%%%%% BODY OF PAPER %%%%%%%%%%%%%%%%%%

\section{Introduction}
The $\Lambda$CDM model is built assuming a perfectly isotropic and homogeneous spatial geometry.\footnote{In our current work, the models: $\Lambda$CDM, $w$CDM, and $w_0w_a$CDM are spatial curvature $\kappa$ dependent.} However, persistent tensions in the estimation of cosmological parameters, most notably the $H_0$ and $S_8$ discrepancies, have boosted interest in extensions that relax these fundamental spacetime symmetries. In fact, a better understanding of the impact that either local inhomogeneities or anisotropies - or both - have at larger scales may hold the key to reconciling the implications of data sets testing the local expansion rate and those that are more sensitive to the global cosmic structure. Observations have revealed small variations in the intensity of cosmic microwave background radiation (CMB) across different directions. These anisotropies are thought to be connected to the origins of structure formation in the universe. Consequently, alternative cosmological models are required to elucidate these phenomena. Among these models, Bianchi-type cosmological models have garnered significant attention \citep{akarsu2019constraints,pradhan2011accelerating,amirhashchi2018probing,sharif2009exact}. The Bianchi metrics \citet{ellis2006bianchi,bianchi1928lezioni} are classified as a set of homogeneous and anisotropic cosmological models which can be seen as generalizations of the Friedmann-Lemaître-Robertson-Walker (FLRW) cosmological model whenever specific symmetries are relaxed. In particular, the Bianchi type-I model \citep{akarsu2019constraints} extends the zero-curvature FLRW model, known as ``flat geometry'', while the Bianchi type-V model generalizes the FLRW model with negative curvature, i.e., open universe \citep{singh2008bianchi}. On the one hand, since due to their anisotropic character, these models violate the Copernican principle, they can be used to determine the consistency of the FLRW model and study the effect of anisotropies on cosmological perturbations. Consequently, the Bianchi family of cosmological models has become an active field of research these days. On the other hand, the consideration of dark energy (DE) sources embedded in Bianchi universes may provide the necessary pressure anisotropy to sustain or even amplify small deviations from isotropy in the late-time universe, offering a compelling alternative to the standard $\Lambda$CDM model. 
\\
\\In the current work, we shall explore the Bianchi Type V universes and confront them with a comprehensive suite of the latest observations, including BAO, cosmic chronometers, Type Ia Supernovae, and redshift-space distortion measurements. Our goal, therefore, is to determine whether such a DDE (i.e., taking into account the generalized EoS parametrization, $w_{\rm DE} = w_0 +w_a(1-a)$, known as Chevallier-Polarski-Linder (CPL) parametrization \citep{chevallier2001accelerating,linder2003exploring})  sector can better describe the expansion history and large-scale structure than the standard $\Lambda$CDM framework. While some models invoke an inherently anisotropic dark energy sector to explain cosmic deviations \cite{BeltranJimenez:2008iye, Koivisto:2008ig,Verma:2024lex}, our approach here focuses on the feasibility of isotropic DDE embedded in Bianchi Type V geometries. In this context, the evolution of the shear scalar, which characterizes the degree of anisotropy, will be proved to be deeply coupled to the expansion history, the latter being governed by a non-constant EoS for the dark energy fluid. Indeed, by considering a DDE component, we investigate whether the corresponding expansion rate is capable of sustaining non-trivial anisotropic signatures that would otherwise decay too rapidly, as widely known in the standard $\Lambda$CDM scenario. This allows us to test whether the interplay between spatial curvature, characteristic of Bianchi-V models, and a DDE fluid provides a better fit to the observed 'fine-grained' features of the late-time expansion data from BAO and cosmic chronometers. 

In the realm of DDE models \citep{ adame2025desi, rezaei2024evidence,alam2025beyond,zhao2012examining,di2025cosmoverse,alam2025beyond}, one common approach has consisted on using a time-varying equation-of-state (EoS) parameter, \( w = p/\rho \), to ascertain the true nature of dark energy. Evidence for  DDE has been examined using various cosmological observations from the 9-year WMAP and Planck data for both flat and non-flat \(\Lambda\)CDM models \citep{hinshaw2013nine}. According to this work, the value of the EoS at 95\% level of confidence (C.L) $-1.71<w<-0.34$ for a flat universe and $w>-2.1$ for non-flat geometry based on WMAP measurements. Taking into account the CPL, observations point towards $w_0 = -1.34 \pm 0.18$ and $w_a = 0.85\pm 0.47$ using \textit{WMAP+eCMB+BAO+H(z)} datasets for flat geometry as optimal fit parameters. Recent observational measurements from DESI BAO 2024 \citep{adame2025desi} provide insight that the time-varying dark energy EoS, parametrized by $w_0$ and $w_a$, is well constrained using the combined datasets of DESI with either CMB or SNIa. Individually, these datasets prefer $w_0 > -1$ and $w_a < 0$, showing evidence for DDE that challenges the standard model of cosmology with different cosmological data combinations. 

The study of DDE models in the context of Bianchi-type metrics has been considered by several authors
\citet{amirhashchi2013phantom,amirhashchi2014interacting,amirhashchi2014interacting, amirhashchi2014viscous, amirhashchi2018probing, pradhan2015accelerating,yadav2012lrs,yadav2016transitioning}. Recently, in \cite{amirhashchi2018probing}, authors have constrained dark energy models' anisotropies using the recent datasets of the Hubble parameter and the new release of SNIa data, and their Joint Light-curve analy\-sis (JLA).  They show that the combination of these two datasets is not enough to place tighter constraints on measurements of the anisotropy parameter $\Omega_{\sigma}$. They reported a best fit value of order $\Omega_{\sigma} \sim 10^{-4}$, the available low redshift data were found to provide only weak constraints on anisotropies. 
% In this manuscript is to study how DDE, when embedded in a Bianchi Type V space-time, affects both the cosmic expansion and the growth of matter perturbations, so confrontation with recent measurements becomes possible. To do so, we have considered the time-varying EoS with the Bianchi Type V Universes. 
% The development of large-scale structure in Bianchi Type V backgrounds through cosmic perturbations has been addressed with time varying variable Newtonian  $G$ and cosmological $\Lambda$  for bulk viscous fluid as presented in \cite{abebe2023perturbations}.  \adlcd{Please rephrase this. A time varying $\Lambda$ sounds very weird. Also, instead of {\it time varying variable Newtonian  $G$} I would say {\it time varying effective gravitational constant $G_{\rm eff}$}}
\\
\\
% with comparisons made to non-flat isotropic $\Lambda$CDM, $w$CDM, and $w_0w_a$CDM for detailed cosmological analysis \textcolor{red}{[it is not clear with this wording which cases one is studying]}.\ssa{check now?} \textcolor{red}{[First, LCDM does not have a {\it varying equation of state}, so this needs to be rephrased. Second, is the non-flat isotropic $\Lambda$CDM a Bianchi V model? I think we are mixing what spacetime(s) we study and what are the DE models we put on the rhs of the Equations of Motion. Please rephrase the whole paragraph.]}
We organize the manuscript as follows. In Sec. \ref{back}, we present the derivation of the background field equations. In this section, we also introduce the generalized equation for the Hubble parameter using the parameterized EoS $w_0$–$w_a$ for DDE, which we shall employ in the MCMC simulations in Sec. \ref{resultanddiscussion}. In Sec. \ref{perturbations}, we derive the full set of evolution equations for linear cosmological scalar perturbations using the 1+3 covariant formalism. We present the evolution equations for the density contrast $\delta_m(z)$, the growth factor $\mathcal{D}_m(a)$ to facilitate the study of structure formation. The detailed results, along with a comprehensive statistical analysis of the work, are demonstrated in Sec.  \ref{resultanddiscussion}. Finally, in Sec .~\ref {disc}, we present our conclusions. Throughout the manuscript, unless indicated, we use $8\pi G= c =1$ units and $(-,+,+,+)$ spacetime signature.

\section{Background equations}\label{back} 
% In this paper, we consider the DDE behavior to study the cosmic evolution of a homogeneous and anisotropic universe whose geometry is described by the Bianchi type $-V$ cosmological model.  In this instance, the following expression provides t
The line element of Bianchi type-V in a synchronous comoving coordinate system takes the form
\begin{align}\label{metric}
{\rm d}s^2 = -{\rm d}t^2 + A(t)^2 {\rm d}x^2 + {\rm e}^{2 \alpha x } \left[ B(t)^2 {\rm d}y^2 + C(t)^2 {\rm d}z^2\right]~.
\end{align}
where $\alpha$ is a constant. Thus, Einstein's field
equations (EFEs) can be written as 
%\all{Please look at the following modifications of $T_{\mu}^\nu$}
\begin{align}
R_\mu^\nu - \frac{1}{2}g^\nu_{\mu} R =  T_\mu^\nu\;, %T^{m}_{ij} + T^{\rm DE}_{ij} ~,
\end{align}
where $R_\mu^\nu$ denotes the Ricci tensor, $R$ the Ricci scalar, and  $ T_\mu^\nu$ the energy-momentum tensor  can be expressed as
\begin{align}
    T_\mu^\nu = \left(\rho_{tot}+P_{tot}\right)u_\mu u^\nu + g^\nu_{\mu} P_{tot} \;, %- \Pi_\mu^\nu
\end{align}
where $u^\nu = (-1,\vec{0})$ is the comoving four-vector velocity and obey $u_\nu u^\nu = -1$,  $\rho_{tot}$ is the total cosmic fluid, and $P_{tot}$ is the pressure of the total cosmic fluid.
% and $\Pi_{\mu}^\nu$ is an anisotropic trace-free tensor which captures the direction of the pressure deviation inherent in Bianchi models. \adlcd{What has happened with $\Pi_\mu^\nu$ later? Is it zero? By assumption or as a result?} 
 For the case of the energy-momentum tensor of the matter, which includes both baryonic and dark matter as a single effective fluid, and Dark Energy (DE), respectively, being presented by 
\begin{align*}\label{Tij}
T^{(m)\nu}_{\mu} &= \mbox{diag}[-\rho_m, 0,0,0] = \mbox{diag}[-1, 0,0,0]\rho_m~,\\
T^{({\rm DE})\nu}_{\mu} & = \mbox{diag}[-\rho_{\rm DE}, p_{\rm DE},p_{\rm DE},p_{\rm DE}]\nonumber\\& 
= \mbox{diag}[-1, w_{\rm DE},w_{\rm DE},w_{\rm DE}]\rho_{\rm DE}~. 
\end{align*}
%\textcolor{red}{[If $m$ is dark or baryonic (dust) matter, $w^m=0$ and we can save space and lighten the equations above and below. Please amend.]} 
where  $\rho_m$ and $p_m$ are the energy density and the pressureless, respectively, of the matter (dark matter+baryonic) fluid which are related to each other via the EoS $w_{m}=p_m/\rho_m = 0$ and $\rho_{\rm DE}$ and $p_{\rm DE}$ are the energy density and pressure of the DE fluid, respectively, whose EoS is $w_{\rm DE} = p_{\rm DE}/\rho_{\rm DE}$.
%\adlcd{Please use DE, not cursive and capital, instead of $de$. Please change where appropriate throughout the section.}
Thus, the spatial diagonal components, the temporal components, and the off-diagonal constraints of the EFEs for the metric \eqref{metric} become 
\begin{eqnarray}
&\frac{\ddot{B}}{B} +  \frac{\ddot{C}}{C} +  \frac{\dot{B}}{B}\frac{\dot{C}}{C} - \frac{\alpha^2}{A^2}  =  - p_{\rm DE}~,\label{Gtt}\\
&\frac{\ddot{A}}{A} +  \frac{\ddot{C}}{C} +  \frac{\dot{A}}{A}\frac{\dot{C}}{C} - \frac{\alpha^2}{A^2}  =  -  p_{\rm DE}~,\label{Gxx}\\
&\frac{\ddot{A}}{A} +  \frac{\ddot{B}}{B} + \frac{\dot{A}}{A}\frac{\dot{B}}{B}  -  \frac{\alpha^2}{A^2}   = - p_{\rm DE} ~, \label{Gyy}\\
&\frac{\dot{A}}{A}\frac{\dot{B}}{B} + \frac{\dot{A}}{A}\frac{\dot{C}}{C} + \frac{\dot{B}}{B}\frac{\dot{C}}{C} - \frac{3\alpha^2}{A^2} = \rho_{m} + \rho_{\rm DE}~,\label{Gzz}\\
&\frac{2\dot{A}}{A} -\frac{\dot{B}}{B}- \frac{\dot{C}}{C} =0~.\label{Gxy}
\end{eqnarray}
%
% \adlcd{Equations 4-6 are incorrect on their right-hand sides: First, $w_x$ has not been defined since above we defined $w^{DE}_{x,y,z}$; and second, $w_x p^{DE}$ is probably $w^{DE}_x \rho^{DE}$. Please revise. Also, where we have said or assumed that $w^{DE}_x = w^{DE}_y = w^{DE}_z\equiv w^{DE}$?}
%
The covariant conservation of the energy-momentum tensor is $\nabla^i T_{ij}=0$. Since the matter and dark-energy components are considered to be separately  conserved, one obtains 
\begin{equation}\label{dotRhoEq1}
\dot{\rho}^m + \Theta (1+w^m)\rho^m =0~,
%\end{equation}
\;\;{\rm and}\;\; 
% \begin{equation}
\dot{\rho}_{\rm DE} +\Theta(1+w_{\rm DE})\rho_{\rm DE}= 0~,
\end{equation}
where $\Theta=\nabla_j u^j$
is the expansion scalar. We also define the mean Hubble parameter by $H=\Theta/3$.
% these equations may equivalently be written as 
% \begin{equation}\label{dotRhoEq}
% \dot{\rho}^m + 3H (1+w^m)\rho^m =0~,
% \end{equation}
% and 
% \begin{equation}
% \dot{\rho}^{DE} +3H(1+w^{DE})\rho^{DE}= 0~,
% \end{equation}
% \adlcd{This is not the usual way of writing the conservation of each fluid. Each fluid is separately conserved in our case. Please rewrite. Also, H has not been defined yet. I guess this equation should have $3H\rightarrow\Theta\equiv \nabla_i u^i$ since H has not been defined yet.}
On the other hand, the shear tensor is defined as
\begin{align}
\label{sigmadefin}
&\sigma_{ij}  =\frac{1}{2}
\left( h^k_i \nabla_k u_j + h^k_j \nabla_k {u}_i\right)  - \frac{1}{3} \Theta h_{ij}\;,
\end{align}
% \adlcd{The $\dot{u}$ above is wrong. Also, in the expression above there is $1/2$ prefactor missing multiplying the first two terms. Have the results below taken into account those mistakes?}
where $h_{ij}= g_{ij} + u_i u_j$ is the projection tensor and $\Theta$ is the expansion scalar. Then, the contraction of the shear tensor gives 
\begin{align}
\sigma_{ij}\sigma^{ij} = \frac{1}{3} \left[ \left(\frac{\dot{A}}{A} - \frac{\dot{B}}{B}\right)^2 +  \left( \frac{\dot{B}}{B} - \frac{\dot{C}}{C}\right)^2 + \left( \frac{\dot{C}}{C} - \frac{\dot{A}}{A}\right)^2 \right]\;.
\end{align}
% \textcolor{purple}{[GO:the tonsorial indices in this equation are missing]}\ssa{check now?} \adlcd{Brackets above are wrong. The expression above has other mistakes: The minus must be plus and the 2/3 is 1/3. Please revise carefully.}
 We define the volume $V$ and the average/mean Hubble parameter as 
\begin{align}
    &V\equiv a^3\equiv\sqrt{|-g_{ij}|} = ABC~,\\
    &H\equiv\frac{1}{3}\Theta =
    \frac{1}{3}\left(\frac{\dot{A}}{A} + \frac{\dot{B}}{B} + \frac{\dot{C}}{C}\right)= \frac{1}{3}(H_x + H_y+H_z)=\frac{\dot{a}}{a}~,
\end{align}
where $\dot{A}/A= H_x$. $\dot{B}/B= H_y$ and $\dot{C}/C= H_z
$ are usually referred to as the Hubble directional parameters in the $x, y$ and $z$ directions, respectively.
It has been shown by \cite{alfedeel2018generalized,abebe2023perturbations,yadav2012lrs}, the field equations \eqref{Gtt}-\eqref{Gzz} can be reduced to a  system of first-order coupled differential equations for the metric variables $A,B$ and $C$ as:   
% from \eqref{Gxx}, \eqref{Gxx} from \eqref{Gyy} and \eqref{Gyy} from \eqref{Gzz} 
%yields the following relations: 
%
% \adlcd{Substracting equations do not give integration constants. Something else needs to be done. What?}
\begin{align} \label{ABC}
\frac{\dot{A}}{A}  = \frac{\dot{B}}{B} + \frac{k_1}{ABC}\;, \quad
\frac{\dot{B}}{B}  = \frac{\dot{B}}{B} + \frac{k_2}{ABC}\;, \quad
\frac{\dot{C}}{C}  = \frac{\dot{A}}{A} + \frac{k_3}{ABC}~.
\end{align}
where $k_1,k_2$, $k_3$ are integration constants satisfying $k_1+k_2+k_3 =0$. Note that in the particular case of imposing the Bianchi type-V condition $A^2=BC$, one obtains that $A=a$, which further implies $k_2=k_3=k_1$. In the general case,  integrating  equation \eqref{ABC} with respect to time $t$ gives explicit expressions for the metric coefficients $B$ and $C$ as follows:
\begin{align}
A(t) & = a(t)~,\label{Aformula} \\
 B(t) & = m_1 a(t) \exp \left[ -\frac{k_2+k_3}{3}  \int_{t_0}^t \frac{{\rm d}\tilde{t}}{a^3(\tilde{t})} \right]\;,\label{BFormula}\\
 C(t) & = \frac{a(t)}{m_1 } \exp\left[ \frac{k_2+k_3}{3} \int_{t_0}^t \frac{{\rm d}\tilde{t}}{a^3(\tilde{t})} \right] ~,\label{CFomula}
\end{align}
which clearly satisfies that $ABC=a^3$. Here we have defined $t_0$ as an arbitrary reference cosmic time, and $m_1$ is an arbitrary positive constant. Now, substituting Eqs. \eqref{Aformula}-\eqref{CFomula} into Eq. \eqref{sigmadefin}, the shear scalar is given by \citep{alfedeel2018generalized,abebe2023perturbations}
\begin{align}
    \label{sigma} 
    \sigma^2 & \equiv \frac{1}{2} \sigma^{ij} \sigma_{ij}~ % \nonumber \\% 
     = \frac{1}{6} \left[ H_{x}-H_{y})^{2}+(H_{y}-H_{z})^{2}+(H_{z}-H_{x})^{2}\right]    \nonumber\\
     & = \frac{\sigma_0^2}{a^6}~,
\end{align}
where $\sigma_0^2 = k_1^2$ is a constant that is related to the universe anisotropies. 
% {$\Omega_\sigma=\sigma_0^2/3H^2$ as it will be shown later.
Since $\Omega_\sigma = \sigma_0^2/3H^2 a^6$, the shear contribution decreases rapidly during the cosmic expansion and tends to zero at late times $(z\to 0)$.
This behavior follows directly from the evolution of $\sigma^2$ in Eq. \eqref{sigma}. In the isotropic FLRW limits, the expansion rates $H_x=H_y=H_z$, imply $\sigma_0=0$, $\Omega_\sigma=0$. Hence, the FLRW universe is recovered as the shear vanishes.

% This quantity measures the deviation from the perfect cosmological isotropy ($\Omega_\sigma=0$). For anisotropic cosmological models such as Bianchi Type-I, it is expected that  $\Omega_\sigma$  decreases with time and dies out as redshift $z$ approaches zero \adlcd{Is it expected or is it a result? We have not talked about Type-I and now this model appears out of the blue. If we ant to mention anisotropic models in general, then there is no need to mention Type-I explicitly}.  In the FRW model is obtained when the expansion rates \adlcd{\sout{$x,y$, and $z$} $H_x$, $H_y$, $H_z$} are equal. This implies the choice of $k_1 = k_2=0$ \adlcd{and $k_3$?} and similarly $\sigma_0=0\to \Omega_\sigma=0$, and $a=a(t)$ is the universe average scale factor. \adlcd{The last nine lines just to say that in FLRW $\sigma=0$}

% \textcolor{orange}{\textbf{Alnadief, Just add justification $\Omega_\sigma \ge 0$} based on the work, the whole point of the physics of this paper depends on this physical parameter and needs a strong motivation. 
% \begin{itemize}
%     \item \url{https://arxiv.org/pdf/1712.02072}
%     \item \url{https://arxiv.org/pdf/1802.04251}
% \end{itemize}
% }{

% where the shear scalar is given by $ \sigma^{2}=\frac{1}{6} \left( H_{x}-H_{y})^{2}+(H_{y}-H_{z})^{2}+(H_{z}-H_{x})^{2}\right)$, \adlcd{You called them $H_x$, $H_y$ and $H_z$ before. This expression is (12), no need to repeat it.}
The anisotropic density parameter is defined as \(\Omega_{\sigma}=\sigma^{2}/ 3H^{2}\), given that $\sigma^{2}$ is a sum of squared differences in expansion rates, it is inherently non-negative. Additionally, the mean Hubble rate $H^{2}$ remains strictly positive in an expanding universe.  Consequently, $ \sigma^{2}\ge 0 \Leftrightarrow \Omega_{\sigma}\ge 0$.
 This condition is a direct physical requirement derived from the kinematical structure of Bianchi type-V space-times %\adlcd{We are studying Type-V, why then Type-I here?}.
The shear contribution functions as a positive-definite effective energy density, generally decaying as $\sigma^{2}\propto V^{-2}$, where $V$ represents the volume scale factor.  Recent literature on anisotropic cosmology and anisotropic inflation presents similar treatments and interpretations of shear energy as a non-negative component \cite{pan2018}.

The summation of the field equations \eqref{Gtt}-\eqref{Gzz} can be simplified to the generalized Friedman equation for Bianchi type-V as demonstrated below (recall that $\alpha$ was defined in Eq.(\ref{metric}))
% am{references needed here. This is problematic since it does not allow a negative curvature contribution, as $\alpha^2$ is always positive}: \ssa{Explained below Eq. \eqref{GFreidmanEq}}
\begin{align}
%\label{F1}
%2\left( \frac{\ddot{a}}{a}\right) + \frac{4L^2}{3a^6}  &= -p^{m} - p^{de}=-\frac{1}{3}(\rho+3p)~,\\
3H^2  &= \rho_{m} + \rho_{\rm DE} +\frac{\sigma^2_0}{a^6} + \frac{3\alpha^2 }{a^2} ~,\label{F2}
\end{align}
and
\begin{align}\label{GFreidmanEq}
2\left( \frac{\ddot{a}}{a}\right) + \left(\frac{\dot{a}}{a}\right)^2 + \frac{\sigma^2_0}{a^6} -  \frac{\alpha^2 }{a^2} = -p_{m} - p_{\rm DE}~.
\end{align}
We define the following dimensionless density, curvature, and anisotropy parameters, see \citep{prasad2021exact}:
\begin{align}\label{density-defintion}
&\Omega_{m0} = \frac{1}{3H^2_0} \rho_{m0}\;,~ \Omega_{\sigma0} =  \frac{\sigma^2_0}{3H^2_0}\;,
\Omega_\alpha =  \frac{3\alpha^2}{H^2_0a^2}\;,~\nonumber\\& \Omega_{k0} = -\frac{\kappa}{3H^2_0a^2}\;,~\Omega_{\rm DE 0} = \frac{\rho_{\rm DE0}}{3H^2_0} \;.
\end{align}
In this work, the parameter $\alpha$ characterizes the spatial curvature of the Bianchi type$-V$ spacetime. As discussed in \citep{goliath1999homogeneous}; we adopt the relation $\kappa = -3\alpha^2$ so that $\Omega_\alpha  \equiv \Omega_{k0} $. In the isotropic limit $(\sigma=0)$, the model reduces to the non-flat FLRW universe. This choice recovers the corresponding Friedmann equations for a  homogeneous and isotropic universe as
\begin{align}\label{normal-friedmann}
    3H^2 = \rho_m  + \rho_{\rm DE} - \frac{\kappa}{a^2}.
\end{align}
%\ssa{check it now}
%\adlcd{Once again a misconception in the previous statement: It is for non-spatially flat ($k\neq0$) FLRW metric where the GR Friedmann equation has a $-\Omega_k^0/a^2$ term, not in LCDM. One can have LCDM in spatially-flat FLRW and then there is no such a term.} 
%

%
%
%\adlcd{There is no need to copy the equation above.}
%We are assuming a non-interacting system of two fluids; Dark energy and Dark matter, with $w^m=0$. 
%Thus, \adlcd{{\it Thus} is not the right connector here. It seems that conservation equations need LCDM to be true.} 

\noindent Accordingly, the conservation equation in each fluid (for matter, dark energy, and shear) can be re-expressed in  compact form as
\begin{equation}\label{rhomEq}
\dot{\rho}_i +3H(1+w_i)\rho_i =0\;,
\end{equation}
%\adlcd{I mentioned that already} 
where $\rho_i$ represents the energy density for matter, DE, and shear fluid ($\rho_\sigma \equiv \sigma^2/a^6$), and the corresponding EoS parameters $w_i = w_m = 0$ for matter fluid, $w_i = w_{\rm DE}$ for DE, and $w_i = w_\sigma = 1$  for the shear fluid, respectively. Integrating Eq. \eqref{rhomEq} gives $\rho_m  \propto a^{-3}$, $\rho_\sigma  \propto a^{-6}$ (which seems an effective stiff fluid)  and $\rho_{\rm DE}  \propto a^{-3(1+w_{\rm DE})}$, where $a$ is the average scale factor of the universe.
% \footnote{ The conservation equations: for matter fluid \begin{equation}
% \dot{\rho}^m +3H(1+w^m)\rho^m =0\;,
% \end{equation} for dark energy \begin{equation}
% \dot{\rho}^{de} +3H(1+w^{de})\rho^{de} =0\;,
% \end{equation} and for shear fluid becomes \begin{equation}
% \dot{\rho}^\sigma +3H(1+w^{\sigma})\rho^\sigma =0\;.
% \end{equation}}
As mentioned earlier, we consider the parametrization for the dark energy EoS: $w_{de} = w_0 + w_a(1+a)$ as discussed in \cite{moffat2025dynamical,linder2003exploring, du2025cosmological,oliveira2025omega0omegaacdm}. 
The general form of the Friedman equation in the parametrization for the dark energy EoS takes the form\footnote{The  Friedman equation for $\Lambda$CDM is recovered for the case of $w_a = 0 = \Omega_\sigma, w_0 = -1$  and  Bianchi Type-V, for the case of $w_a = 0, w_0 = -1$;  $w$CDM is recovered provided $w_a = 0 = \Omega_\sigma, w_0 \neq -1$  and  $w$Bianchi Type-V, for the case of $w_0 \neq -1, w_a = 0$.  Finally   $w_0w_a$CDM  is recovered whenever $w_{0,a} \neq0, \Omega_\sigma = 0$.  }
\begin{align}\label{Friedmann}
H^2(z) =& H_0^2\bigg[ \Omega_{m0} (1+z)^3 + \Omega_{\kappa 0} (1+z)^2 
+ \Omega_{\sigma0}(1+z)^6\nonumber\\&+\Omega_{\rm DE0} (1+z)^{3(1 + w_0 + w_a)} {\rm exp}{\left(\frac{-3w_a z}{1+z}\right)} \bigg]\;.
\end{align}
\\
\\
The CMB and large‑scale structure indicate that any present‑day anisotropy must be very small, so that the late‑time universe is well approximated by an almost‑FLRW geometry. In this regime, even Bianchi-V cosmologies admit an effective average scale factor and Hubble rate, with the shear and curvature contributions entering as small corrections to the background expansion. Consequently, the standard FLRW definitions of comoving, luminosity, and angular‑diameter distances remain applicable, while anisotropies affect them only indirectly through the modified Hubble function and the effective curvature term. This justifies using the same late‑time distance indicators as in FLRW analyses - such as those inferred from BAO, Type Ia supernovae, and cosmic chronometers - while explicitly tracking how a non‑zero shear fluid and negative spatial curvature alter predictions for $D_A(z),D_M(z),D_V(z)$ and $D_H(z)$ at low and intermediate redshifts. A similar approach has also been considered in the recent works \cite{jalalzadeh2024observational,sarmah2025observational,prasad2021exact}.  In this context, the angular diameter distance, the radial BAO distance, and the volume-averaged BAO distance measurements can be given by
\begin{eqnarray}\label{DA}
&&D_A(z) = \frac{D_M(z)}{1+z}\;,\\
&&\label{DH}
D_H(z) = \frac{c}{H(z)}\;,\;{\rm and}\\
&&\label{DV}
D_V(z) =\left[ D_M^2(z)\,\frac{c z}{H(z)} \right]^{1/3} \;,
\end{eqnarray}
respectively. The sound horizon at the drag $r_d$ epoch is given by 
 \begin{equation}
r_d=\int_{z_d}^{\infty}\frac{c_s(z)}{H(z)}{\rm d}z~,   
 \end{equation}
 where $z_d$ is the redshift at drag epoch and $c_s(z)$ is the sound speed of the photon-baryon fluid.
%  See my paper \url{https://doi.org/10.1140/epjc/s10052-020-7961-3}
% \begin{eqnarray}
%     \dot{w_{X}} = (w_{X}+1)(w_{X}-c_s^2)
% \end{eqnarray}
\section{Perturbation equations}\label{perturbations} 
This section employs the 1+3 covariant and gauge-invariant perturbation formalism to derive the evolution equations of the matter density contrast. This formalism, developed initially in \cite{ hawking66, ellis1989covariant,ellis1990density,dunsby1992cosmological}, examines perturbations and structure formation of the universe. Later in the work \citep{abebe2012covariant,abebe2013large,ntahompagaze2018study,sahlu2020scalar,sahlu2025structure} in different aspects of modified gravity theory. By taking into account the 1+3 approach, we study the structure growth of the universe in the anisotropic Bianchi type-V universes. For further details on the 1 + 3 covariant formalism and its usefulness, see \citet{bruni1992gauge,de2015constraining,sahlu2025constraining,sahlu2023confronting}.   As presented in \citep{ellis1989covariant}, the Raychaudhuri equation regulates the dynamics of expansion in a cosmic context and is essential for understanding the genesis of singularities; it is expressed as follows:
% \begin{equation}
%     \dot{\Theta} = -\frac{1}{3}\Theta^2 -2\sigma^2 - \Lambda - \frac{1 }{2}(1+3w) \rho_m + \tilde{\nabla}^{a} \dot{u}_i\;.
% \end{equation}
\begin{equation}
    \dot{\Theta} = -\frac{1}{3}\Theta^2 -2\sigma^2 - \frac{1 }{2}(1+3w_{de}) \rho_{de} - \frac{1 }{2} \rho_m + \tilde{\nabla}^{a} \dot{u}_i\;.
\end{equation}
Following \cite{ellis1989covariant,dunsby1992cosmological}, we define the covariant gauge-invariant gradient variables that characterize perturbations for matter energy density $D^m_a$,  volume expansion $Z_a$, and the shear contributions $\Sigma_a$ with the 1 + 3 covariant formalism as
\begin{align}\label{XX}
D^m_{a}\equiv\frac{a\nabla_{a}\rho_m}{\rho_m}\;,\quad  Z_{a}\equiv a\nabla_{a} \Theta\;,\quad\mbox{and}\quad  \Sigma^\sigma_{a}\equiv a\nabla_{a}\sigma\;.
\end{align}
As described in \cite{ellis1990density, dunsby1992cosmological}, for a non-interacting fluid, the conservation equations {for each fluid} are given as
\begin{align}
&\dot{\rho}_{i} = -\Theta (1+w)\rho_i + (1+w)\rho_i \tilde{\nabla}^a \Psi_a \;
\\ &(1+w)\rho_i\dot{u}^a  = -\tilde{\nabla}_a p_i +\dot{\Psi}_a + (3c^2_s-1)\frac{\Theta}{3} \Psi_a - \Pi_a 
\end{align}
where $\Psi_a = \frac{q_a}{(1+w)\rho_i}$ and $\Pi_a = \frac{\tilde{\nabla}^b\pi_{ab}}{(1+w)\rho_i}$ stand for the heat flux ($q_a$) and 
the stress energy-momentum tensor ($\pi_{ab}$), respectively. Also, the sound speed $c_{s,i}^2 = \delta p_i / \delta \rho_i$ is significant because it correlates the perturbed pressure with the perturbed energy density for each fluid, whereas the time derivative of the fluid EoS satisfies $\dot{w}_i = (w_i + 1)(w_i - c_{s,i}^2)$. For non-interacting fluids, we shall assume that the equation-of-state parameter is time independent, hence $\dot{w}_i = 0$, see the work  \citep{ballesteros2010dark} for further. In this situation, the sound speed is equal to the equation-of-state parameter, $w_i = c_{s,i}^2$.
\\
\\From Eq. \eqref{XX}, the  corresponding first-order time derivative of these gradient variables  for matter and shear fluids are given by the following equations:
\begin{align}
\dot{D}^m_a  & - w\Theta D^m_a %- Y_a
+ (1+w) Z_a=0~,\\ 
\dot{Z}_a  & + \frac{2}{3}\Theta Z_a +
\bigg[ \frac{1}{2}(1+3w) \rho_m + \frac{w\dot{\Theta}}{(1+w)}  + 
\frac{w\rho_m }{(\rho_m+p_m)} \nabla{^2} \bigg] D^m_a \nonumber\\&+ 4\sigma \Sigma_a  %+ X_a
=0\;, \\
\dot{\Sigma}_a & + \Theta \Sigma_a + \sigma Z_a -  \frac{w}{(1+w)}  \sigma  \Theta D^m_a 
%- \sigma^b_a \Sigma_b
=0\;.
\end{align}
Since the scalar quantities are responsible for the formation of large-scale structures, the so-called scalar decomposition technique, which is broadly discussed in  \cite{ellis1989covariant, dunsby1992cosmological, abebe2012covariant,abebe2013large,sahlu2025structure}, is pertinent at this stage. Hence, we define the following scalar quantities %\adlcd{ The latter is redundant here}\ssa{}
\begin{align}
\label{scalardefination}
 \Delta_m \equiv a \nabla^{a} D^m_a\;, \quad Z\equiv a\nabla^{a}Z_a\;,\quad \mathcal{S}\equiv a \nabla^{a}\Sigma_a\;.
\end{align}
By employing the definitions in \eqref{scalardefination}, the evolution equations in these scalar variables  are
\begin{align}
&\dot{\Delta}_m-  w\Theta \Delta_m +(1+w) Z = 0\;,\\&
\dot{Z}  + \frac{2}{3}\Theta Z +
\bigg[\frac{1}{2}(1+3w)\rho_m + \frac{ w}{(1+w)}\left(\dot{\Theta}+\frac{2\alpha^2}{a^2}\right)   + \frac{w}{(1+w)}\nabla^{2} \bigg]\Delta_m  \nonumber\\& +4\sigma \mathcal{S}=0 \\&
\dot{\mathcal{S}}  + \Theta \mathcal{S} + \sigma Z  - \frac{w}{(1+w)}   \sigma \Theta \Delta_m =0\;,
\end{align}
%\adlcd{In Eq. (39) above there is something missing to join the first and the second line}

By performing now the usual harmonic Fourier space transformation \citep{ellis1989covariant, dunsby1992cosmological, abebe2012covariant,abebe2013large,sahlu2025structure}, the evolution of the perturbations in the $k^{\rm th}$ mode yield
\begin{align}
  &\dot{\Delta}^k_m -
w \Theta \Delta^k_m + (1+w) Z^k =0\;,\label{Deltakdt}\\
& \dot{Z}^k + \frac{2}{3}\Theta Z^k 
+\bigg[ \frac{1}{2}(1+3w)\rho_m +    \frac{w}{(1+w)}\left(\dot{\Theta}+ \frac{2\alpha^2 -k^2}{a^2}\right)\bigg]\Delta^k_m 
\nonumber\\& +  4\sigma \mathcal{S}^k=0\;,\label{Zkdt}\\&
\dot{\mathcal{S}}^k+ \Theta \mathcal{S}^k + \sigma Z^k  
-  \frac{w}{(1+w)} \sigma \Theta \Delta^k_m=0\;.\label{Sigmakdt}  
\end{align}
Furthermore, by differentiating the system once again with respect to time, we obtain \eqref{Deltakdt}-\eqref{Sigmakdt} expressed as follows: 
    \begin{align}
& \ddot{\Delta}^k_m  =    \left(w\Theta -\frac{2}{3}\Theta + \frac{\dot{w}}{w+1} \right) \dot{\Delta}^k_m + \Big[ \frac{2}{3} 
w\Theta^2  + \frac{1}{2}(1+3w)(1+w)\rho_m  \nonumber\\&    +2 w \dot{\Theta} + w\frac{2\alpha^2 -k^2}{a^2}    + \frac{\dot{w}_m}{w+1} \Theta \Big] \Delta^k_m + 4(1+ w) \sigma  \mathcal{S}^k~, \label{second1}  \\
&\ddot{\mathcal{S}}^k  = (-\dot{\Theta} + 4\sigma^2) \mathcal{S}^k - \Theta 
\dot{\mathcal{S}}^k - \frac{\sigma \Theta }{1+w} (5/3 - w) \dot{\Delta}^k_m  \nonumber\\& +
 \sigma  \bigg\{ \frac{1}{2}(1+3 w) \rho_m  + \frac{w}{(1+w)} \bigg[ 2\dot{\Theta} + \frac{2}{3} \Theta^2  + \frac{2\alpha^2 -k^2}{a^2} \bigg] \bigg\}\Delta^k_m ~.\label{second2}
\end{align}
By assuming the matter component makes a significant contribution to structure formation, $w= 0$, and the adiabatic perturbation dynamics $\dot{w}_m = 0$.  Then Eqs. \eqref{second1} and \eqref{second2} reduce, respectively, to 
\begin{align}
&\ddot{\Delta}_m =    -\frac{2}{3}\Theta \dot{\Delta}_m + \frac{1}{2}\rho_m \Delta_m + 4\sigma  \mathcal{S}~, \label{xxx} \\
&\ddot{\mathcal{S}}  = \left[3H^2 +2\sigma^2 + \frac{1 }{2}(1+3w_{\rm DE}) \rho_{\rm DE}  + \frac{\rho_m}{2} + 4\sigma^2\right] \mathcal{S}\nonumber\\& - \Theta 
\dot{\mathcal{S}} - \frac{5\sigma \Theta }{3} \dot{\Delta}_m +
 \frac{\sigma \rho_m }{2} \Delta_m ~.\label{xx}
\end{align}
% For further investigation of the growth structure, we transform the time-dependent evolution equation to redshift space as presented in \citep{sahlu2025structure}. For any time derivative function $Q(t)$, it will be transformed into redshift space via the following relation:
% \begin{align}
% &\dot{Q}(t)= -(1+z)H(z) Q'(z)~, \nonumber\\& \ddot{Q}(t)=  (1 + z)H^2(z) Q'(z) + (1 + z)^2 H^2(z) Q''(z)   \nonumber\\& + (1 + z)^2 H(z) H'(z) Q'(z)~, 
% \end{align}
%\adlcd{Q1. What is $\Lambda$ above?}
%
In what follows,  we introduce the fractional matter density fluctuations $\delta_m(z)$, %,  as widely considered in other literature \citep{sahlu2025structure,sahlu2020scalar} 
\begin{align}
    \delta_m(z) = \frac{\Delta_m(z)}{\Delta_m(z_{in})} \;.
\end{align}
% This indicates that in the linear cosmological theory, valid at sufficiently early times and sufficiently large spatial scales, when the initial fluctuations in the matter energy density $\Delta_m(z_{in}) = 10^{-5}$ are much less than unity, the density fluctuations $\Delta_m(z)$ evolve independently of the spatial scale $\kappa$, see \citep{abebe2013large}.
% \begin{align}
% & (1 + z)H^2\Delta_m'(z) + (1 + z)^2 H^2 \Delta_m''(z) + (1 + z)^2 H H' \Delta_m'(z) =  \nonumber\\&  2(1+z)H^2 {\Delta}'_m + \frac{1}{2}\rho_m \Delta_m + 4\sigma  \mathcal{S}~, \label{xxx} \\
% &(1 + z)H^2\mathcal{S}'(z) + (1 + z)^2 H^2 \mathcal{S}''(z) + (1 + z)^2 H H' \mathcal{S}'(z) =  \nonumber\\&   = (3H^2 +2\sigma^2 + \Lambda + \frac{\rho_m}{2} + 4\sigma^2) \mathcal{S} +3(1+z) H^2 
% {\mathcal{S}}' \nonumber\\&+5(1+z) H^2\sigma{\Delta}'_m +
%  \frac{\sigma \rho_m }{2} \Delta^k_m ~.\label{xx}
% \end{align}
%%%%%%%%%%%%%%%%%%%%%%%%%%%%%%%%%%%%%%%%%%%%%%%%%%%%%%%%%%%%%%%%%%%%
Transforming the equations into redshift space, as presented in \citep{sahlu2025structure}, the second-order evolution Eqs. \eqref{xxx} and \eqref{xx} are given by\footnote{where $\Omega_m = \frac{\Omega_{m0}}{h^2(z)}(1+z)^3\,,\; \Omega_\sigma = \frac{\Omega_{\sigma0}}{h^2(z)}(1+z)^6\,, \Omega_{\rm DE} = \frac{\Omega_{\rm DE0}\mathcal{F}(z)}{h^2(z)}\;,$  $\mathcal{F}(z) = (1+z)^{3(1 + w_0 + w_a)} {\rm exp}{\left(\frac{-3w_a z}{1+z}\right)}$, $h(z) = H(z)/H_0$, and $'$ holds for the derivative with respect to redshift.}
 \begin{align}
&(1+z)^2\,\delta_m''(z)+ (1+z)\bigg[(1+z)\frac{h'}{h} - 1\bigg]\,\delta_m'(z)
- \tfrac{3}{2}\Omega_{m}\,\delta_m(z)\nonumber\\&
- \frac{4\sigma}{H_0^2h^2}\,\mathcal{S}(z)
= 0~, \label{xxx_collected}\\
&(1+z)^2\,\mathcal{S}''(z)
+ (1+z)\bigg[(1+z)\frac{h'}{h} - 2\bigg]\,\mathcal{S}'(z)
- \frac{1}{h^2}\Bigg(3+ 3\Omega_{\rm DE}  \nonumber\\& +\tfrac{3\Omega_{m}}{2} + 18\Omega_{\sigma}\Bigg)\,\mathcal{S}(z)
- 5(1+z)\sigma\,\delta_m'(z)
- \tfrac{\sigma 3\Omega_{m}}{2}\,\delta_m
= 0~\;, \label{xx_collected}
\end{align} 
The above Eqs. \eqref{xxx_collected} and \eqref{xx_collected} represent  
coupled second-order ordinary differential equations of the growth of the matter contrast, $\delta_m(z)$. 
In most cases, different works  \citep{ntahompagaze2018study,abebe2012covariant,abebe2013large,sahlu2020scalar} consider the quasi-static approximations to reduce the system of equations to a single, easily solvable equation for matter growth. This motivation is also discussed in the recent work in \citep{sahlu2025structure, sahlu2025constraining,sahlu2026observational}. By taking into account this approach,  we adopted this approximation, wherein the first- and second-order derivatives of the anisotropic variable $\mathcal{S}$ are neglected, i.e., $\ddot{\mathcal{S}}, \dot{\mathcal{S}} \approx  0$, 
the matter density fluctuations evolution equations given as
\begin{align}\label{growth2}
 (1 + z)^2\delta ''_m  =
 -(1+z)\left[ 5 \Psi -1  + (1 + z) \frac{h'}{h} \right]  \delta'_m  + \frac{3}{2}\Omega_m\left( 1-\Psi \right) \delta_m
\end{align}
where,  
\begin{eqnarray}
\Psi = \frac{4\Omega_\sigma  }{ 1+6 \Omega_\sigma + \Omega_{de} + 0.5  \Omega_m }.
\end{eqnarray}
\begin{figure}
    \includegraphics[width=1.1\linewidth]{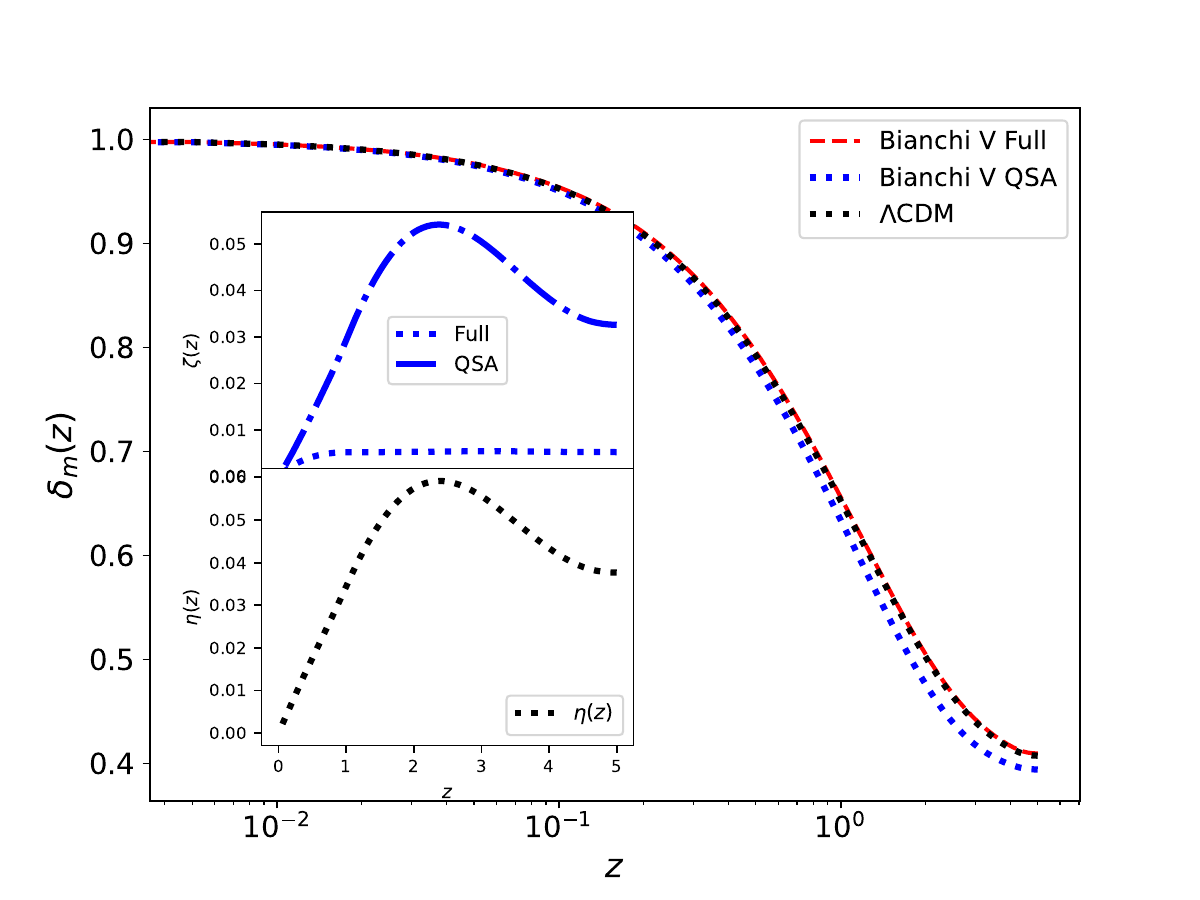}
    \caption{Density contrasts evolution as per the resolution of the full system Eqs. \eqref{xxx_collected} - \eqref{xx_collected} and the quasi-static approximation \eqref{growth2}. The inner panels represent  the evolution of $\zeta(z)$ Eq. \eqref{varaince} and $\eta(z)$ Eq. \eqref{relative}. For illustrative purposes we use paradigmatic values of $\Omega_{m0} = 0.315$, $\Omega_{k0} = 0.045$, $\Omega_{\sigma0} = 10^{-4}$.   }
    \label{fig:q1}
\end{figure}
% \adlcd{$\Omega_{\Lambda}$ is not the quantity to appear in the previous equation}
However, before considering the quasi-static approximation for further analysis, we take into account investigating the relative difference of $\delta(z)$ between  $\Lambda$CDM  and the Bianchi type-V model by defining the dimensionless parameter $\zeta(z)$ as \citep{sahlu2025structure}
\begin{eqnarray} \label{varaince}
    \zeta (z) = \bigg| \frac{\delta^{\rm\Lambda CDM}(z)-\delta^{\rm BIV}_{\rm {full,QSA}}(z)}{\delta^{\rm  \Lambda CDM}(z)}\bigg|\;\;,
\end{eqnarray}
where $\delta^{\rm BIV}_{\rm {full}}$ and $\delta^{\rm BIV}_{\rm {QSA}}$ refer to the density contrast of the model for Bianchi-V (BIV) obtained for the full system (full) or quasi-static (QSA) approximation, respectively, since $\zeta(z)$ helps us to determine how the full system and the quasi-static system depart from the  $\Lambda$CDM predictions.  
Similarly, to compare the quasi-static findings with the full system results, we introduce the dimensionless parameter   
\begin{eqnarray}\label{relative}
    \eta(z) = \bigg|\frac{\delta
    ^{\rm BIV}_{\rm {QSA}}(z)-\delta^{\rm BIV}_{\rm {full}}(z)}{\delta^{\rm BIV }_{\rm {full}}(z)}\bigg|\;,
\end{eqnarray}
and $\eta(z)$ determines how the quasi-static approximation is effective. The numerical results of the density contrast for the full system evolution and the quasi-static together with pertinent relative differences $\zeta (z)$ and $\eta(z)$ as per Eqs. the \eqref{varaince} and \eqref{relative}, respectively,  are presented in Fig. \ref{fig:q1}. 
From this plot, we observe only minimal deviations between the full system and the quasi-static approximation across the redshift in the range  $z = 0~\text{to}~ 5$. {Particularly at late times ($z<1$), the results of density contrast are overlapping, 
the relative difference remains below approximately 3\%, indicating excellent agreement}; henceforth, we consider the quasi-static approximation for further investigations in the current work. 
\\
\\
In the GR approach, in the linear regime $\delta_m\ll1$, the linear growth of fluctuations as a function of time is straightforward, indicating the composition and expansion rate of the Universe, while $\delta_m \ge1$ indicates gravitational collapse due to a gravitational instability into bound structures like galaxies.   As widely discussed in \citep{linder2003cosmic}, the growth factor $\mathcal{\bar{D}}(a)$ is the ratio of the perturbation amplitude of the density contrast at some scale factor relative to some initial scale factor 
\begin{eqnarray} \label{growthfactor}
    \mathcal{\bar{D}}(a) = \frac{\delta_m(a)}{\delta_m(a = 1)}\;.
\end{eqnarray}
The normalized density contrast $\delta_m(a)$ plays a key role in cosmic structure formation. It starts
small, growing through gravitational instability to form galaxies and clusters. Using  Eq. \eqref{growthfactor} in Eq. \eqref{growth2}, the evolution of the growth factor  is given as
\begin{align}
(1 + z)^2\mathcal{\bar{D}} ''_m  =
 -(1+z)\left[ 5 \Psi -1  + (1 + z) \frac{H'}{H} \right]  \mathcal{\bar{D}}'_m  + \frac{3}{2}\Omega_m\left( 1-\Psi \right) \mathcal{\bar{D}}_m\;. \label{densitycontarstistopic1}
\end{align}
%  Now, the redshift space transform second-order evolution equation is given by
% \begin{align}\label{growth2}\ssa{}
%  (1 + z)^2\mathcal{\bar{D}} ''_m  =
%  -(1+z)\big\{ 5 \Psi -1  + (1 + z) \frac{H'}{H} \big\}  \mathcal{\bar{D}}'_m  + \frac{3}{2}\Omega_m\big\{ 1-\Psi \big\} \mathcal{\bar{D}}_m
% \end{align}
%  \sout{It is often normalized to today's value of the density contrast $\delta(z =0) = 1$ and is governed by a differential equation involving the Hubble parameter and
% matter density. This factor shows how initial density perturbations grow over time due to gravity, influencing the
% formation of large-scale structures.}\adlcd{The previous sentence is not needed in an article.}\adlcd{Q. Where in the text Figure 1 is mentioned? One cannot have a figure which is not mentioned in the text.} 
Then, the growth rate $f(a)$ is defined as 
\begin{align}
    f(a) = \frac{{\rm d}\ln(\mathcal{\bar{D})}}{{\rm d}\ln(a)}\;, %= -(1+z)\frac{{\delta'}(z)}{{\delta}(z)} 
\end{align}
quantifies the growth of cosmic structures, which is sensitive to the derivative of the logarithm of the growth function with respect to the logarithm of the cosmic scale. In a straightforward calculation,  we obtained the first-order derivative of the growth rate $f(z)$ from equation \eqref{growth2} as given by 
\begin{align}
 &(1+z)f'= f^2 -\left[ 5 \Psi -2  + (1 + z) \frac{h'}{h}  \right]f  - \frac{3}{2}\Omega_m\left( 1-\Psi \right)\,. 
\end{align} 

%\adlcd{Any comment on Figure 1? Is the QSA approximation a good approximation?}

In the next sections, we provide a detailed statistical analysis of the  $\Lambda$CDM,  $w$CDM, and  $w_0w_a$CDM together with the Bianchi Type V model, $w$Bianchi Type V model, and $w_0w_a$Bianchi Type V models in order to assess the statistical evidence for DDE resorting to measurements sensitive to either cosmic expansion history and growth of large-scale structures.
%\adlcd{DDE acronym is used, nonetheless in Section 4 once again it is not used and we say "dynamical dark energy" in several places. Please check carefully using a printed copy and make annotations with pen.}

 % \adlcd{Q. What is the $T_{\mu\nu}$ in the latter Bianchi Type V model that you have jsut mentioned? I refer to my previous comment on the confusion of the different cases that are studied}\ssa{I answered this part in Sec. 2, please look at it again.}
 
\section{Results and discussion}\label{resultanddiscussion}
In order to constrain the values of the cosmological parameters 
we employ the Python libraries, including EMCEE \cite{foreman2013emcee, hough2020viability,hough2026kosmulator} and GetDist \cite{lewis2019getdist}
by resorting to the following cosmological data sets:
\begin{enumerate}
    \item   BAO distances and correlation measurements from Data Release 2 (DR2) \cite{andrade2025validation, abdul2025desi} of the Dark Energy Spectroscopic Instrument (DESI) Survey, we refer to this dataset as \textit{DESI DR2 BAO}. \\
\item {Supernovae Type Ia (SNIa)}  dataset compilations, which we have considered as follows: 
\begin{itemize}
    \item \textit{PPS}: we use the SNIa distance moduli measurements from the Pantheon+ sample \cite{brout2022pantheon+}, consisting of 1701 light curves of 1550 distinct SNIa in the redshift range $z \in [0.001, 2.26]$, we refer to this dataset as \textit{ PantheonP + SH0ES}.
    \item \textit{DESY5} data \cite{collaboration2024dark}, a photometrically classified SNIa with redshifts in the range $0.1 < z < 1.13$, complemented by 194 historical low-redshift SNe Ia (also present in the $PPS$ sample), spanning $0.025 < z < 0.1$,  we refer to this dataset as \textit{ DESY5}.
\item \textit{Union3}: we consider the latest Union compilation of 2087 cosmologically useful SNIa from 24 datasets \cite{rubin2023union}. We refer to this dataset as \textit{ Union3}.
\end{itemize}

\item Hubble parameter $H(z)$ measurements, derived from observational Hubble parameter data. This comprises 30 data points obtained from the relative ages of massive, early-time, passively evolving galaxies, known as cosmic chronometers (CC) \cite{moresco2020setting,qi2023model}. We refer to this dataset as \textit{ CC}.\\

\item The redshift-space distortion data, labeled {RSD}, from the VIMOS Public Extragalactic Redshift Survey (VIPERS) and SDSS collaborations. A total of 66 data points for measurements of redshift-space distortion of ${f}\sigma_8$ have been collected and summarized in \cite{kazantzidis2018evolution,skara2020tension}, covering the redshift interval $0.001 \leq z \leq 1.944$.  We refer to this dataset as \textit{ RSD}.
%\adlcd{ Important: why don't we use $\{f, \sigma_8\}$ separate data?}\ssa{}
\end{enumerate}
To obtain more robust constraints on the cosmological parameters, and considering the inconsistency between the \textit{PantheonPlus +SH0ES}  and \textit{DESI DR2 BAO}  datasets, as presented in \cite{afroz2025hint}, which indicates a violation of the distance duality relation, we shall resort to the following combined-data analyses:
\begin{enumerate}
\item[{\it i)}] \textit{DESI DR2 BAO + CC + RSD}, 
\item[{\it ii)}] \textit{DESI DR2 BAO + CC + Union3 + RSD},  
\item[{\it iii)}] \textit{DESI DR2 BAO + CC + DESY5 + RSD}, and 
\item[{\it iv)}]  \textit{ PantheonP + SH0ES  + CC + RSD},
\end{enumerate}
in an effort to improve the precision and constraints on the models under study, allowing for a more comprehensive picture of the Universe. 
 {Subsequently, we have been using the combined datasets mentioned above to constrain different cosmological parameters: $H_0$, $\Omega_{m0}\;, \Omega_{k0} \;,r_d\;,  M_{bs}$, $S_8$, and $\Omega_{\sigma0}$. Table \ref{tab1:cosmo_constraints} presents such results for $\Lambda$CDM and Bianchi Type V; whereas in Table \ref{tab2:cosmo_constraints} we do it for $w$CDM and $w$Bianchi-V;  and in Table \ref{tab3:cosmo_constraints} for $w_0w_a$CDM and $w_0w_a$Bianchi-V  models.

\subsection{Results for $w$ and $w_0-w_a$ values }

In the DDE models, the key parameters constrained in this paper, $w_0$ and $w_a$, are derived from the parameterized EoS,  which describes the time evolution of dark energy, where the results are summarized in Table \ref{tab2:cosmo_constraints}  -\ref{tab3:cosmo_constraints} for the case  $w_{de}= w$  and  $w_{\rm DE} = w_0 + w_a(1 - a)$) respectively.  For instance: 

i) For the  $w$CDM model, the MCMC contour plots are also presented in Fig.~\ref{fig:x1}, and the plot clearly shows the posterior distributions of $w>-1$  at 68\%  C.L. for all joined datasets with the sole exception of \textit{PantheonP + SH0ES + CC + RSD}; 

ii)  For the $w_0w_a$CDM model, we find that the significance of the tension with $\Lambda$CDM ($w_0 = -1$, $w_a=0$) becomes $1.89\sigma$, $2.12\sigma$, $2.73\sigma$, and $1.50\sigma$ for these four cases across each dataset (see Fig. \ref{fig:w0wacdm}); 

iii)  For the same datasets, in the $w$Bianchi model, the corresponding MCMC contour plot is shown on the right panel of Fig.~\ref{fig:wBianchi}.
From this plot, we notice that the  68\%  posterior distributions of $w>-1$  for all catalogs except  \textit{PantheonP+ CC  + RSD}.

iv)  In the same manner, for the $w_0w_a$ Bianchi Type V models (see Table \ref{tab3:cosmo_constraints}), we find that
 the significance of the tension with $\Lambda$CDM ($w_0 = -1$, $w_a=0$) becomes $1.90\sigma$, $1.86\sigma$, $1.79\sigma$, and $1.40\sigma$ (see Fig. \ref{fig:w0waBianchi}).
\\
\\
 % \adlcd{I would not say conditions, but statistical bounds, since they are not a priori conditions (priors) which have been imposed on the parameters}
 % \\
 % \\
  Overall, the results show that $w_0 > -1$ and $w_a < 0$ appear across all joint datasets, except for \textit{PantheonP + SH0ES + CC + RSD}. The EoS evolution for the DDE models under consideration is shown in Figs. \ref{fig:CPL1} and \ref{fig:CPL2} for $w_0w_a$CDM and $w_0w_a$Bianchi Type V models respectively. Our results for the parameterized EoS indicate that the evidence of DDE in the Bianchi Type V universe is consistent with recent observations \citep{adame2025desi}, making it a good candidate beyond the $\Lambda$CDM model. 
  \begin{figure}
      \includegraphics[width=1.0\linewidth]{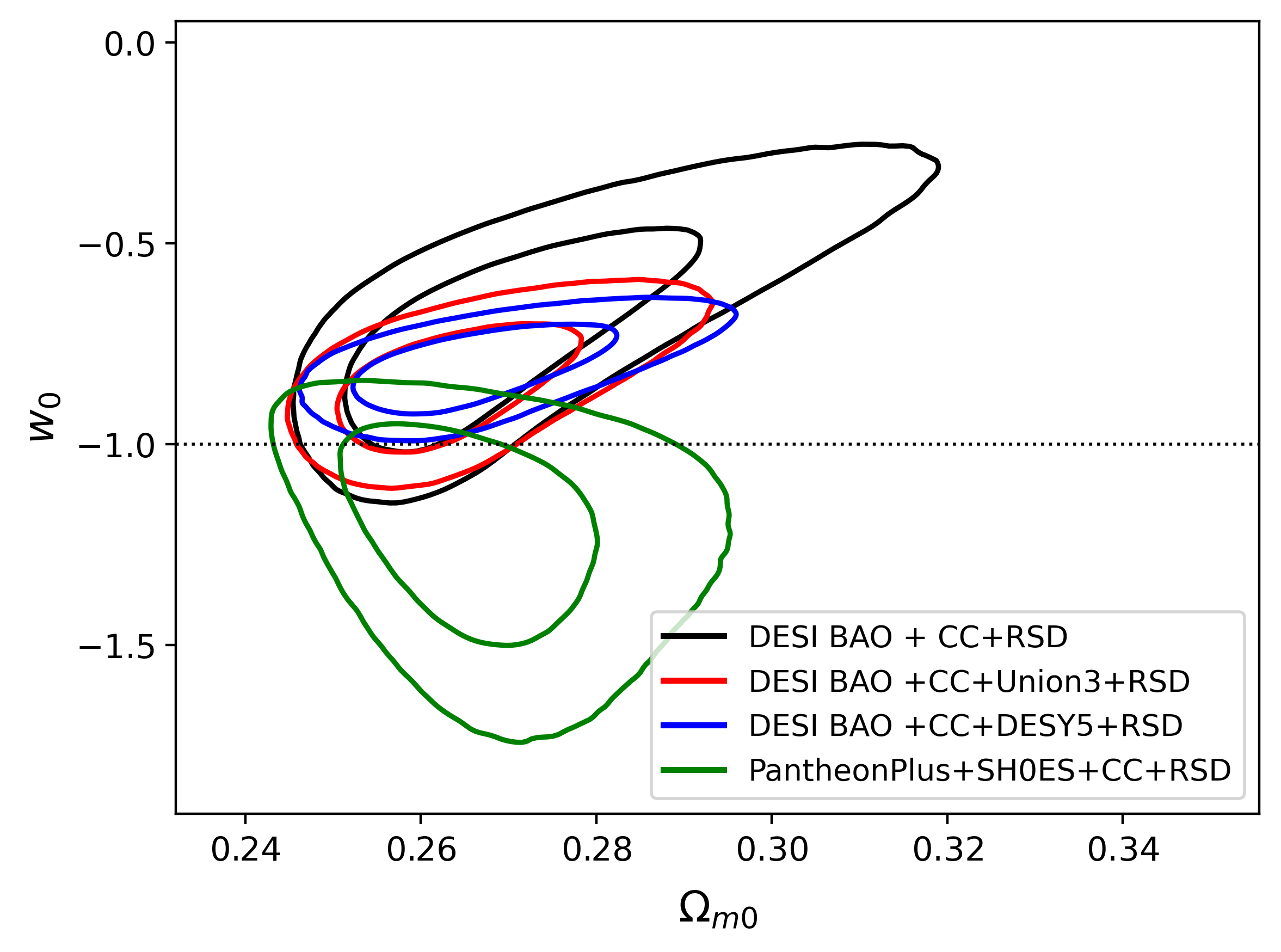 }
      \caption{ Posterior distributions of the $w$CDM model at 68\% and 95\% C.L. for the four joint datasets. The EoS parameter is $w>-1$, meaning that the model       
      consistently represents a quintessence cosmic phase at 68\% C.L., except for the combination \textit{PantheonP + SH0ES + CC + RSD}}
      \label{fig:x1}
  \end{figure}
  \begin{figure}
      \includegraphics[width=1.0\linewidth]{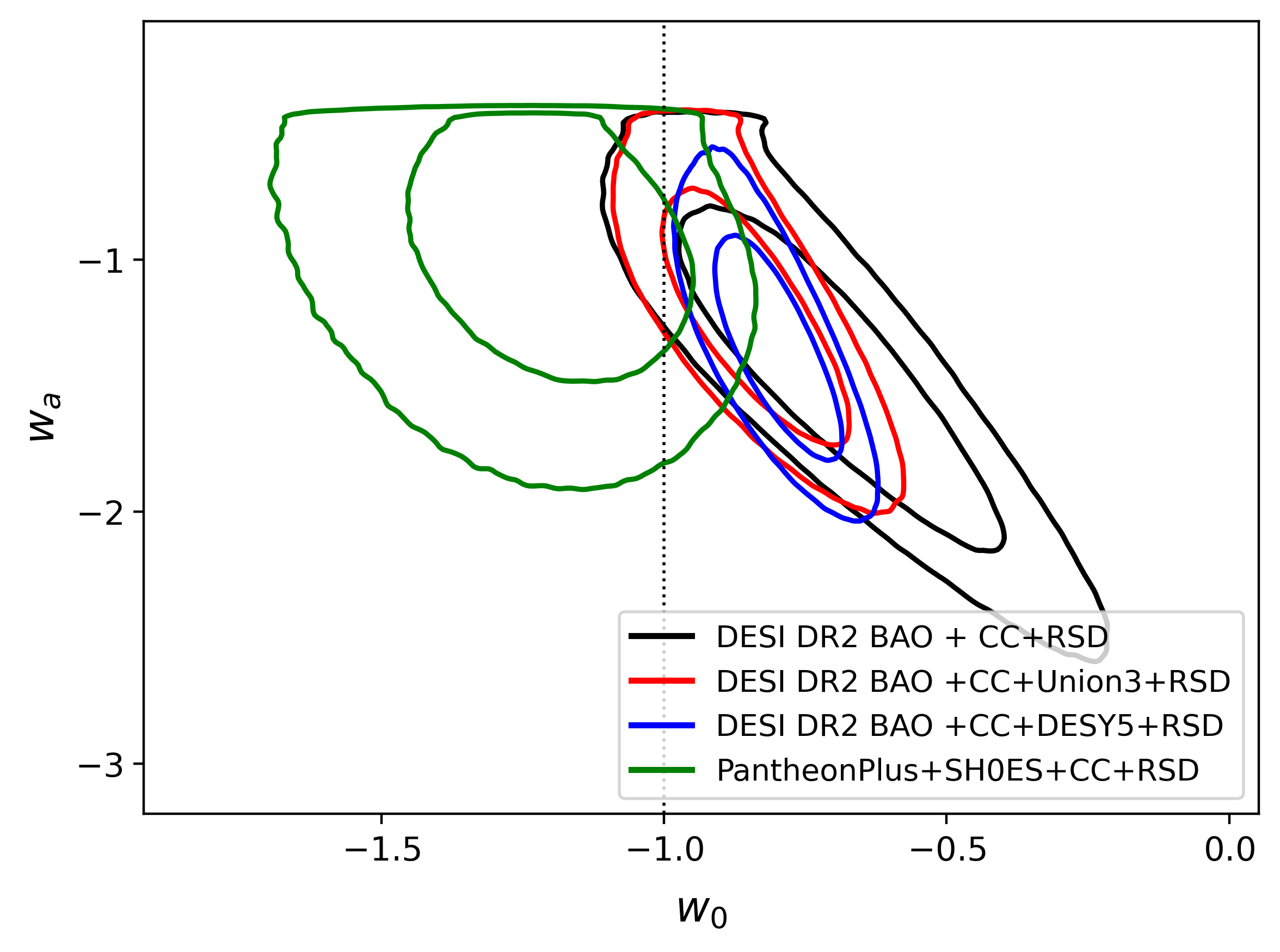}
      \caption{Posterior distributions  of the $w_0w_a$CDM model at 68\% and 95\% C.L for the four joint datasets. Since the EoS parameter satisfies $w_0>-1$, the model consistently favors the quintessence phase of the universe at 68\% C.L, with the sole exception of the \textit{PantheonP + SH0ES + CC + RSD} dataset. }
      \label{fig:w0wacdm}
  \end{figure}
   \begin{figure}
      \includegraphics[width=1.0\linewidth]{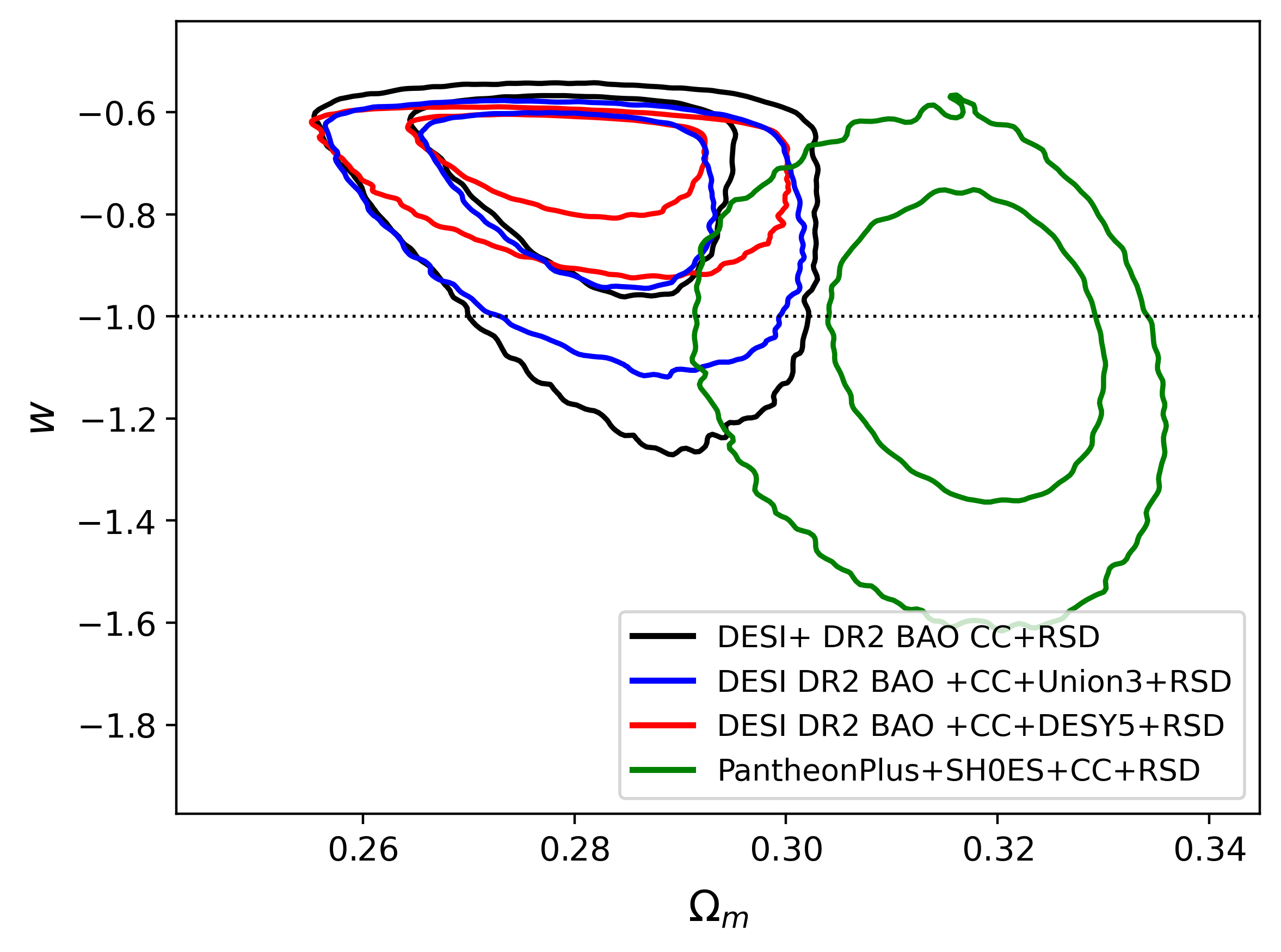}
      \caption{  Posterior distributions of the $w$Bianchi Type V  model at 68\% and 95\% C.L for the four joint datasets. Since the favored EoS parameter is $w>-1$, the model consistently represents the quintessence phase of the universe at 68\% C.L., with the sole exception of the  \textit{PantheonP + SH0ES + CC + RSD} dataset. }
      \label{fig:wBianchi}
  \end{figure}
   \begin{figure}
      \includegraphics[width=1.0\linewidth]{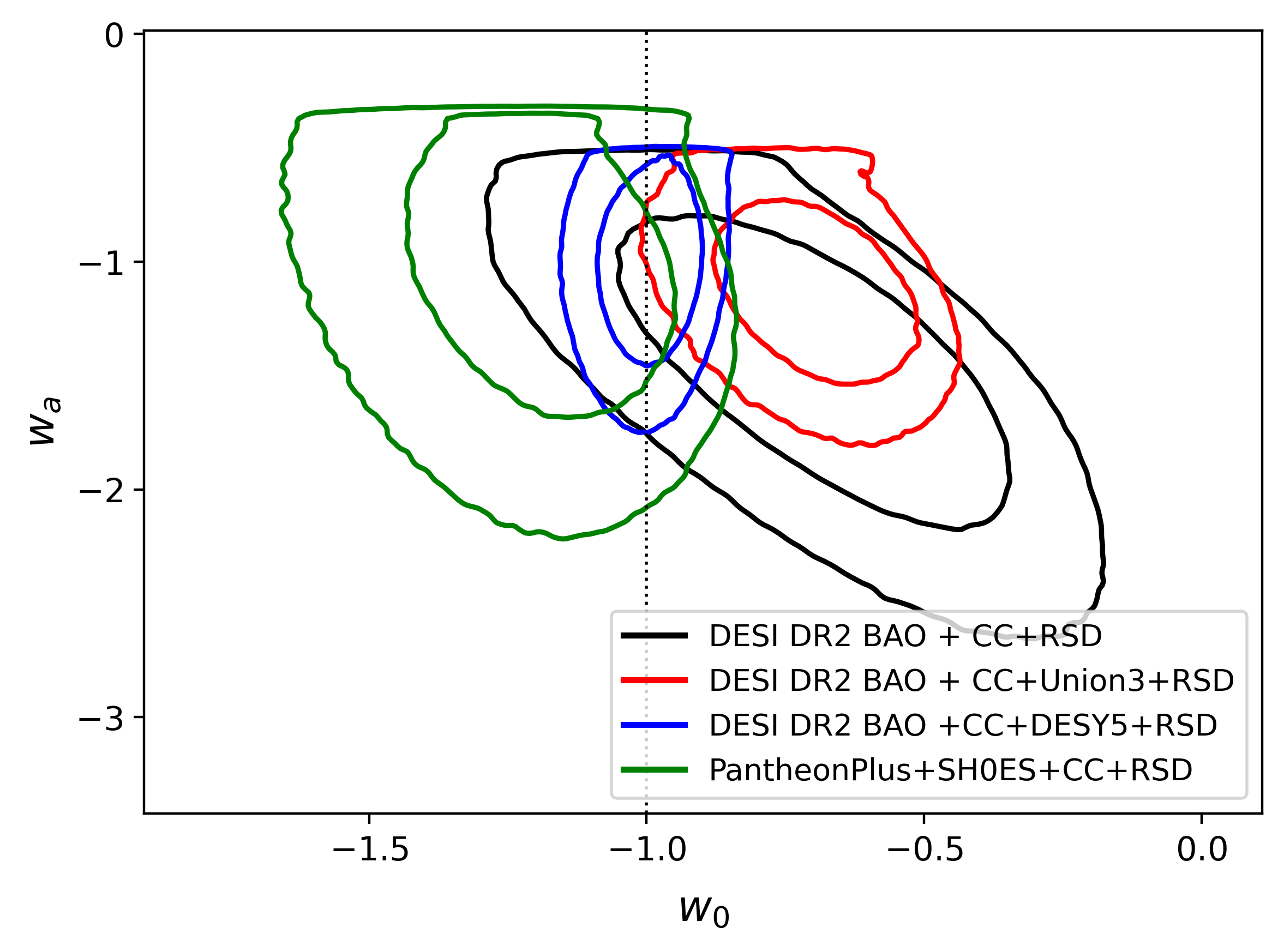}
\caption{  Posterior distributions of  the $w_0w_a$Bianchi Type V  model at 68\% and 95\% C.L for the four joint datasets. Since the favored EoS parameter is $w_0>-1$,  the model consistently represents the quintessence phase of the Universe at 68\% C.L., except for the \textit{DESI DR2 BAO + CC + DESY5 + RSD} and  \textit{PantheonP + SH0ES + CC + RSD} datasets. } 
      \label{fig:w0waBianchi}
  \end{figure}
   \begin{figure}
      \includegraphics[width=1.0\linewidth]{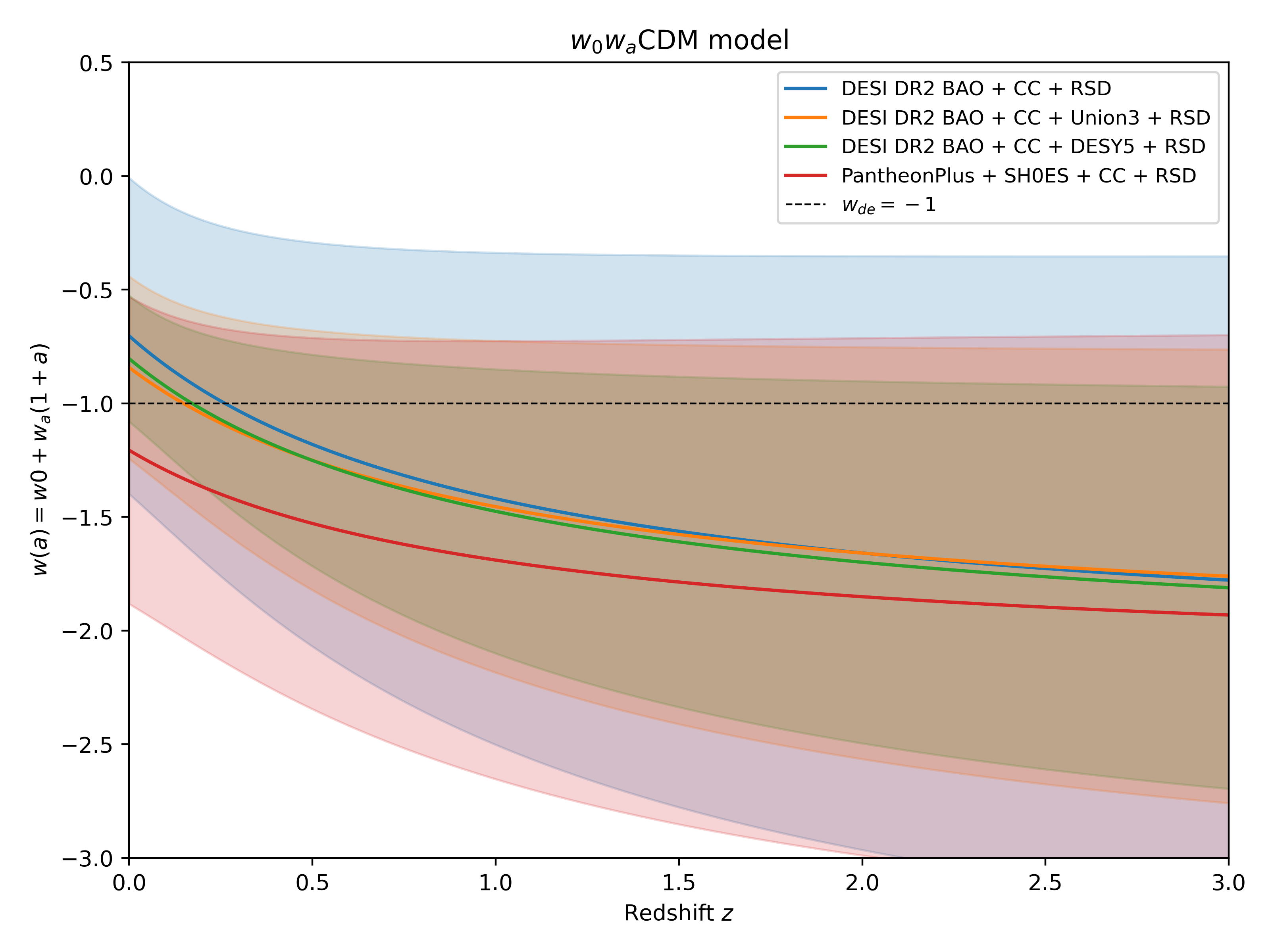}
       \caption{EoS parameter diagram for the $w_0w_a$CDM model,  where the values of the parameters are taken from Table \ref{tab3:cosmo_constraints}.  The red, blue, green, and orange curves  are represent the \textit{DESI DR2 BAO + CC +RSD}, \textit{DESI DR2 BAO + CC + Union3 +RSD} , \textit{DESI DR2 BAO + CC + DESY5 +RSD}, and \textit{PantheonP+ CC +RSD}, respectively. }
       \label{fig:CPL1}
   \end{figure}
   \begin{figure}
      \includegraphics[width=1.0\linewidth]{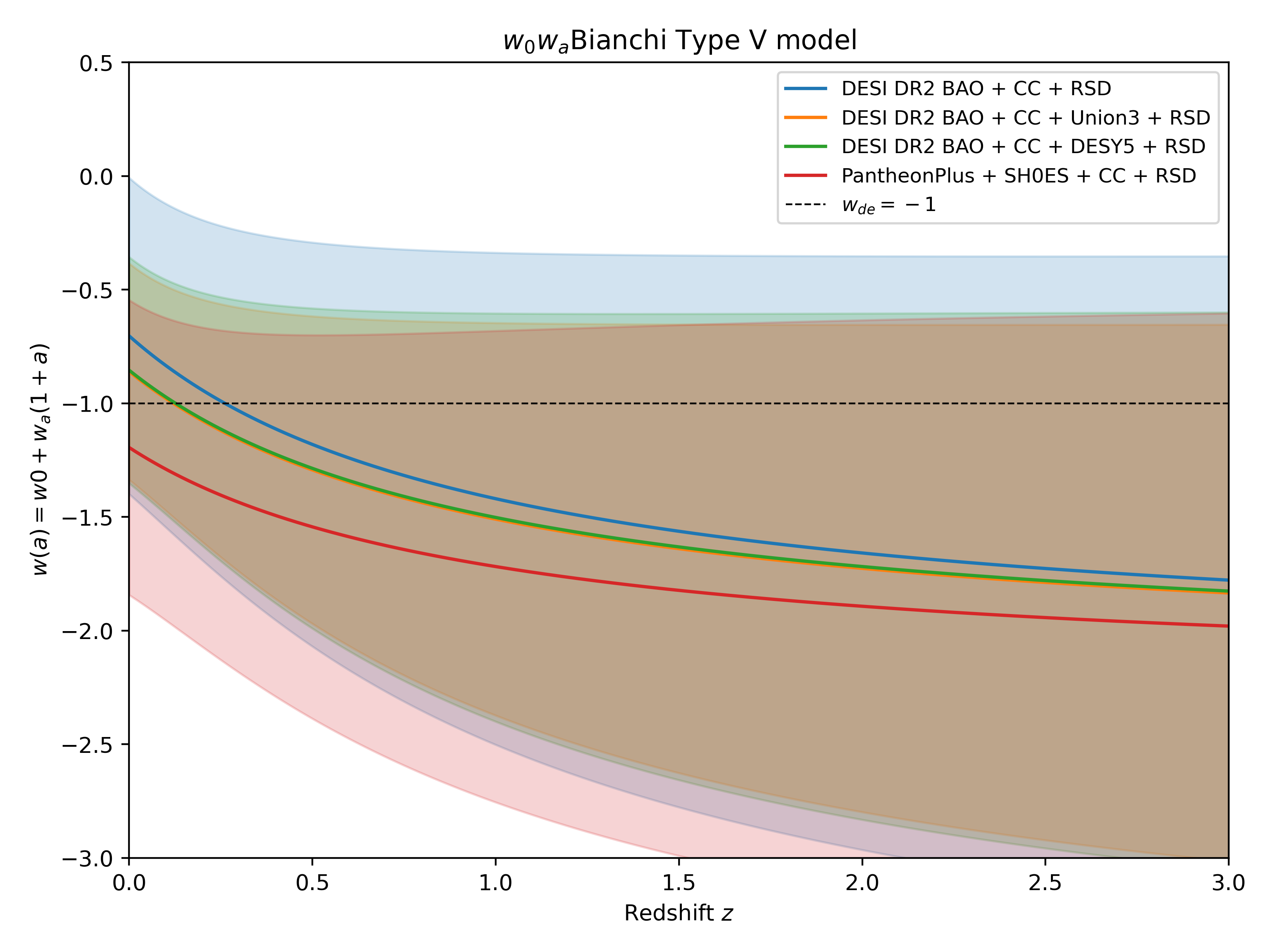}
       \caption{EoS parameter diagram for the  $w_0w_a$Bianchi Type V model, where the values of the parameters are taken from Table \ref{tab3:cosmo_constraints}.  The red, blue, green, and orange curves  are represent the \textit{DESI DR2 BAO + CC +RSD}, \textit{DESI DR2 BAO + CC + Union3 +RSD}, \textit{DESI DR2 BAO + CC + DESY5 +RSD}, and \textit{PantheonP+ CC +RSD}, respectively. }
       \label{fig:CPL2}
  \end{figure} 
% When $w_a =0$, the non-constant equation of state parameter is recovered $w_{de} = w_{0}$, , the values of $w_{de}>-1$ describe the phantom crossing of the universe for all datasets in the $w$Bianchi-V model as mentioned before.
\begin{table*}
    \centering
    \begin{tabular}{|l|l|l|l|l|l|l|l|}
        \hline
      &CL & $H_0$ & $\Omega_{m0}$ & $\Omega_{k0}$ & $\Omega_{\sigma0}$ & $r_d/M_{abs}$ & $S_8$ \\
         \hline
         \multicolumn{8}{|c|}{\textbf{ $\Lambda$CDM}} \\
        \hline
        \multicolumn{8}{|c|}{\textit{ DESI DR2 BAO  + CC + RSD}} \\
           \multicolumn{8}{|c|}{\textit{}} \\
       % \hline
       & 68\% &$ 68.875^{+1.707}_{-1.685}$& $ 0.316^{+0.005}_{-0.005}$ & $ -0.029^{+0.028}_{-0.027}$ &--- & $ 146.740^{+3.453}_{-3.298}$ & $ 0.789^{+0.024}_{-0.024}$ \\
        & 95\% & $ 68.875^{+3.370}_{-3.322}$ & $ 0.316^{+0.010}_{-0.010}$ & $ -0.029^{+0.057}_{-0.053}$ &---& $ 146.740^{+6.944}_{-6.350}$ & $ 0.789^{+0.048}_{-0.047}$ \\
%        & best-fit &68.875 & 0.316 & $-0.029$ &---& 146.728  & 0.789 \\
        \hline
        \multicolumn{8}{|c|}{\textit{ DESI DR2 BAO   +CC + Union3+RSD}} \\
         \multicolumn{8}{|c|}{\textit{}} \\
        & 68\% &$ 69.302^{+1.443}_{-1.419}$ & $ 0.317^{+0.005}_{-0.005}$ & $ -0.006^{+0.025}_{-0.024}$  &--- & $ 144.617^{+2.928}_{-2.849}$& $ 0.795^{+0.024}_{-0.024}$
 \\
        & 95\% &  $ 69.302^{+2.852}_{-2.779}$ & $ 0.317^{+0.010}_{-0.010}$ & $ -0.006^{+0.050}_{-0.047}$ & ---& $ 144.617^{+5.860}_{-5.494}$ & $ 0.795^{+0.048}_{-0.047}$
 \\
  %      & best fit & 69.319 & 0.317 & -0.007 & & 144.625 & 0.794 \\
        \hline\\
        % & $\chi^2$ & $\chi^2_{\text{red}}$ & AIC & $\Delta$AIC &  BIC & \Delta BIC& \\
        % \hline
       
        \multicolumn{8}{|c|}{\textit{ DESI DR2 BAO   +CC + DESY5 + RSD}} \\
           \multicolumn{8}{|c|}{\textit{}} \\
        & 68\% & $69.031^{+0.211}_{-0.210}$ & $0.321^{+0.005}_{-0.005}$ & $0.063^{+0.019}_{-0.019}$ &--- & $ 141.646^{+0.631}_{-0.622}$
 & $0.811^{+0.025}_{-0.025}$ \\
        & 95\% & $69.031^{+0.415}_{-0.415}$ & $0.321^{+0.011}_{-0.011}$ & $0.063^{+0.037}_{-0.036}$ &---&$ 141.646^{+1.250}_{-1.218}$ & $0.811^{+0.025}_{-0.025}$ \\
     %   & best-fit & 69.031 & 0.321 & 0.063 & & 141.653 & 0.811 \\
        \hline
        \multicolumn{8}{|c|}{\textit{PantheonP+SH0ES +CC+ RSD}} \\
           \multicolumn{8}{|c|}{\textit{}} \\
        & 68\% &  $ 71.807^{+0.873}_{-0.861}$  &  $ 0.324^{+0.006}_{-0.006}$  & $ 0.010^{+0.027}_{-0.026}$ &---& $ -19.297^{+0.025}_{-0.025}$  &  $ 0.821^{+0.028}_{-0.027}$ \\
        & 95\% &  $ 71.807^{+1.729}_{-1.687}$ & $ 0.324^{+0.012}_{-0.012}$  & $ 0.010^{+0.053}_{-0.051}$  &---&   $ -19.297^{+0.049}_{-0.049}$ &  $ 0.821^{+0.055}_{-0.053}$ \\
     %   & best fit & & & & & & \\
        \hline
        % D5 & 68\% & & & & & & \\
        % & 95\% & & & & & & \\
        % & best fit & & & & & & \\
        % \hline
         \multicolumn{8}{|c|}{\textbf{Bianchi-V}} \\
        \hline
         \multicolumn{8}{|c|}{\textit{ DESI DR2 BAO   + CC + RSD}} \\
         \multicolumn{8}{|c|}{\textit{}} \\
       & 68\% & $67.367^{+1.730}_{-1.708}$ & $ 0.286^{+0.008}_{-0.009}$ & $0.112^{+0.039}_{-0.038}$ & $0.000244^{+0.000300}_{-0.000301}$ &
           $146.883^{+3.461}_{-3.303}$ & $0.747^{+0.030}_{-0.030}$ \\
        & 95\% & $67.367^{+3.413}_{-3.366}$ & $0.286^{+0.016}_{-0.018}$ & $0.112^{+0.079}_{-0.074}$ & $0.000244^{+0.000594}_{-0.000594}$ & $146.883^{+7.005}_{-6.385}$ & $0.747^{+0.060}_{-0.059}$ \\
        %& best fit & 67.367 & 0.286 & 0.112 & 0.000244  & 146.883  & 0.747  \\
        \hline
         \multicolumn{8}{|c|}{\textit{ DESI DR2 BAO   + CC + Union3 + RSD}} \\
         \multicolumn{8}{|c|}{\textit{}} \\
        & 68\% & $68.432^{+1.418}_{-1.410}$ & $0.287^{+0.008}_{-0.009}$ & $0.117^{+0.032}_{-0.031}$ & $0.000265^{+0.000290}_{-0.000287}$ & $144.420^{+2.895}_{-2.802}$ & $0.747^{+0.030}_{-0.029}$ \\
        & 95\% & $68.432^{+2.822}_{-2.758}$ & $0.287^{+0.016}_{-0.018}$ & $0.117^{+0.064}_{-0.061}$ & $0.000265^{+0.000576}_{-0.000565}$ & $144.420^{+5.818}_{-5.445}$ & $0.747^{+0.058}_{-0.058}$ \\
       % & best-fit & 68.432 & 0.287 & 0.177 & $0.000265$ & 144.420 & 0.747 \\
        \hline
         \multicolumn{8}{|c|}{\textit{ DESI DR2 BAO   + CC + DESY5 + RSD}} \\
         \multicolumn{8}{|c|}{\textit{}} \\
        & 68\% & $68.666^{+0.219}_{-0.220}$ & $0.289^{+0.008}_{-0.009}$ & $0.160^{+0.023}_{-0.022}$ & $0.000435^{+0.000285}_{-0.000278}$ & $142.020^{+0.631}_{-0.627}$ & $0.759^{+0.029}_{-0.029}$ \\
        & 95\% & $68.666^{+0.433}_{-0.433}$ & $0.289^{+0.016}_{-0.018}$ & $0.160^{+0.046}_{-0.043}$ & $0.000435^{+0.000570}_{-0.000544}$ & $142.020^{+1.249}_{-1.236}$ & $0.759^{+0.057}_{-0.058}$ \\
       % & best-fit & 68.666 & 0.289 & 0.160 &$0.000435$ & 142.020 & 0.759 \\
        \hline
          \multicolumn{8}{|c|}{\textit{ DESI DR2 BAO   + CC + DESY5 + RSD}} \\
         \multicolumn{8}{|c|}{\textit{}} \\
        & 68\% & $ 71.894^{+0.864}_{-0.864}$&   $ 0.317^{+0.007}_{-0.009}$ &$ 0.051^{+0.031}_{-0.029}$ &  $ 0.002419^{+0.000529}_{-0.000339}$ &  $ -19.291^{+0.025}_{-0.025}$
 & $ 0.810^{+0.029}_{-0.031}$ \\
        & 95\% &  $ 71.894^{+1.722}_{-1.700}$ & $ 0.317^{+0.013}_{-0.022}$  &  $ 0.051^{+0.063}_{-0.056}$ &$ 0.002419^{+0.001324}_{-0.000568}$ &  $ -19.291^{+0.049}_{-0.049}$  &  $ 0.810^{+0.055}_{-0.063}$ \\
       % & best fit & & & & & & \\
        \hline
    \end{tabular}
    \caption{Constraints on cosmological parameters for various data combinations for $\Lambda$CDM and  Bianchi-V models. $H_0$ values are provided in ${\rm km\,s^{-1}\,Mpc^{-1}}$, $rd$ and $M_{abs}$ in Mpc  units.}
    %\adlcd{Does $r_d/M_{bs}$ has units?} \adlcd{Why for some catalogues we give 68\% and 95\% and for others 68\% and 95\%?}}
    \label{tab1:cosmo_constraints}
\end{table*}
\begin{table*}
    \centering
    \begin{tabular}{|l|l|l|l|l|l|l|l|l|}
        \hline
      &CL & $H_0$ & $\Omega_{m0}$ & $\Omega_{k0}$ & $\Omega_{\sigma0}$ & $r_d/M_{bs}$ & $w_0$ &  $S_8$ \\
        \hline
         \multicolumn{8}{|c|}{\textbf{$w$CDM}} \\
        \hline
                \multicolumn{8}{|c|}{\textit{ DESI DR2 BAO   + CC + RSD}} \\
         \multicolumn{8}{|c|}{\textit{}} \\
       &68\% & $ 66.485^{+1.853}_{-1.816}$  & $ 0.286^{+0.007}_{-0.007}$ &  $ 0.054^{+0.055}_{-0.068}$ & & $ 146.910^{+3.463}_{-3.306}$ &  $ -0.854^{+0.104}_{-0.105}$
  &  $ 0.758^{+0.031}_{-0.031}$\\
&95\% & $ 66.485^{+1.853}_{-1.816}$
 & $ 0.286^{+0.014}_{-0.014}$
&  $ 0.054^{+0.104}_{-0.164}$ & & $ 146.910^{+6.976}_{-6.383}$ &  $ -0.854^{+0.204}_{-0.197}$&  $ 0.758^{+0.062}_{-0.059}$\\
%&best-fit &  66.652 &  0.325 & 0.054 & & 146.910 & -0.752 &  0.844\\
\hline\\
        \multicolumn{8}{|c|}{\textit{ DESI DR2 BAO   + CC +Union3 + RSD}} \\
         \multicolumn{8}{|c|}{\textit{}} \\
      & 68\% & $ 68.023^{+1.441}_{-1.427}$ & $ 0.330^{+0.006}_{-0.006}$ & $ -0.167^{+0.067}_{-0.083}$ & &  $ 144.417^{+2.886}_{-2.799}$
& $ -0.762^{+0.066}_{-0.069}$ &  $ 0.841^{+0.029}_{-0.028}$\\
        & 95\% & $ 68.023^{+2.855}_{-2.803}$ & $ 0.330^{+0.013}_{-0.013}$ & $ -0.167^{+0.122}_{-0.174}$ &&   $ 144.417^{+5.817}_{-5.437}$ &  $ -0.762^{+0.120}_{-0.139}$ &  $ 0.841^{+0.056}_{-0.055}$ \\
   %     & best fit & 68.023  & 0.330 & $-0.167$ & & 144.417 & -0.762 &  0.841 \\
        \hline
\\
        \multicolumn{8}{|c|}{\textit{ DESI DR2 BAO   +CC + DESY5+RSD}} \\
         \multicolumn{8}{|c|}{\textit{}} \\
& 68\% & $ 68.371^{+0.241}_{-0.241}$ &$ 0.288^{+0.007}_{-0.007}$ & $ 0.031^{+0.058}_{-0.065}$  & &$ 142.454^{+0.655}_{-0.653}$
 &$ -0.812^{+0.056}_{-0.060}$
 & $ 0.764^{+0.027}_{-0.027}$ \\
&95\% & $ 68.371^{+0.478}_{-0.472}$ & $ 0.288^{+0.013}_{-0.014}$ &$ 0.031^{+0.109}_{-0.138}$ && $ 142.454^{+1.299}_{-1.288}$ & $ -0.812^{+0.108}_{-0.122}$ & $ 0.764^{+0.054}_{-0.052}$\\
%&best-fit & 68.351  & 0.333 & $-0.031$  && 142.543 & $-0.712$  &  0.853  \\
\hline\\
      \multicolumn{8}{|c|}{\textit{ PantheonP  + SH0ES + CC  + RSD}}\\
         \multicolumn{8}{|c|}{\textit{}} \\
       & 68\% & $ 71.767^{+0.871}_{-0.859}$&$ 0.283^{+0.013}_{-0.014}$ & $ 0.098^{+0.103}_{-0.141}$
 &&  $ -19.295^{+0.025}_{-0.025}$  &   $ -1.004^{+0.169}_{-0.210}$
&  $ 0.735^{+0.027}_{-0.026}$\\
        & 95\% & $ 71.767^{+1.724}_{-1.685}$ &  $ 0.283^{+0.007}_{-0.007}$ &   $ 0.098^{+0.166}_{-0.261}$ &&  $ -19.295^{+0.049}_{-0.049}$
  &  $ -1.004^{+0.262}_{-0.400}$

 &   $ 0.735^{+0.054}_{-0.052}$ \\
%        & & & & & & & & & \\
        % \hline\\
        % D5 & & & & & & & & & \\
        % & & & & & & & & & \\
        % & & & & & & & & & \\
 \hline
         \multicolumn{8}{|c|}{\textbf{$w$ Bianchi-V}} \\
        \hline
                \multicolumn{8}{|c|}{\textit{ DESI DR2 BAO   + CC + RSD}} \\
         \multicolumn{8}{|c|}{\textit{}} \\
       &68\% & $ 66.207^{+1.889}_{-1.797}$  & $ 0.282^{+0.009}_{-0.009}$ & $ 0.044^{+0.139}_{-0.166}$ & $ 0.000552^{+0.000432}_{-0.000374}$
&$ 146.820^{+3.463}_{-3.294}$ &  $ -0.73^{+0.107}_{-0.203}$ &  $ 0.754^{+0.031}_{-0.031}$\\
&95\% &$ 66.207^{+3.898}_{-3.474}$  &$ 0.282^{+0.018}_{-0.019}$
 &$ 0.044^{+0.213}_{-0.242}$& $ 0.000552^{+0.000817}_{-0.000524}$
 &   $ 146.820^{+6.981}_{-6.345}$
&  $ -0.73^{+0.154}_{-0.458}$&   $ 0.754^{+0.062}_{-0.061}$\\
\hline\\
        \multicolumn{8}{|c|}{\textit{ DESI DR2 BAO   + CC + Union3 + RSD}} \\
         \multicolumn{8}{|c|}{\textit{}} \\
      & 68\% & $ 67.803^{+1.455}_{-1.431}$ & $ 0.281^{+0.009}_{-0.009}$ & $ 0.044^{+0.120}_{-0.152}$& $ 0.000683^{+0.000589}_{-0.000646}$
& $ 144.261^{+2.906}_{-2.806}$ &  $ -0.756^{+0.102}_{-0.147}$ &  $ 0.747^{+0.030}_{-0.030}$\\
        & 95\% & $ 67.803^{+2.882}_{-2.805}$ & $ 0.281^{+0.017}_{-0.019}$ &   $ 0.044^{+0.194}_{-0.238}$&$ 0.000683^{+0.001029}_{-0.001215}$
 & $ 144.261^{+5.833}_{-5.434}$ & $ -0.756^{+0.149}_{-0.301}$ & $ 0.747^{+0.059}_{-0.059}$\\
        \hline\\
                \multicolumn{8}{|c|}{\textit{ DESI DR2 BAO   + CC + DESY5  + RSD}} \\
         \multicolumn{8}{|c|}{\textit{}} \\
 & 68\% &  $ 68.26^{+0.24}_{-0.23}$& $ 0.279^{+0.009}_{-0.009}$ & $ 0.116^{+0.178}_{-0.228}$ & $ 0.000758^{+0.000481}_{-0.000552}$
 & $ 142.715^{+0.673}_{-0.671}$ & $ -0.694^{+0.076}_{-0.105}$ &  $ 0.748^{+0.030}_{-0.030}$ \\
&95\% & $ 68.26^{+0.48}_{-0.45}$ & $ 0.279^{+0.017}_{-0.018}$ & $ 0.116^{+0.292}_{-0.365}$ & $ 0.000758^{+0.000865}_{-0.001076}$ & $ 142.715^{+1.359}_{-1.320}$ & $ -0.694^{+0.108}_{-0.217}$ &  $ 0.748^{+0.058}_{-0.059}$ \\
\hline
          \multicolumn{8}{|c|}{\textit{ PantheonP  + SH0ES + CC  + RSD}} \\
         \multicolumn{8}{|c|}{\textit{}} \\
        & 68\% & $ 71.896^{+0.869}_{-0.868}$ &  $ 0.317^{+0.008}_{-0.010}$ & $ 0.089^{+0.087}_{-0.119}$  &  $ 0.002419^{+0.000529}_{-0.000339}$ &  $ -19.293^{+0.025}_{-0.025}$ & $ -1.068^{+0.180}_{-0.210}$ & $ 0.811^{+0.029}_{-0.031}$ \\
        & 95\% & $ 71.896^{+1.718}_{-1.697}$ & $ 0.317^{+0.014}_{-0.022}$  &  $ 0.089^{+0.151}_{-0.281}$ &$ 0.002419^{+0.001324}_{-0.000568}$ &   $ -19.293^{+0.049}_{-0.050}$ & $ -1.068^{+0.322}_{-0.441}$  &  $ 0.811^{+0.057}_{-0.065}$ \\
      \hline\\
    \end{tabular}
    \caption{Constraints on cosmological parameters for various data combinations for  $w$CDM and $w$Bianchi-V models. $H_0$ values are provided in ${\rm km\,s^{-1}\,Mpc^{-1}}$, $rd$ and $M_{abs}$ in Mpc  units.}
    \label{tab2:cosmo_constraints}
\end{table*}
  \begin{table*}
    \centering
    \begin{tabular}{|l|l|l|l|l|l|l|l|l|l|l|}
        \hline
        & CL & $H_0$& $\Omega_{m0}$ & $\Omega_{k0}$ & $\Omega_{\sigma0}$ & $r_d/M_{abs}$ &  $w_{\rm DE}/w_0$~~~~~~~~~~~~ &$w_a$ & $S_8$ \\
        \hline
      \multicolumn{9}{|c|}{\textbf{$w_0w_a$ CDM}}\\
\hline\\
 \multicolumn{8}{|c|}{\textit{ DESI DR2 BAO   +CC + RSD}} \\
         \multicolumn{8}{|c|}{\textit{}} \\
 & 68\% & $ 64.780^{+2.038}_{-2.014}$ & $ 0.269^{+0.018}_{-0.011}$&  $ 0.254^{+0.033}_{-0.034}$ &
 & $ 147.160^{+3.458}_{-3.316}$ & $ -0.703^{+0.191}_{-0.173}$
 &  $ -1.433^{+0.421}_{-0.460}$
  &   $ 0.706^{+0.034}_{-0.031}$\\
        & 95\% &$ 64.780^{+4.057}_{-3.943}$ & $ 0.269^{+0.042}_{-0.018}$ & $ 0.254^{+0.066}_{-0.071}$ & & $ 147.160^{+6.972}_{-6.361}$
 & $ -0.703^{+0.388}_{-0.322}$
&  $ -1.433^{+0.773}_{-0.919}$&     $ 0.706^{+0.070}_{-0.058}$ \\
        \hline
         \multicolumn{8}{|c|}{\textit{ DESI DR2 BAO   + CC + Union3 + RSD}} \\
         \multicolumn{8}{|c|}{\textit{}} \\
        & 68\%& $ 67.388^{+1.468}_{-1.437}$ & $ 0.264^{+0.011}_{-0.008}$ &$ 0.238^{+0.027}_{-0.028}$
 & &$ 144.330^{+2.879}_{-2.793}$ & $ -0.842^{+0.107}_{-0.103}$ &  $ -1.225^{+0.323}_{-0.327}$  &   $ 0.686^{+0.025}_{-0.024}$\\
&95\% &$ 67.388^{+2.900}_{-2.816}$ & $ 0.264^{+0.025}_{-0.015}$ &   $ 0.238^{+0.053}_{-0.057}$ &&  $ 144.330^{+5.778}_{-5.445}$ & $ -0.842^{+0.213}_{-0.197}$
  &  $ -1.225^{+0.607}_{-0.636}$
 & $ 0.686^{+0.050}_{-0.048}$\\
        \hline\\
         \multicolumn{8}{|c|}{\textit{ DESI DR2 BAO   +CC + DESY5+RSD}} \\
         \multicolumn{8}{|c|}{\textit{}} \\
        &68\% & $ 68.033^{+0.535}_{-0.539}$
  &  $ 0.267^{+0.011}_{-0.009}$
& $ 0.234^{+0.026}_{-0.028}$ &
& $ 142.707^{+0.682}_{-0.668}$& $ -0.803^{+0.072}_{-0.073}$
&  $ -1.344^{+0.300}_{-0.286}$
 & $ 0.689^{+0.045}_{-0.045}$\\
&95\% & $ 68.033^{+0.535}_{-0.539}$ & $ 0.267^{+0.024}_{-0.016}$
&$ 0.234^{+0.049}_{-0.058}$ & 
&  $ 142.707^{+1.346}_{-1.315}$& $ -0.803^{+0.140}_{-0.143}$& $ -1.344^{+0.599}_{-0.544}$
& $ 0.689^{+0.023}_{-0.023}$\\
%&best-fit & 69.031 & $ 0.321$ & $ 0.063$& &  141.653& & & 0.811\\
\hline
 \\
  \multicolumn{8}{|c|}{\textit{ PantheonP + SH0ES +CC +RSD}} \\
         \multicolumn{8}{|c|}{\textit{}} \\
        &68\% &   $ 71.656^{+0.870}_{-0.859}$
  & $ 0.267^{+0.010}_{-0.009}$  & $ 0.265^{+0.041}_{-0.043}$
  && $ -19.299^{+0.025}_{-0.025}$
& $ -1.206^{+0.159}_{-0.191}$ & $ -0.982^{+0.332}_{-0.379}$& $ 0.687^{+0.028}_{-0.028}$  \\
        & 95\%&   $ 71.656^{+1.724}_{-1.685}$  &    $ 0.267^{+0.024}_{-0.017}$ & $ 0.265^{+0.077}_{-0.086}$& & $ -19.299^{+0.049}_{-0.049}$
 &  $ -1.206^{+0.289}_{-0.401}$ &  $ -0.982^{+0.534}_{-0.811}$
&  $ 0.687^{+0.056}_{-0.055}$ \\
%        & best fit&  $L(\theta) = -34.197$& & & & & & & \\
        \hline\\
         
        % D5 & 68\% & & & & & & & &  \\
        % &95\% & & & & & & & &  \\
        % & best-fit & & & & & & & &  \\
        % \hline
        %\hline
       \multicolumn{9}{|c|}{\textbf{$w_0w_a$ Bianchi-V}}\\
\hline
 \\
  \multicolumn{8}{|c|}{\textit{ DESI DR2 +CC  +RSD}} \\
         \multicolumn{8}{|c|}{\textit{}} \\
& 68\% & $ 64.741^{+2.241}_{-2.131}$   & $ 0.271^{+0.023}_{-0.018}$&  $ 0.254^{+0.033}_{-0.034}$ & $ 0.000425^{+0.000456}_{-0.000298}$
&  $ 147.164^{+3.464}_{-3.312}$ & $ -0.689^{+0.224}_{-0.240}$
 &  $ -1.462^{+0.425}_{-0.476}$
  &  $ 0.707^{+0.040}_{-0.038}$\\
        & 95\% &  $ 64.741^{+4.533}_{-4.118}$ & $ 0.271^{+0.054}_{-0.036}$ & $ 0.254^{+0.066}_{-0.071}$ &  $ 0.000425^{+0.000989}_{-0.000405}$
& $ 147.164^{+6.974}_{-6.385}$
 & $ -0.689^{+0.420}_{-0.483}$
& $ -1.462^{+0.791}_{-0.983}$&    $ 0.707^{+0.081}_{-0.075}$ \\
 \hline\\
 \\
  \multicolumn{8}{|c|}{\textit{ DESI DR2 +CC Union3 +RSD}} \\
         \multicolumn{8}{|c|}{\textit{}} \\
        & 68\%&  $ 67.021^{+1.451}_{-1.430}$ &  $ 0.260^{+0.009}_{-0.007}$  &$ 0.182^{+0.046}_{-0.049}$
 & $ 0.000738^{+0.000516}_{-0.000458}$
&  $ 144.434^{+2.873}_{-2.781}$ &  $ -0.697^{+0.110}_{-0.125}$ &   $ -1.131^{+0.266}_{-0.269}$ &  $ 0.680^{+0.024}_{-0.023}$\\
&95\% &$ 67.021^{+2.870}_{-2.801}$ & $ 0.260^{+0.021}_{-0.012}$ &   $ 0.182^{+0.087}_{-0.102}$ &$ 0.000738^{+0.001008}_{-0.000692}$&  $ 144.434^{+5.738}_{-5.401}$&$ -0.697^{+0.203}_{-0.262}$
  &   $ -1.131^{+0.530}_{-0.557}$
 & $ 0.680^{+0.048}_{-0.044}$\\
  \hline\\
 \\
  \multicolumn{8}{|c|}{\textit{ DESI DR2  BAO +CC DESY5 +RSD}} \\
         \multicolumn{8}{|c|}{\textit{}} \\
        &68\% & $ 68.360^{+0.258}_{-0.255}$ &   $ 0.265^{+0.014}_{-0.012}$
 &  $ 0.227^{+0.025}_{-0.026}$ &  $ 0.000389^{+0.000430}_{-0.000431}$ &  $ 144.683^{+1.026}_{-0.968}$
&$ -0.994^{+0.062}_{-0.067}$ &   $ -1.016^{+0.277}_{-0.300}$

 &$ 0.684^{+0.030}_{-0.031}$\\
&95\% &  $ 68.360^{+0.509}_{-0.501}$&  $ 0.265^{+0.031}_{-0.024}$ &$ 0.227^{+0.048}_{-0.053}$ & $ 0.000389^{+0.000851}_{-0.000855}$ &   $ 144.683^{+2.068}_{-1.850}$ &  $ -0.994^{+0.118}_{-0.136}$ &  $ -1.016^{+0.467}_{-0.615}$ 
&  $ 0.684^{+0.059}_{-0.061}$\\
 \\
  \multicolumn{8}{|c|}{\textit{ PantheonP + SH0ES +CC +RSD}} \\
         \multicolumn{8}{|c|}{\textit{}} \\
        &68\% &$ 71.844^{+0.875}_{-0.862}$ &  $ 0.303^{+0.014}_{-0.013}$ &  $ 0.214^{+0.039}_{-0.041}$ &    $ 0.002227^{+0.000463}_{-0.000328}$
 & $ -19.296^{+0.025}_{-0.025}$ &    $ -1.194^{+0.153}_{-0.182}$ &  $ -1.048^{+0.411}_{-0.467}$ & $ 0.765^{+0.033}_{-0.034}$  \\
        & 95\%&  $ 71.844^{+1.740}_{-1.697}$  & $ 0.303^{+0.035}_{-0.026}$  & $ 0.214^{+0.076}_{-0.082}$  & $ 0.002227^{+0.001125}_{-0.000588}$& $ -19.296^{+0.050}_{-0.050}$  &  $ -1.194^{+0.279}_{-0.383}$ & $ -1.048^{+0.665}_{-0.986}$  &  $ 0.765^{+0.065}_{-0.069}$\\
        \hline
    \end{tabular}
    \caption{Constraints on cosmological parameters for various data combinations for  $w_0w_a$CDM and $w_0w_a$Bianchi-V models. $H_0$ values are provided in  ${\rm km\,s^{-1}\,Mpc^{-1}}$, $rd$ and $M_{abs}$ in Mpc  units. }
    \label{tab3:cosmo_constraints}
\end{table*}
\begin{figure}
    \includegraphics[width=1.0\linewidth]{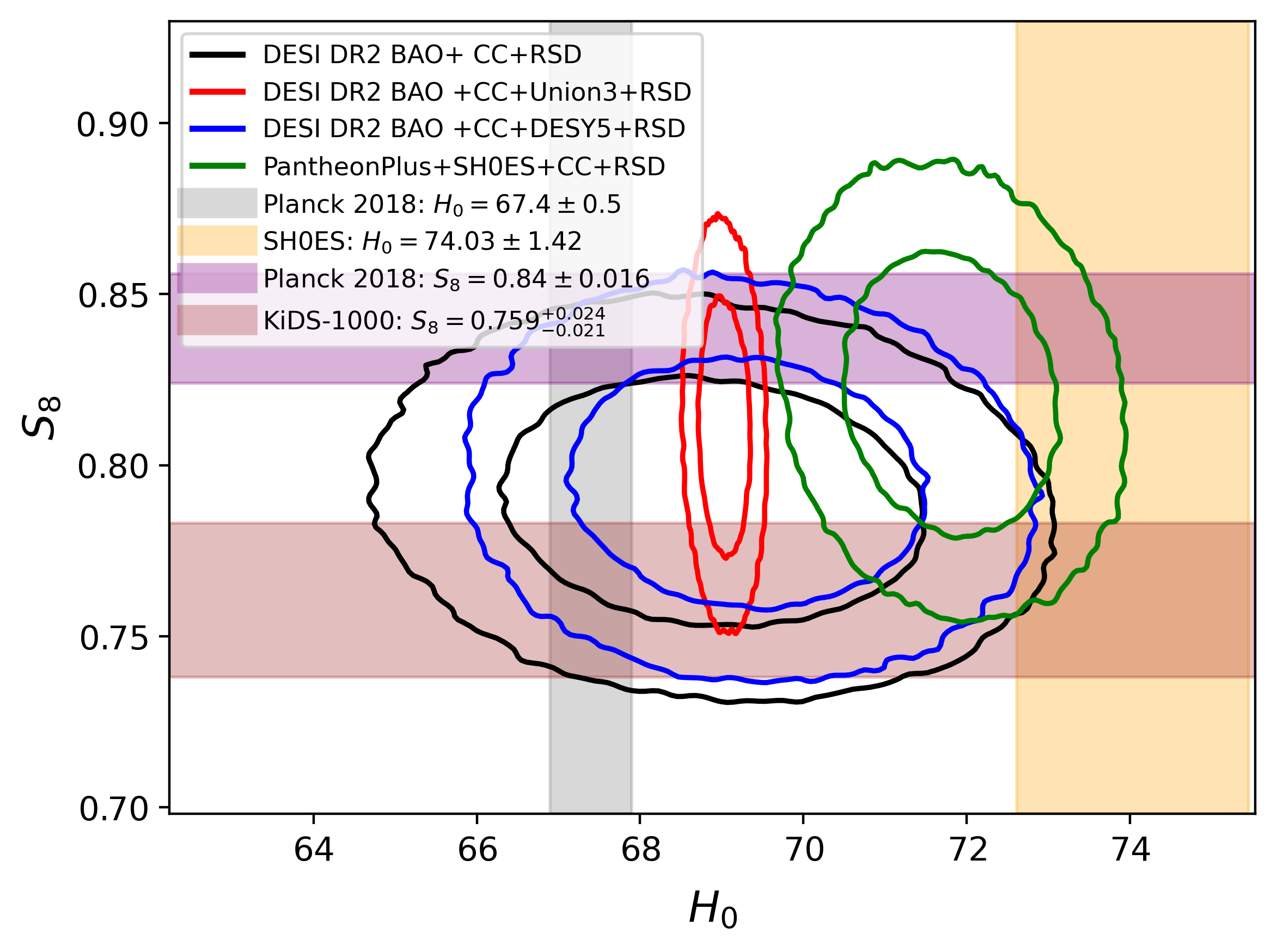 }
    \caption{ $S_8$-$H_0$ diagram for  \(\Lambda\)CDM model. $H_0$ values are provided in ${\rm km\,s^{-1}\,Mpc^{-1}}$ units. The $H_0$ tensions are \({0.44\sigma,\,0.67\sigma,\,2.51\sigma,\,2.48\sigma}\) with Planck 2018 data,  and \({1.42\sigma,\,1.50\sigma,\,3.38\sigma,\,1.00\sigma}\) with SH0ES (\(H_0 = 74.03 \pm 1.42\)) measurements.  The \(S_8\) tension between the  \(\Lambda\)CDM model and  Planck~2018 ($S_8 = 0.834^{+0.016}_{-0.016}$  have a difference \(0.89\sigma,\, 0.78\sigma,\, 0.77\sigma,\, 0.23\sigma\), and \(0.57\sigma, 0.68\sigma,\, 1.55\sigma,\, 1.06\sigma\) with KiDS-1000 data ($S_8 = 0.759^{+0.024}_{-0.021}$).}
    \label{fig:nonflatx}
\end{figure}
\subsection{Results for $H_0$  and $S_8$ values } 
% \adlcd{As mentioned above, change the title. Also, the units in the title are not relevant}
The inferred values of \(H_0\) and \(S_8\) are taken from Tables \ref{tab1:cosmo_constraints} - \ref{tab3:cosmo_constraints} to compare our models considered with direct and indirect measurements. The discrepancy between the indirect measurements of \(H_0\) (in ~${\rm km\,s^{-1}\,Mpc^{-1}}$ units), including Planck~2018 (\(H_0 = 67.4 \pm 0.5\))~\citep{aghanim2020planck} and DESI-2024 (\(H_0 = 68.52 \pm 0.62\))~\citep{adame2025desi}, and the direct (local) measurements, such as SH0ES (\(H_0 = 74.03 \pm 1.42\))~\citep{riess2019large}, $H0LiCOW$ (\(H_0 = 73.3 \pm 1.8\))~\citep{wong2020h0licow}, and $HST$ (\(H_0 = 73.8 \pm 2.4\))~\citep{riess20113}, remains unresolved. This persistent tension may point toward extensions of $\Lambda$CDM to understand the accelerating expansion of the Universe fully.  In our analysis, the best-fit values of \(H_0\) for the \textit{DESI~ DR2~BAO + CC + RSD}, \textit{DESI~DR2~BAO + CC + Union3 + RSD}, and \textit{DESI~DR2~BAO + CC + DESY5 + RSD} combinations are generally consistent with the indirect measurements. In contrast, for the \textit{PantheonP + SH0ES + CC + RSD} dataset, the inferred \(H_0\) values are closer to local measurements. 
\begin{figure}
    \includegraphics[width=1.0\linewidth]{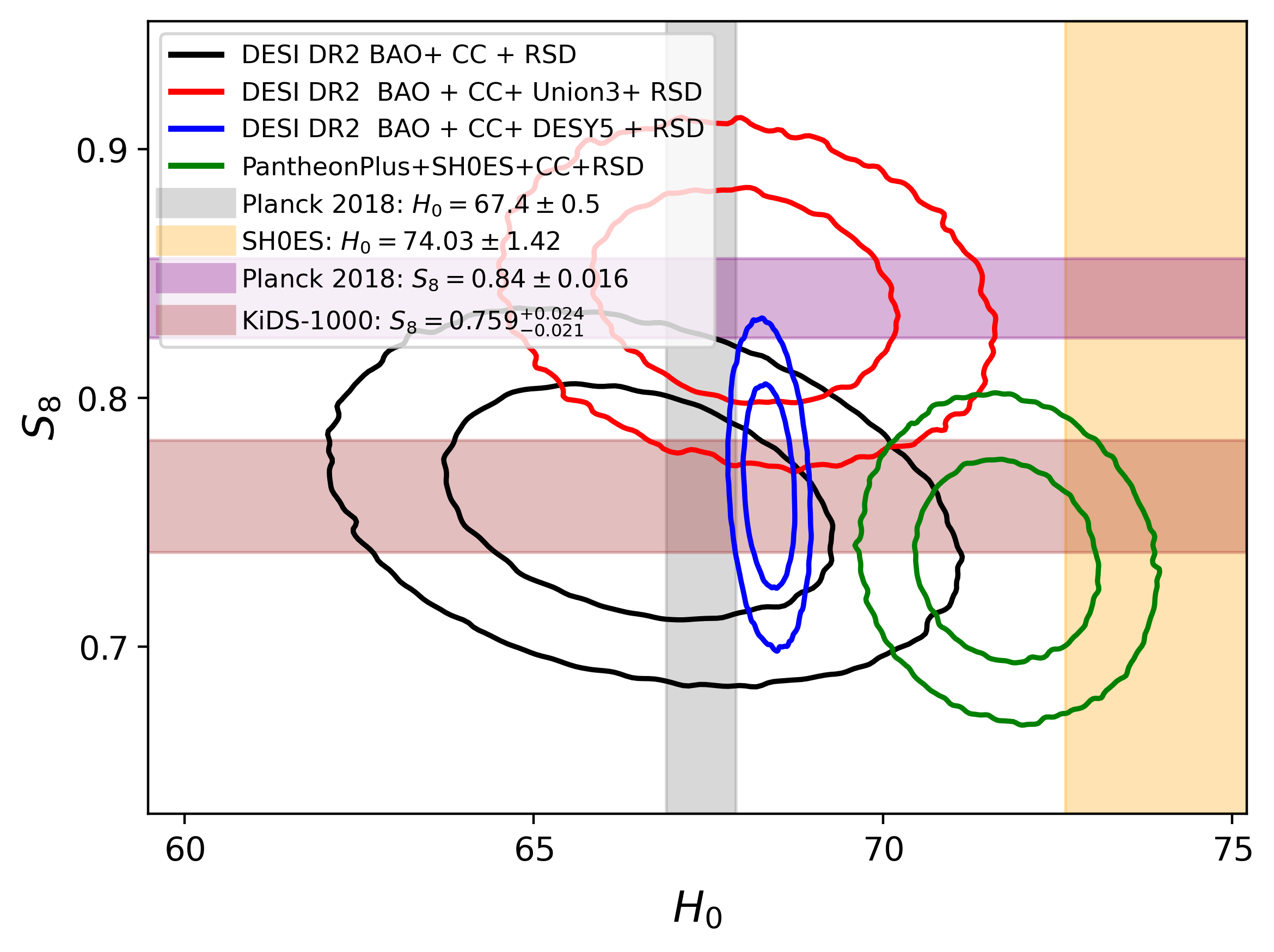  }
    \caption{ $S_8$-$H_0$ diagram for  \(w\)CDM model. $H_0$ values are provided in ${\rm km\,s^{-1}\,Mpc^{-1}}$ units. The $H_0$ tensions are \({0.04\sigma,\, 0.22\sigma,\, 1.38\sigma,\, 2.46\sigma}\) with Planck 2018 data,  and \({2.82\sigma,\, 1.90\sigma,\, 3.79\sigma,\, 1.02\sigma}\) with SH0ES (\(H_0 = 74.03 \pm 1.42\)) measurements.  The \(S_8\) tension between the  \(w\)CDM model and  Planck~2018 ($S_8 = 0.834^{+0.016}_{-0.016}$  have a difference $1.94\sigma, 2.87\sigma,5.18\sigma,2.55\sigma$  and $0.78\sigma, 1.35\sigma,2.18\sigma,1.20\sigma$  with KidS-1000 data ($S_8 = 0.759^{+0.024}_{-0.021}$).}
    \label{fig:nonflatx1}
\end{figure}
\\
\\
Additionally,  the matter clustering parameter, \( S_8 = \sigma_8 \sqrt{\Omega_{m0}/0.3} \), provides key insight into the amplitude of matter density fluctuations and the matter content of the Universe, which underlie the process of structure formation. The corresponding constrained values are listed in Tables~\ref{tab1:cosmo_constraints}, \ref{tab2:cosmo_constraints}, and \ref{tab3:cosmo_constraints} for all models using all combined datasets:   late-time measurements, such as KidS-1000 ($S_8 = 0.759^{+0.024}_{-0.021}$)~\cite{asgari2021kids}, $KiDS-450$ ($S_8 = 0.745^{+0.039}_{-0.039}$)~\cite{hildebrandt2017kids}, $DES~Y1$ ($S_8 = 0.759^{+0.025}_{-0.023}$)~\cite{abbott2018dark}, and $DES~Y3$ ($S_8 = 0.759^{+0.025}_{-0.023}$)~\cite{amon2022dark}, yield lower $S_8$ values compared to early-time measurements such as Planck~2018 ($S_8 = 0.834^{+0.016}_{-0.016}$)~\cite{aghanim2020planck} and the ACT collaboration results ($S_8 = 0.830 \pm 0.043$, $0.840 \pm 0.030$, and $0.846 \pm 0.016$ for ACT, ACT+WMAP, and ACT+Planck, respectively)~\cite{aiola2020atacama}.
\\
\\
Notably, the Planck-2018 and SH0ES measurements of $H_0$ exhibit approximately a \({5.2\sigma}\) tension, and the persistent discrepancy of $S_8$ between the \rm{KiDS}-1000 and Planck-2018 measurements is approximately $2.68\sigma$, indicating an open problem in Cosmology and may hint at new physics beyond the $\Lambda$CDM model. In the following, the analysis has been performed taking into account the model's values of \(H_0\)  and $S_8$ at 95\% C.L., and  for all the above datasets:
\begin{enumerate}
    \item For the  \(\Lambda\)CDM model with Planck 2018 \((H_0 = 67.4 \pm 0.5\, {\rm km\,s^{-1}\,Mpc^{-1}}\), the tensions are \({0.44\sigma,\,0.67\sigma,\,2.51\sigma,\,2.48\sigma}\). With SH0ES data, the tensions are \({1.42\sigma,\,1.50\sigma,\,3.38\sigma,\,1.00\sigma}\).  This indicates that the model's \(H_0\) values have relaxed tension with both measurements for \textit{DESI~DR2~BAO + CC + RSD}, and \textit{DESI~DR2~BAO + CC + DESY5 + RSD} datasets. In contrast,  the tension is still significant for the case of \textit{DESI~DR2~BAO + CC + Union3 + RSD} and \textit{PantheonP + SH0ES + CC + RSD} datasets. For more details, see the diagram Fig. \ref{fig:nonflatx}.    The \(S_8\) tension between the  \(\Lambda\)CDM model and Planck 2018 data have a difference \(0.89\sigma,\, 0.78\sigma,\, 0.77\sigma,\, 0.23\sigma\), and \(0.57\sigma, 0.68\sigma,\, 1.55\sigma,\, 1.06\sigma\) with KiDS-1000 data and the result is presented in Fig. \ref{fig:nonflatx}. From the figure, we note that the model values of \(S_8\) consistently lie between the two measurements, suggesting they may solve the discrepancy. Both tensions ($H_0$ and $S_8$) are alleviated for all joint datasets.
    \item  In a similar manner, the  \(w\)CDM model has tensions of \({0.04\sigma,\, 0.22\sigma,\, 1.38\sigma,\, 2.46\sigma}\) with Planck 2018, and \({2.82\sigma,\, 1.90\sigma,\, 3.79\sigma,\, 1.02\sigma}\) with SH0ES presented in Fig. \ref{fig:nonflatx1} using the above considered combined datasets. 
    \begin{figure}
    \includegraphics[width=1.0\linewidth]{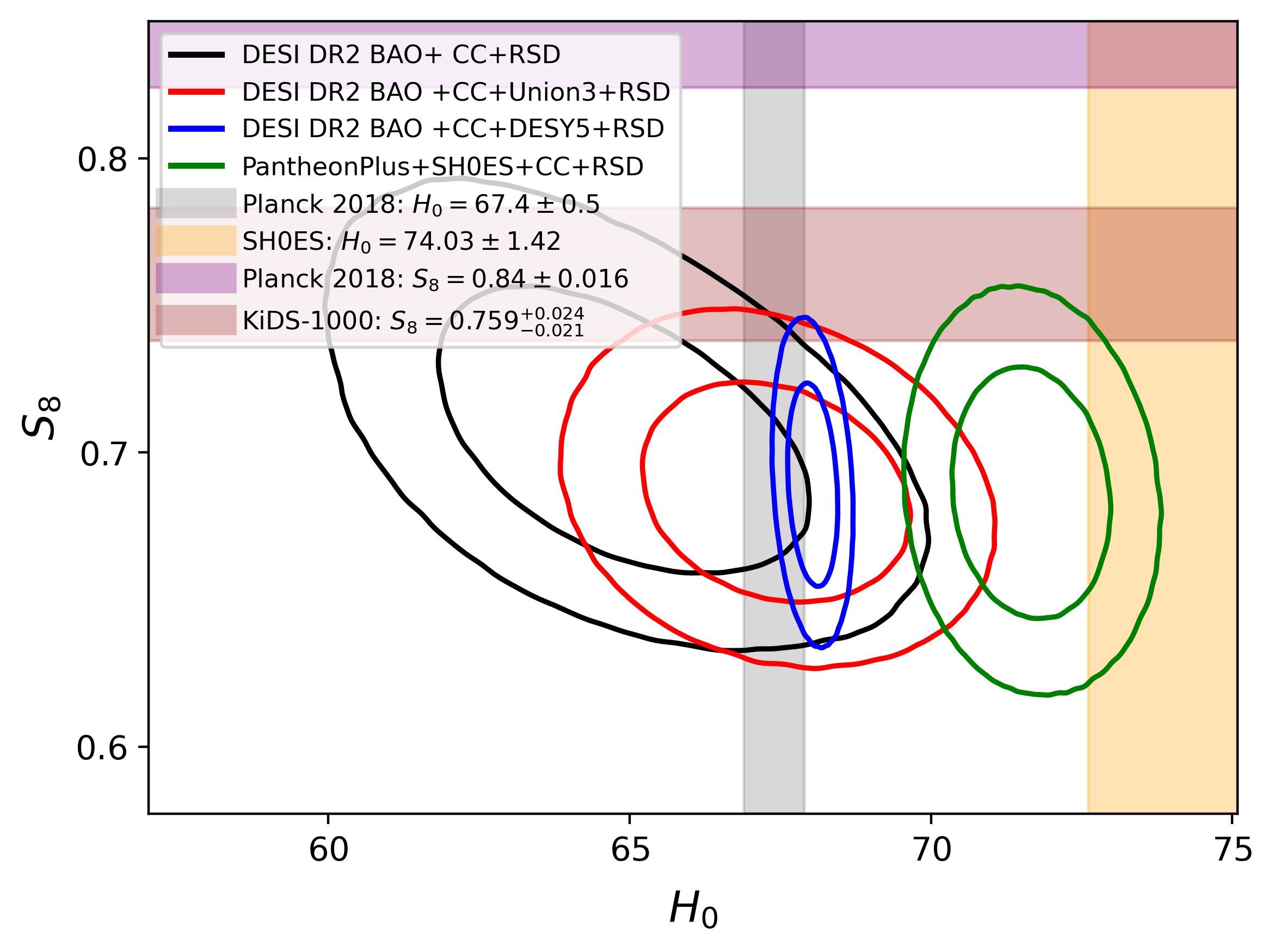}
    \caption{ $S_8$-$H_0$ diagram for the \(w_0w_a\)CDM model. The $H_0$ tensions are \({0.65\sigma,\, 0.00\sigma,\, 0.86\sigma,\, 2.38\sigma}\) with Planck 2018 ($H_0 = 67.4\pm0.5$), and \({2.18\sigma,\, 2.08\sigma,\, 3.95\sigma,\, 1.07\sigma}\)  with SH0ES (\(H_0 = 74.03 \pm 1.42\)) measurements.  The \(S_8\) tension between the  \(w\)CDM model and  Planck~2018 ($S_8 = 0.834^{+0.016}_{-0.016}$  have a difference  $1.21\sigma, 0.12\sigma,1.26\sigma,1.79\sigma$  and $0.02\sigma, 1.36\sigma,0.09\sigma,0.42\sigma$  with KiDS-1000 data ($S_8 = 0.759^{+0.024}_{-0.021}$).}
    \label{fig:nonflatx2w0walcdm}
\end{figure}
\begin{figure}
    \includegraphics[width=1.0\linewidth]{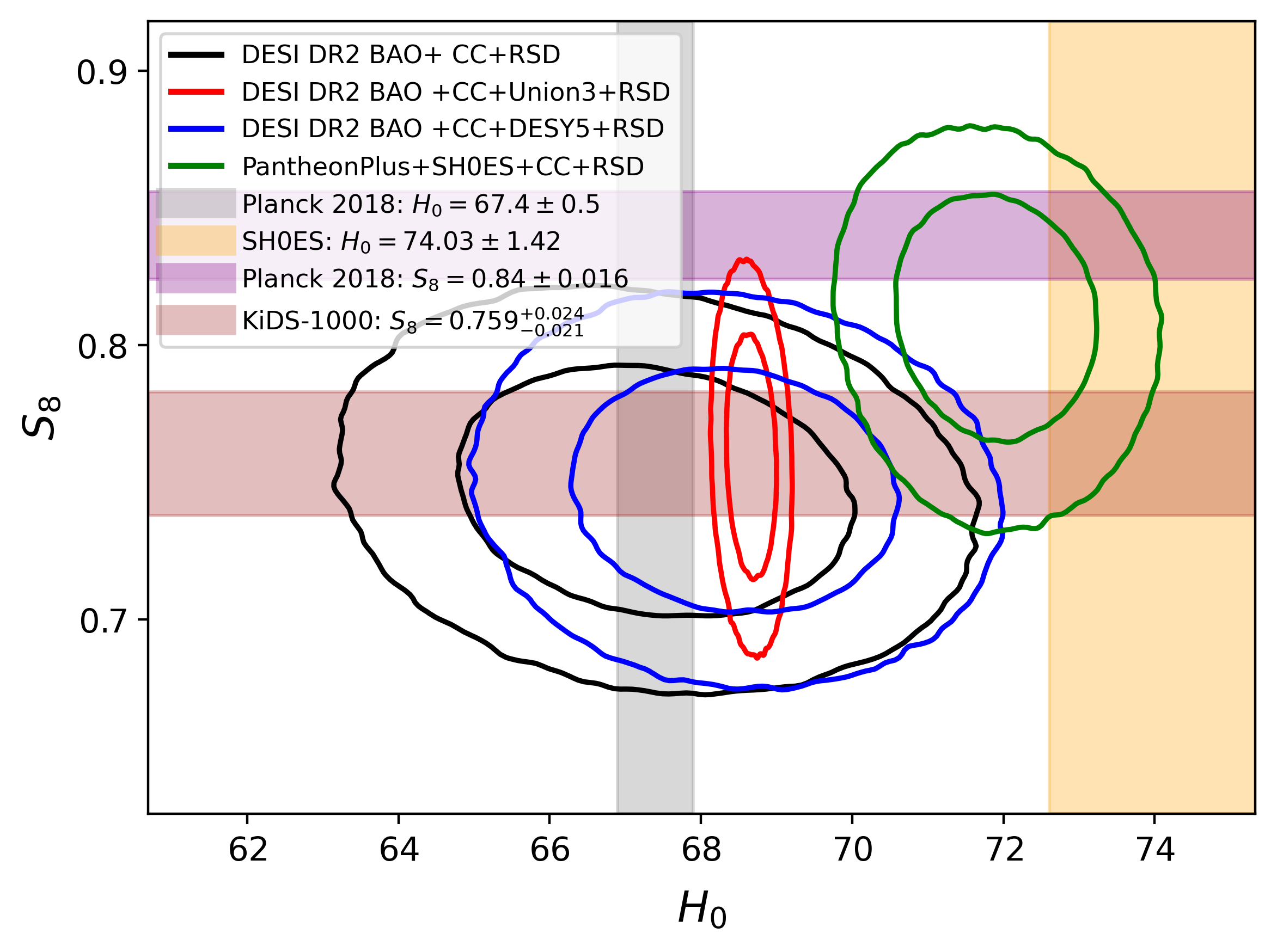  }
    \caption{ $S_8$-$H_0$ diagram for the Bianchi model. $H_0$ values are provided in ${\rm km\,s^{-1}\,Mpc^{-1}}$ units. The $H_0$ tensions are \({0.01\sigma,\, 0.36\sigma,\, 1.91\sigma,\, 2.52\sigma}\) with Planck 2018 data, and \({1.81\sigma,\, 1.79\sigma,\, 3.61\sigma,\, 0.96\sigma}\)  with SH0ES (\(H_0 = 74.03 \pm 1.42\)) measurements.  The \(S_8\) tension between the  \(w\)CDM model and  Planck~2018 ($S_8 = 0.834^{+0.016}_{-0.016}$  have a difference  \(1.41\sigma,1.45\sigma,1.26\sigma,0.42\sigma\), and \(0.19\sigma, 0.20\sigma, 0.03\sigma, 0.88\sigma\)  with KiDS-1000 data ($S_8 = 0.759^{+0.024}_{-0.021}$).}
    \label{fig:Bianchi}
\end{figure}
 \begin{figure}
    \includegraphics[width=1.0\linewidth]{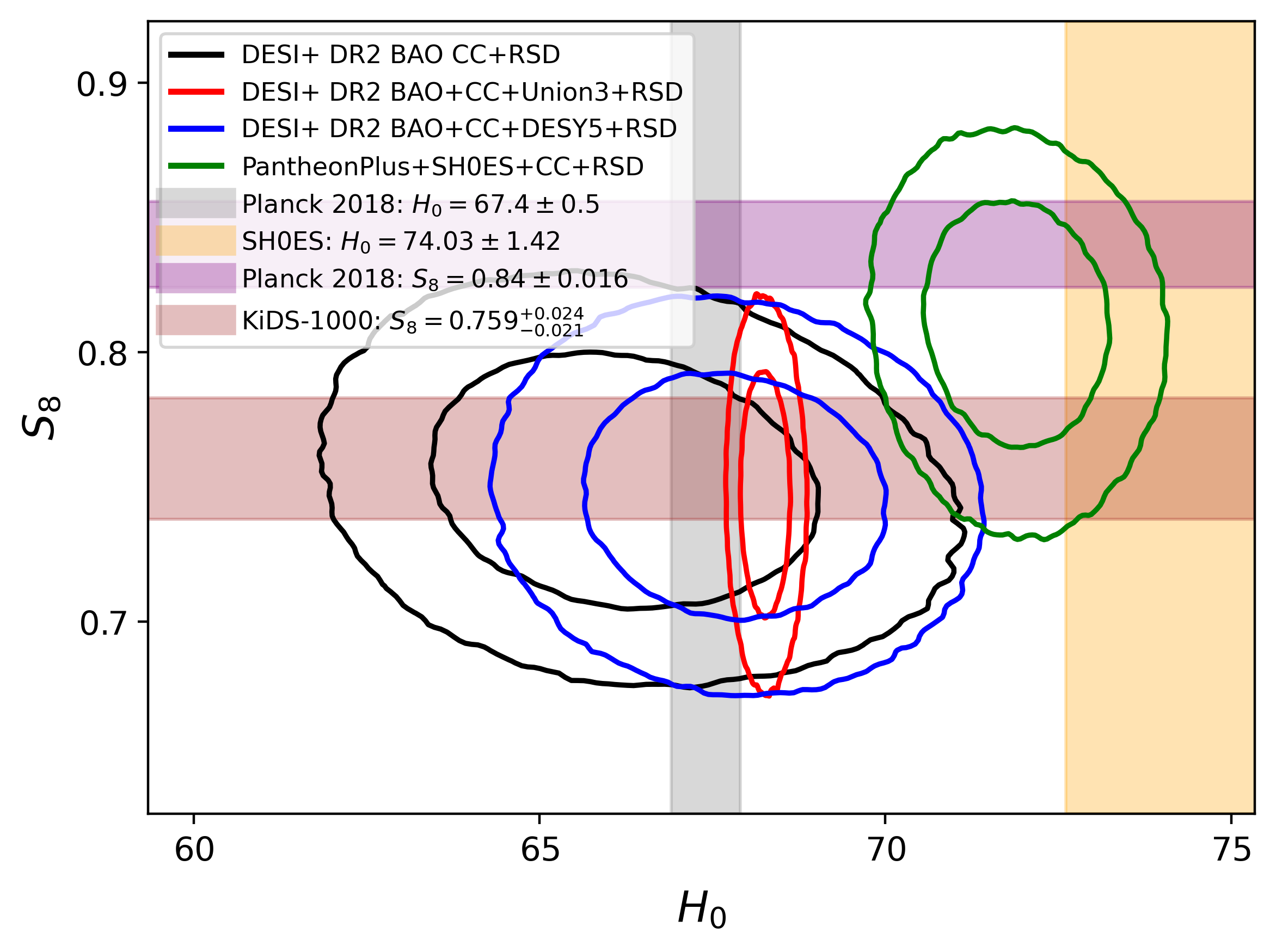  }
    \caption{ $S_8$-$H_0$ diagram for the \(w\)Bianchi model. $H_0$ values are provided in ${\rm km\,s^{-1}\,Mpc^{-1}}$ units.  {The $H_0$ tensions are \({0.01\sigma,\, 0.36\sigma,\, 1.91\sigma,\, 2.52\sigma}\) with Planck 2018 data, and \({1.81\sigma,\, 1.79\sigma,\, 3.61\sigma,\, 0.96\sigma}\)  with SH0ES (\(H_0 = 74.03 \pm 1.42\)) measurements. The \(S_8\) tension between the  \(w\)CDM model and  Planck~2018 ($S_8 = 0.834^{+0.016}_{-0.016}$  have a difference \({1.26\sigma,\, 1.42\sigma,\, 1.43\sigma,\, 0.36\sigma}\) and  \({0.08\sigma,\, 0.19\sigma,\, 0.18\sigma,\, 0.80\sigma}\)  with KidS-1000 data ($S_8 = 0.759^{+0.024}_{-0.021}$).}}
    \label{fig:wBianchix}
\end{figure}
 \begin{figure}
    \includegraphics[width=1.0\linewidth]{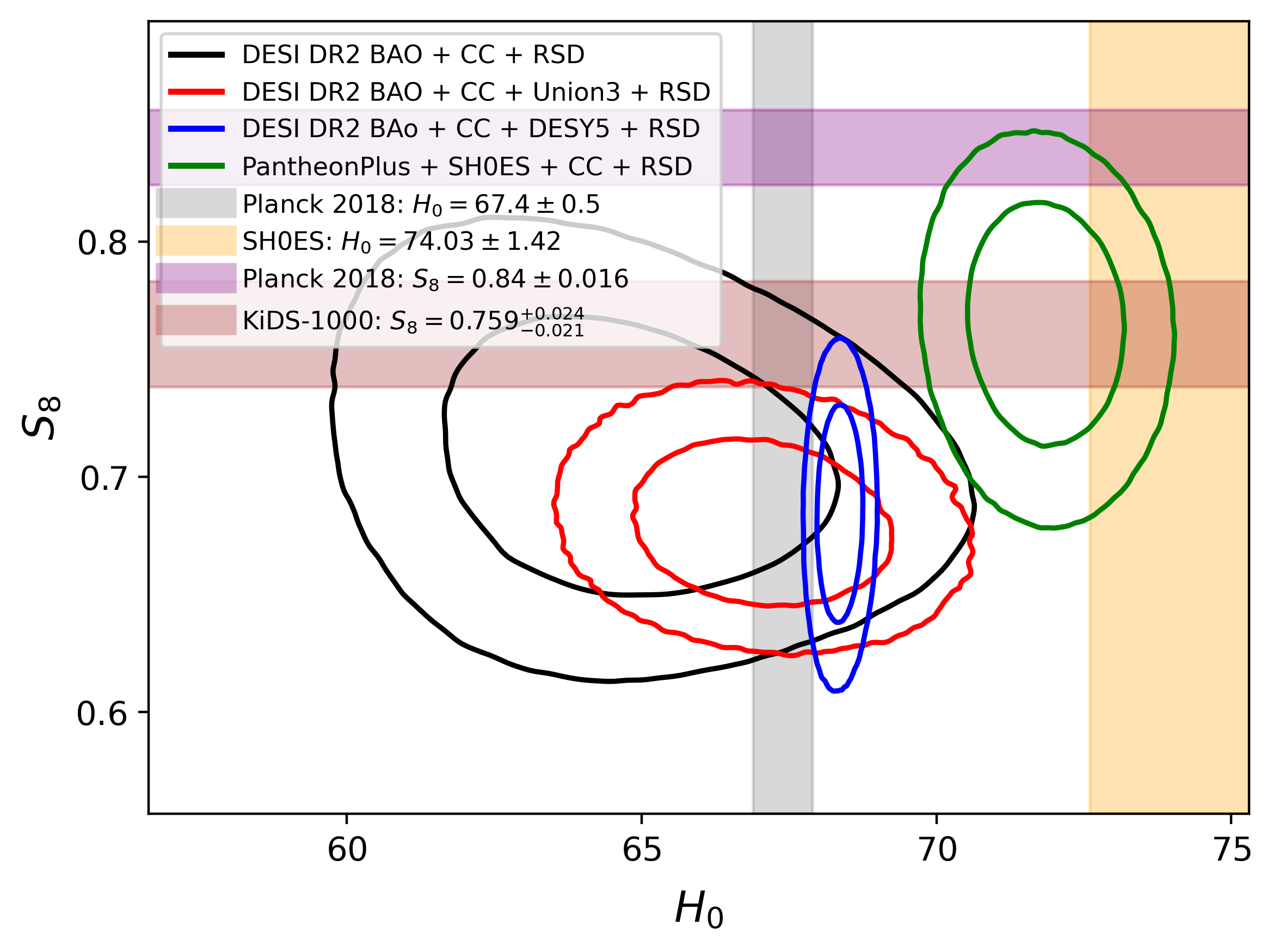  }
    \caption{ $S_8$-$H_0$ diagram for \(w_0w_a\)Bianchi model. {The $H_0$ tensions are  \({0.04\sigma,\, 0.22\sigma,\, 1.38\sigma,\, 2.46\sigma}\) with Planck 2018, and \({2.04\sigma,\, 2.06\sigma,\, 3.76\sigma,\, 0.98\sigma}\)  with SH0ES (\(H_0 = 74.03 \pm 1.42\)) measurements.  The \(S_8\) tension between the  \(w\)CDM model and  Planck~2018 ($S_8 = 0.834^{+0.016}_{-0.016}$  have a difference  \({1.93\sigma,\, 1.97\sigma,\, 1.42\sigma,\, 1.01\sigma}\) and \({0.85\sigma,\, 0.86\sigma,\, 1.17\sigma,\, 0.08\sigma}\)  with KiDS-1000 data ($S_8 = 0.759^{+0.024}_{-0.021}$).}}
    \label{fig:ww0Bianchi}
\end{figure}
    The \(S_8\) tension between the $w$CDM model and Planck 2018 data have a difference $1.21\sigma, 0.12\sigma,1.26\sigma,1.79\sigma$  and $0.02\sigma, 1.36\sigma,0.09\sigma,0.42\sigma$  with KiDS-1000 data. From Fig. \ref{fig:nonflatx1}, we observe that the model values of $H_0$ is favored to Planck 2018 measurements except \textit{PantheonP + SH0ES + CC + RSD}. While the model values of $S_0$ are favored to  KiDS-1000 data measurements except \textit{PantheonP + SH0ES + CC + RSD}.  All these tensions are smaller than the standard ones: $4.39\sigma$ between \textit{Planck} 2018 and SH0ES measurements of $H_0$, and $2.86\sigma$ between \textit{Planck} 2018 and KiDS-1000 measurements of $S_8$. This indicates the $w$CDM model is a potential candidate to resolve both tensions.
\item The model values of $H_0$ at 95\% C.L  taken from  Table \ref{tab3:cosmo_constraints} for the case of the  \(w_0w_a\)CDM model exhibits tensions of \({0.65\sigma, 0.00\sigma, 0.86\sigma, 2.38\sigma}\) with Planck 2018, and \({2.18\sigma,\, 2.08\sigma,\, 3.95\sigma,\, 1.07\sigma}\) with SH0ES using the same datasets considered. From Fig. \ref{fig:nonflatx2w0walcdm}, we note that the $H_0$ value for the \textit{DESI DR2 BAO + CC + RSD} case is smaller. For the \textit{DESI DR2 BAO + CC + Union3 + RSD} and \textit{DESI DR2 BAO + CC + DESY5 + RSD} cases, the $H_0$ values are consistently favored over the \textit{Planck} measurements.  We also find consistently lower values of $S_8$ across all joint datasets, with a large discrepancy (see Fig.~\ref{fig:nonflatx2w0walcdm}). The $S_8$ tension between the $w_0w_a$CDM model and \textit{Planck-2018} is ${1.94\sigma,2.87\sigma,5.18\sigma,2.55\sigma}$, and with KidS-1000 it is ${0.78\sigma,1.35\sigma,2.18\sigma,1.20\sigma}$. This indicates that the  \(w_0w_a\)CDM model performs poorly in alleviating the $S_8$ tensions.  %The model The sigma differences are still lower than the standard $4.39\sigma$.
\item The inferred $H_0$ values for the Bianchi Type model shows tensions of \({0.01\sigma,\, 0.36\sigma,\, 1.91\sigma,\, 2.52\sigma}\) with Planck-2018 data, and \({1.81\sigma,\, 1.79\sigma,\, 3.61\sigma,\, 0.96\sigma}\) with SH0ES measurements. The \(S_8\) tension between the Bianchi Type V model and Planck-2018 are \(1.41\sigma,1.45\sigma,1.26\sigma,0.42\sigma\), while the differences with KidS-1000 are \(0.19\sigma, 0.20\sigma, 0.03\sigma, 0.88\sigma\) were well demonstrated in Fig. \ref{fig:Bianchi} using all considered datasets. From this figure, we observe that all tension values are lower than the standard tension reference for $H_0$ and $S_8$ measurements. This indicates that our hypothetical model may have the potential to alleviate both tensions consistently. 
\item  Using the above considered datasets, the values of $H_0$ for the \(w\)Bianchi model have tensions of \({0.65\sigma,\, 0.02\sigma,\, 0.86\sigma,\, 2.40\sigma}\) with Planck-2018 measurement,   and \({1.98\sigma,\, 1.96\sigma,\, 3.83\sigma,\, 0.98\sigma}\) with SH0ES. Similarly the values of $S_8$  have a deviation of \({1.26\sigma,\, 1.42\sigma,\, 1.43\sigma,\, 0.36\sigma}\) with Planck-2018 and  \({0.08\sigma,\, 0.19\sigma,\, 0.18\sigma,\, 0.80\sigma}\) with KidS-1000 measurements. {Still, all significance differences are lower than our standard references for $H_0$ and $S_8$, as presented in Fig. \ref{fig:wBianchix}.}  From this plot, the values of $H_0$ and $S_8$ at 95\% C.L. have been considered to ignite the tensions. 
\item The significance of $H_0$ for  \(w_0w_a\)Bianchi Type V model is shown in the tensions \({0.64\sigma,\, 0.13\sigma,\, 1.38\sigma,\, 2.46\sigma}\) with Planck 2018, and \({2.13\sigma,\, 2.23\sigma,\, 3.76\sigma,\, 0.98\sigma}\) with SH0ES. For the matter clustering  $S_8$, the significance is given by \({1.93\sigma,\, 1.97\sigma,\, 1.42\sigma,\, 1.01\sigma}\) with Planck-2018 measurement and $KiDS$ \({0.85\sigma,\, 0.86\sigma,\, 1.17\sigma,\, 0.08\sigma}\), see figure \ref{fig:ww0Bianchi}. We noticed that the model is favored in Planck-2018 for $H_0$ and KidS-1000 for $S_8$. While the significance difference is in the range of the standard references, the model is not totally ruled out at this level. 
\end{enumerate}
Overall, the above deviations show tensions below the  \({4.39\sigma}\) range, which represents the probability distribution of  $H_0$ at the 95\% confidence level, together with the corresponding Planck-2018 and SH0ES measurements.  We find that the inferred $H_0$ values for most models fall within an intermediate range between the two measurements. This indicates that DDE, particularly in the  $\Lambda$CDM, $w$CDM, Bianchi~Type~V and $w$Bianchi Type V  models, may serve as a promising candidate for alleviating the Hubble tension and $S_8$ across all combined datasets, thereby igniting speculation about new insights on physics beyond $\Lambda$CDM. From our results (see Table \ref{tab3:cosmo_constraints}), we note that lower values for $S_8$ are consistently obtained for the $w_0w_a$CDM model across all joint datasets. This suggests that the evolving dark energy EoS, parameterized as $w_{\rm de} = w_0 + w_a(1 + a)$, leads to weaker structure growth. Consequently, resolving the \(S_8\) tension introduces a penalty on the model's fit. However, its predictions should still be considered consistent with future observations at more than a 95\% confidence level. In a similar vein, the $w_0w_a$Bianchi  framework shifts toward lower \(S_8\) values. Since the 95\% C.L. of these model values still falls within a standard deviation of $2.68\sigma$, the framework is not entirely ruled out. Notably, when evaluating the parameter space beyond the 95\% threshold, this approach demonstrates an enhanced capability to consistently mitigate both cosmological tensions. Additional high-redshift data, including CMB observations, would be necessary for a more conclusive analysis; however, incorporating CMB data is beyond the scope of this work.

 \begin{figure}
    \includegraphics[width=1.15 \linewidth]{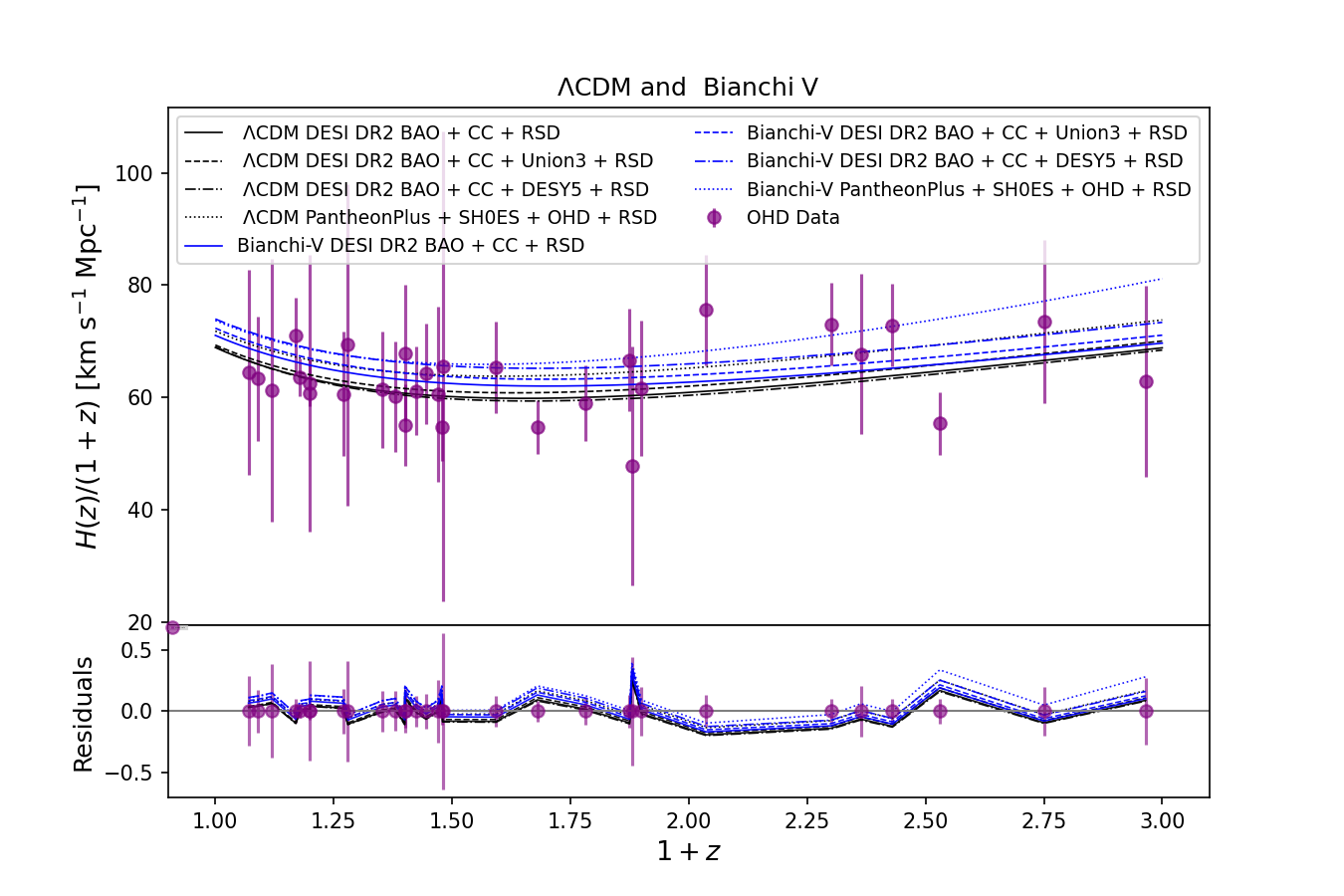}
      \includegraphics[width=1.15\linewidth]{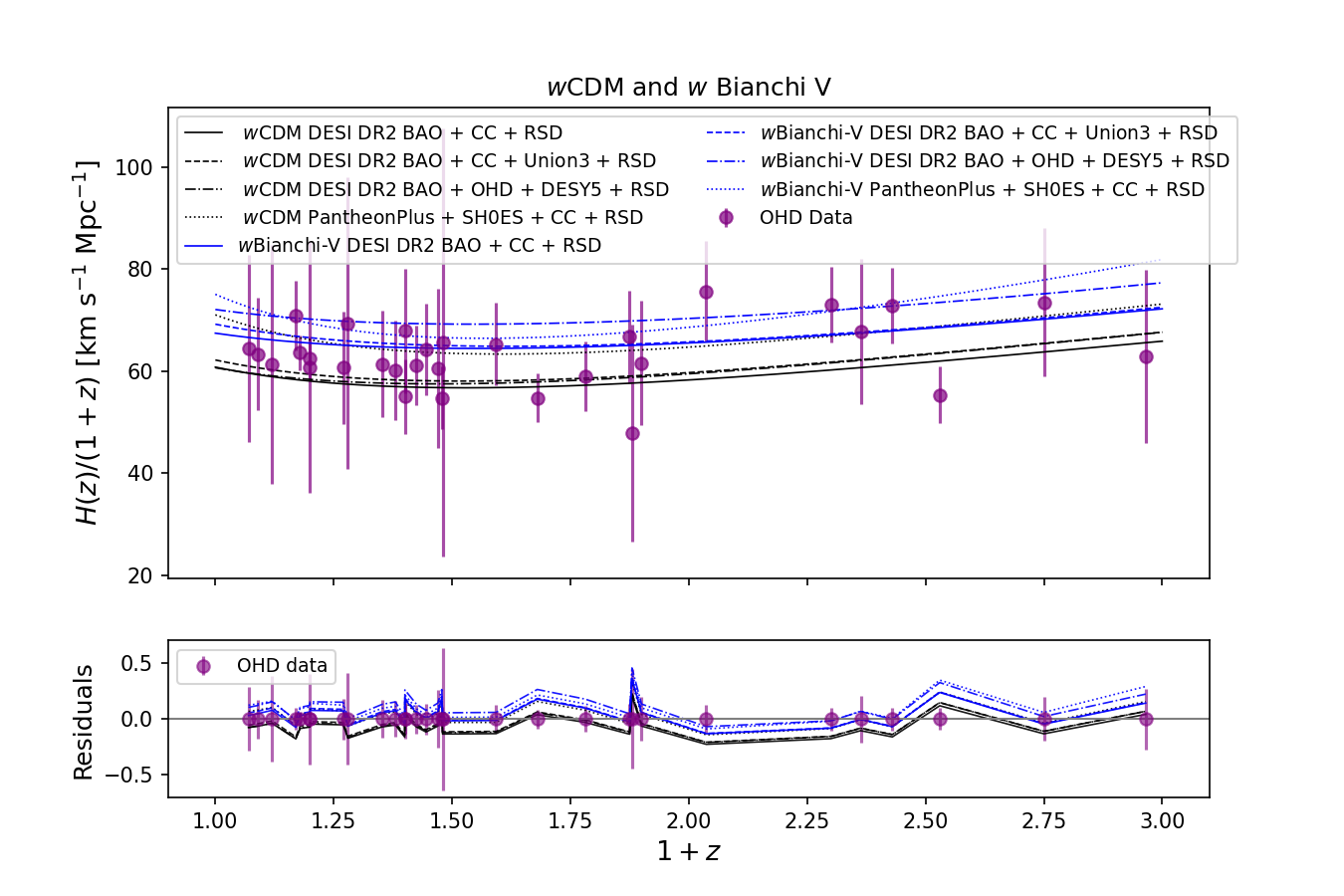}
      \includegraphics[width=1.15 \linewidth]{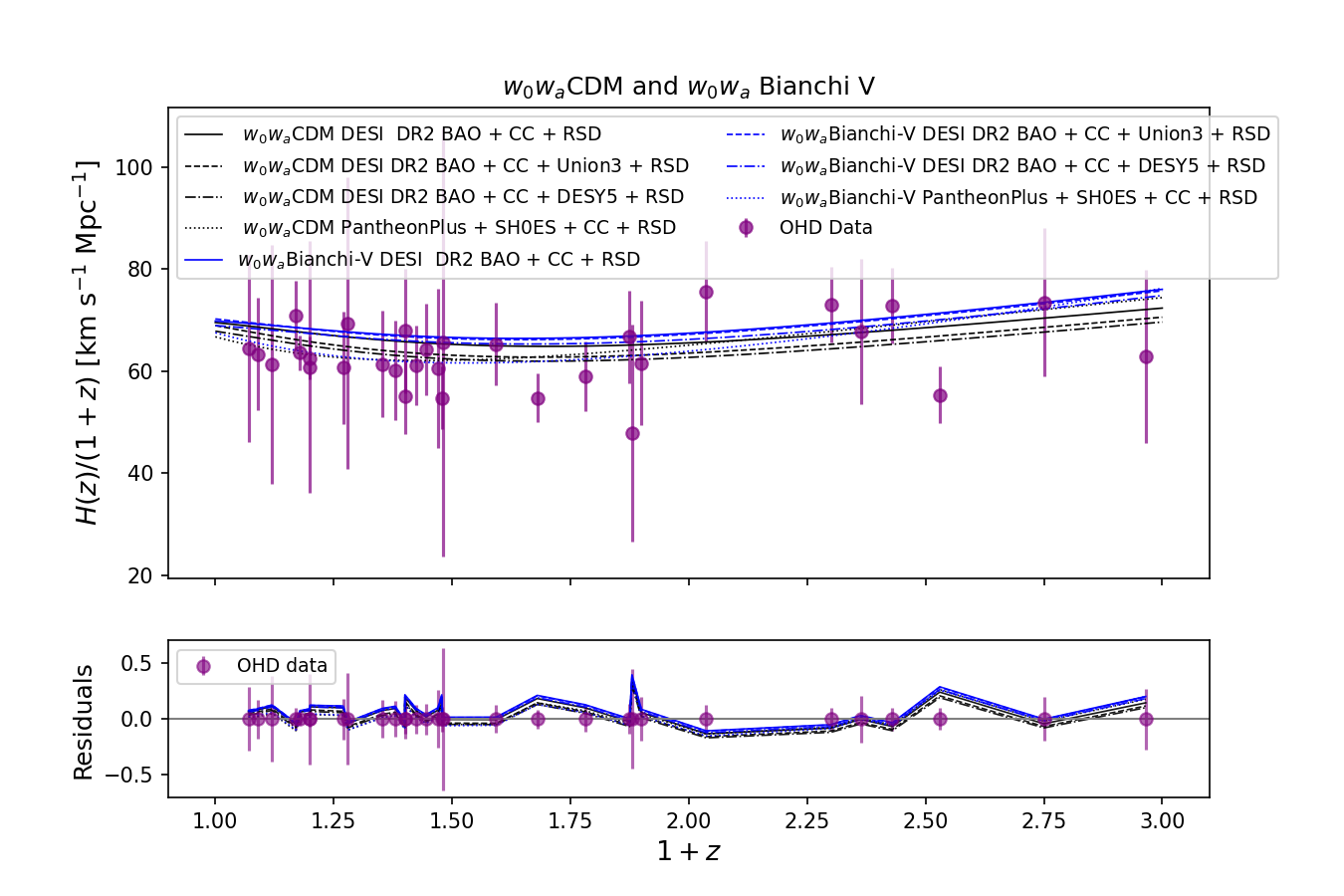}
    \caption{The figure shows the Hubble parameter, $H(z)/(1+z)$, along with the corresponding residuals as per its definition in  Eq. 
\eqref{Residual},
for different cosmological models. The \textit{Upper panel} presents results for the  $\Lambda$CDM and Bianchi Type V models, the \textit{Middle panel} for the  $w$CDM and $w$Bianchi Type V models, and the \textit{Bottom panel} for the  $w_0w_a$CDM and $w_0w_a$Bianchi Type V models. All panels use the best-fit values of the constrained parameters listed in Tables \ref{tab1:cosmo_constraints}, \ref{tab2:cosmo_constraints}, and \ref{tab3:cosmo_constraints} for the combined datasets, respectively.}
    \label{fig:Hubble}
\end{figure}
 \subsection{Results for $\Omega_{\sigma0}$, $\Omega_{k0}$, $\Omega_{m0}$, $r_d$ and $M_{abs}$ values }
 % \adlcd{As mentioned above, change the title. We are not parametrizing these quantities, we are obtaining statistical constraints.}
 
The above Tables \ref{tab1:cosmo_constraints} - \ref{tab3:cosmo_constraints} also show that the constraint value of the Bianchi Type V anisotropy density parameter $\Omega_{\sigma0}$ is of the order of $2.44 \times 10^{-4}$, which is ten times smaller than the expected value of the CMB value $10^{-5}$, as mentioned earlier. All the datasets used here give tight constraints and improve the value of $\Omega_{\sigma0} $, as opposed to the work of \cite{amirhashchi2020constraining}. These tables also show that the constrained values of $\Omega_{k0}$ for all the considered cosmological models provide the geometry of the Universe. The fractional matter density $\Omega_{m0}$ and sound horizon at the drag epoch $r_d$, which originated from BAO, provide valuable information about the dynamical evolution of the Universe (see \citep{de2023sound} for more details). We consistently obtain lower values of $r_d$ across all models using the combined datasets of \textit{DESI DR2 BAO + CC + Union3 + RSD} and \textit{DESI DR2 BAO + CC + DESY5 + RSD} due to the presence of the compilations of SNIa distance moduli. The absolute magnitude $M_{abs}$ is also constrained from Pantheon +SH0ES, which calibrated the distance ladder as presented in \citep{brout2022pantheon+} for all models. 
%\adlcd{The last sentence {\it The same as the absolute...} does not make sense. Please rephrase.}
\\
\\
We now proceeded a step further and showed the numerical results of various physical quantities at various redshifts. Specifically, the Hubble parameter $H(z)$ with the \textit{CC} data presented in Fig. \ref{fig:Hubble}, the fractional density parameters $\Omega_i$, where $i$ represents each fluid,
% = \Omega_{i0}(1+z)^n/(H/H_0)^2$ as shown in Fig. \ref{fig:fractionaldensity}, where $n = w_i$, \adlcd{This thing of $n = w_i$  is not understood. What are the possibilities for $i$?}
the distance measurements with the result $DESI$ $DR2$ $BAO$ measurements (i.e., $D_m/r_d$, $D_H/r_d$, and $D_V/r_d$ in Mpc)   presented in Fig. \ref{fig:distance}, and the growth factor $D(a)$ presented in Fig. \ref{fig:growthfactor} to explain the cosmic dynamics using the best-fit values of these constrained cosmological parameters taken from Tables \ref{tab1:cosmo_constraints} - \ref{tab3:cosmo_constraints}.

\subsection{Numerical Results of $H(z)$}

In Fig.~\ref{fig:Hubble}, we also include lower panels for the residual, defined as  
\begin{equation}
\label{Residual}
\mathrm{R}(z_i) = \frac{H_{\mathrm{model}}(z_i) - H_{\mathrm{obs}}(z_i)}{H_{\mathrm{obs}}(z_i)} \, .
\end{equation}
The \textit{top-panel} of Fig. \ref{fig:Hubble} shows the $H(z)$ diagram for the  \(\Lambda\)CDM (black curves) and Bianchi Type~V (blue curves) models with the residues. The deviation between the two models is consistent at lower and higher redshifts.  The Bianchi Type V model provides a good fit to the CC data at large redshifts, $z > 1$. In contrast, the  \(\Lambda\)CDM does so at a lower redshift $z <0.75$ where their Chi-square values $\chi^2$ is presented later in subsection \ref{stastical} Table \ref{tab:model_comparison}.    The \textit{middle panel} of Fig. \ref{fig:Hubble} shows the  \(w\)CDM (black curves)  and \(w\)Bianchi Type~V  (blue curves) models. The deviation between the two models is very minimal across all models except the blue-dotted curve and black-dotted curve, which are the best fit values as obtained from the  \textit{PantheonP+CC+RSD} datasets, see their $\chi^2$ values in Table \ref{tab:model_comparison} for further information.  The corresponding results also reflect this.  Similarly, in the  \textit{bottom panel} of the diagram, $H(z)$  is presented for the case of  \(w_0w_a\)CDM and \(w_0w_a\)Bianchi Type~V models.   The result indicates that at low redshifts \(w_0w_a\)CDM models (black curves)  has a good fit with CC data, while at high redshifts, the \(w_0w_a\)Bianchi Type~V models (blue curves).  This may arise from the inclusion of the shear term \(\rho_\sigma\), which contributes to the anisotropic expansion of the Universe. Although the present value of \(\Omega_{\sigma0}\) is very small since \(\rho_\sigma \propto a^{-6}\), its effect becomes more pronounced at large redshifts, influencing the early cosmic evolution, as reflected in Fig.~\ref{fig:fractionaldensity}. 

\subsection{Numerical results of fractional density parameters}
As we clearly observe in the fractional energy densities diagram displayed in Fig. \ref{fig:fractionaldensity}, the contribution of the shear fluid has a significant influence on the early universe across all models. According to the \textit{Planck} 2018 measurements, the matter-dark energy equality redshift  is
$z_{\rm eq, Planck} = \left(\frac{\Omega_\Lambda}{\Omega_{m0}}\right)^{1/3} - 1 \approx 0.296,$
since \(\Omega_{m0} = 0.315\) and \(\Omega_\Lambda = 0.685\) \citep{aghanim2020planck}. Due to the dynamical behavior of our dark energy models, our numerical results vary from the Planck value. This means that: i)  the matter-dark energy equality redshift  is \(z_{\rm eq} = 0.321,\; 0.296,\; 0.243,\; 0.271\) for the  \(\Lambda\)CDM model, and ii) 
 \(z_{\rm eq} = 0.356,\; 0.354,\; 0.350,\; 0.290\) for the Bianchi Type~V model, using \textit{DESI  DR2 BAO + CC + RSD}, \textit{DESI  + CC + Union3 + RSD},\textit{DESI  DR2 BAO + CC + DESY5 + RSD}, and \textit{PantheonP + SH0ES + CC + RSD}, respectively. iii) Using the same datasets, the equality redshifts are \(z_{\rm eq} = 0.375,\; 0.365,\; 0.383,\; 0.292\) for the \(w\)CDM model, iv) \(z_{\rm eq} = 0.365,\; 0.367,\; 0.372,\; 0.290\) for the \(w\)Bianchi Type~V model. v) Similarly, \(z_{\rm eq} = 0.375,\; 0.363,\; 0.383,\; 0.383\)  for the \(w_0w_a\)CDM model and \(z_{\rm eq} = 0.365,\; 0.367,\; 0.372,\; 0.399\) , for the \(w_0w_a\)Bianchi Type~V model.

% Here is the final version with "$z_{\rm eq} =$" included directly before every value in the table matrix to make it completely explicit.

 These results indicate that the \textit{Planck} 2018 value, \(z_{\rm eq} = 0.296\), corresponds to a later onset of dark energy domination. In contrast, our results on the equality of matter-dark energy windows in the range $0.290 \leq z_{\rm eq} \leq 3.99 $ suggest that dark energy becomes dominant at an earlier epoch. As seen in Fig. \ref{fig:fractionaldensity}, the influence of the shear fluid is clearly reflected in this behavior, while the \(w_0w_a\) Bianchi Type~V model remains relatively close to the \textit{Planck} result.
 \begin{figure}
    \includegraphics[width=1.0\linewidth]{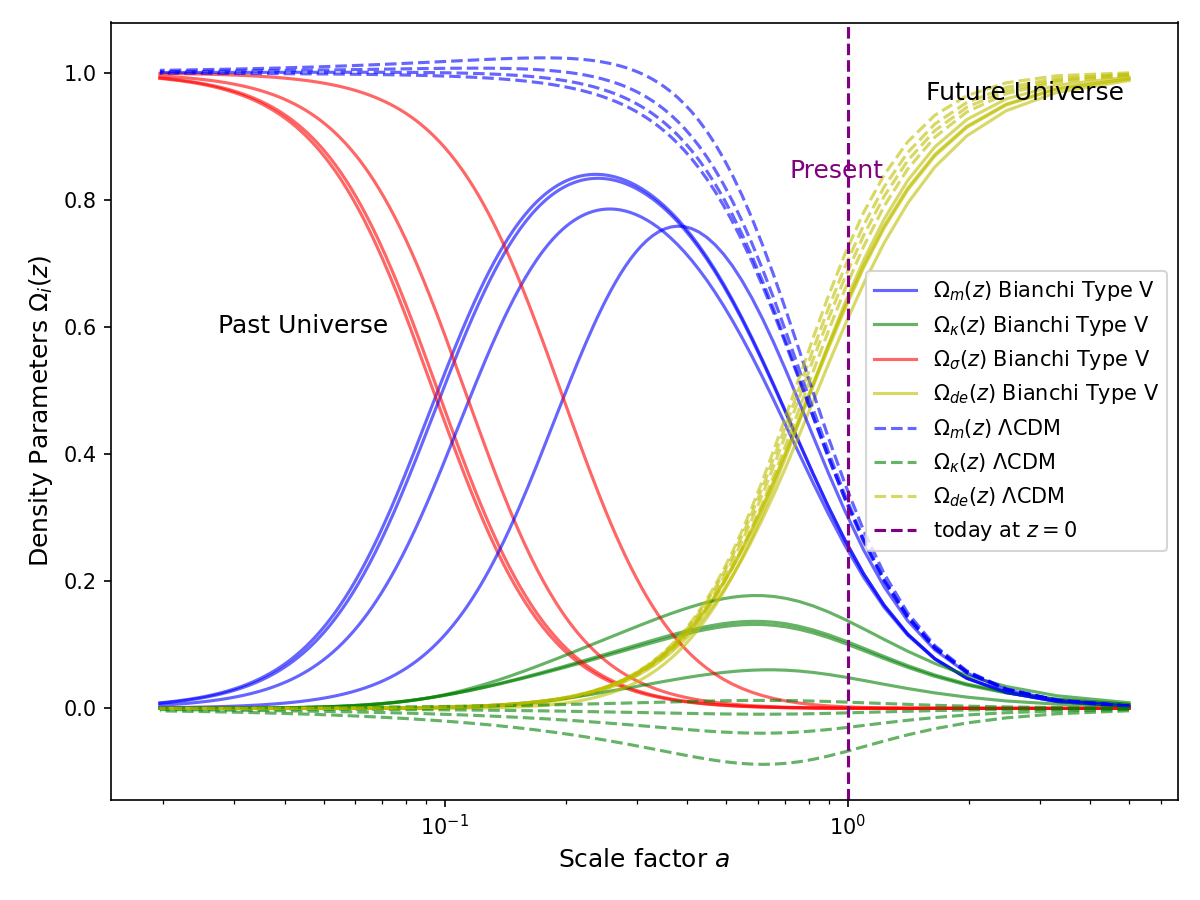}
    \includegraphics[width=1.0\linewidth]{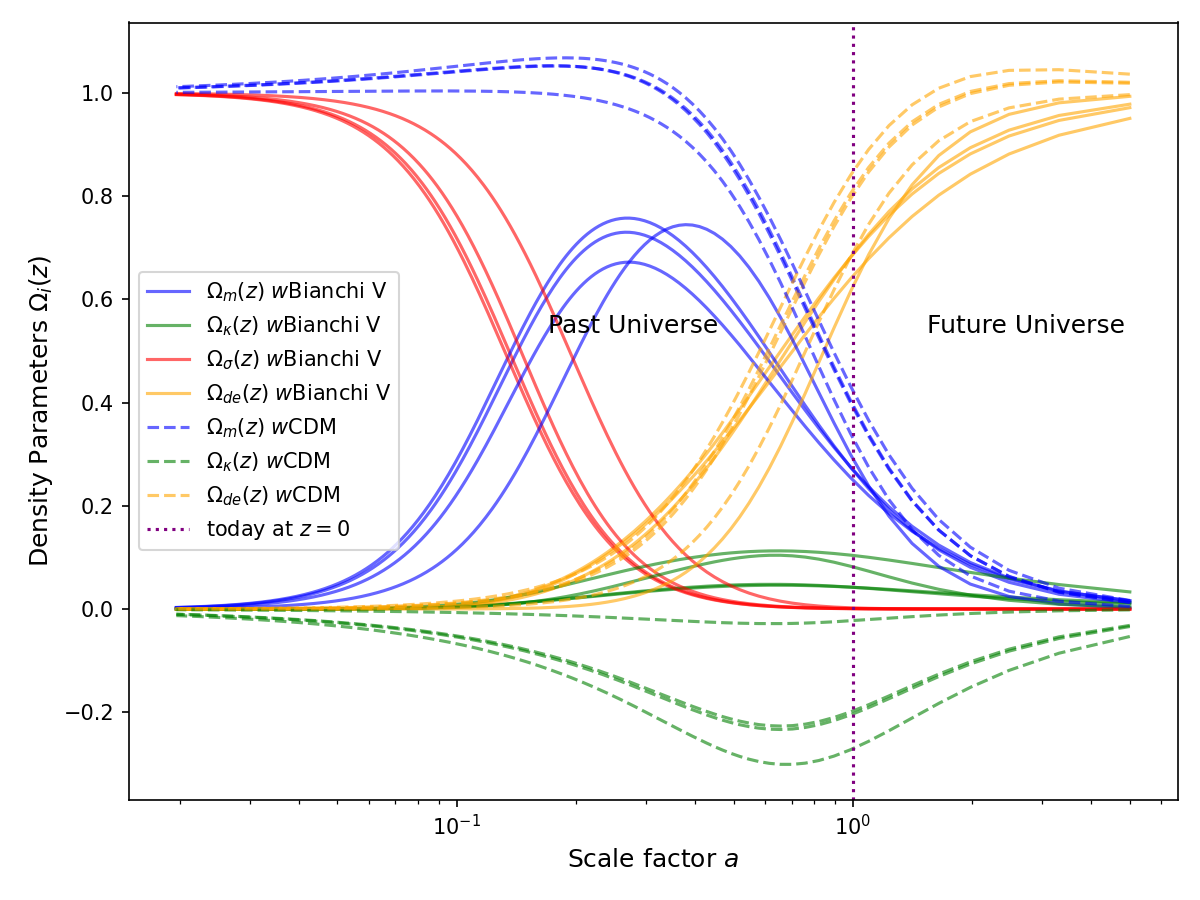}
    \includegraphics[width=1.0\linewidth]{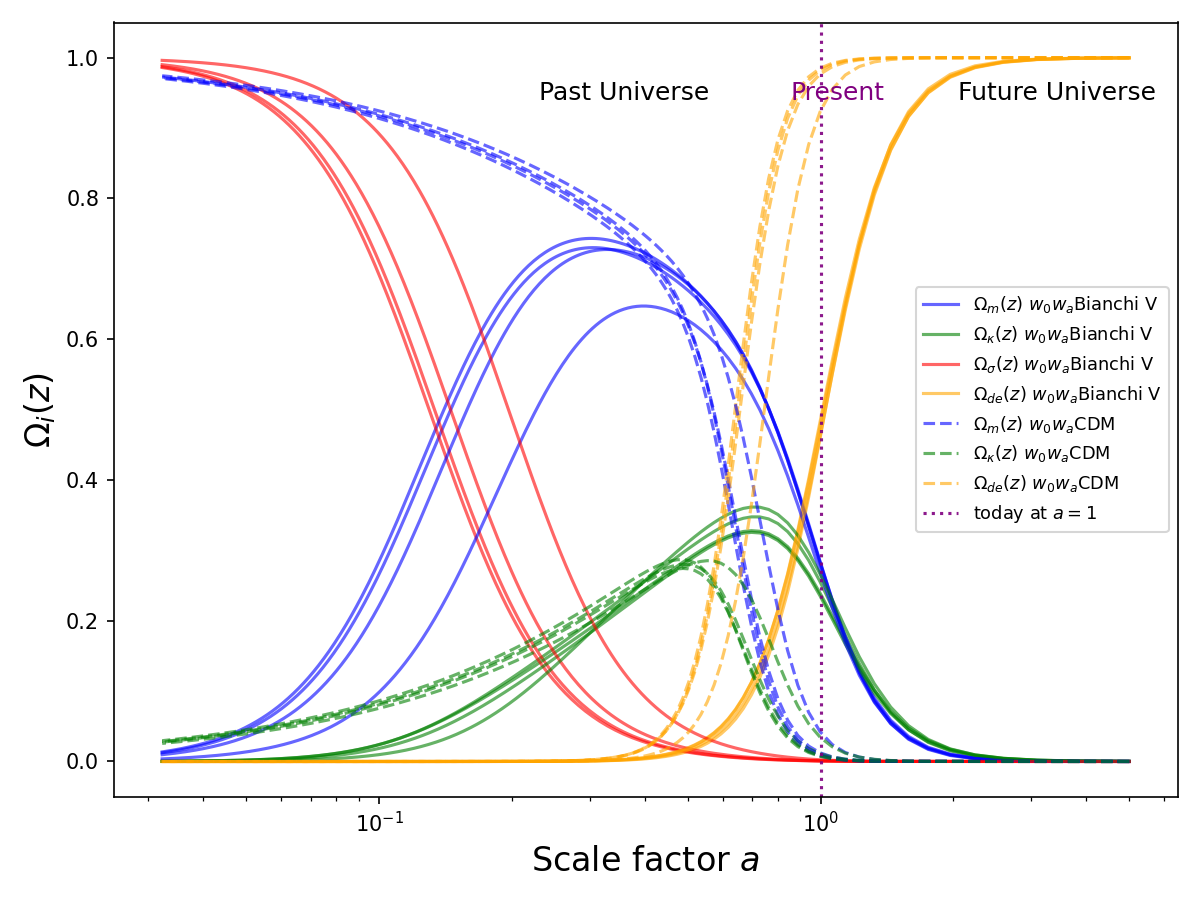}
    \caption{Fractional density parameters for matter fluid $\Omega_{m}(a)$, curvature fluid $\Omega_{k}(a)$, shear fluid $\Omega_{\sigma}(a)$ and dark energy $\Omega_{\rm DE}(a)$ for different cosmological models. The \textit{top panel} presents results for the  $\Lambda$CDM and Bianchi Type V models, the \textit{middle panel}  for the  $w$CDM and $w$Bianchi Type V models, and the \textit{bottom panel} for the  $w_0w_a$CDM and $w_0w_a$Bianchi Type V models. All plots use the best-fit values of the constrained parameters listed in Tables \ref{tab1:cosmo_constraints}, \ref{tab2:cosmo_constraints}, and \ref{tab3:cosmo_constraints} for the combined datasets, respectively.}
    \label{fig:fractionaldensity}
\end{figure}
\begin{figure}
    \includegraphics[width=1\linewidth]{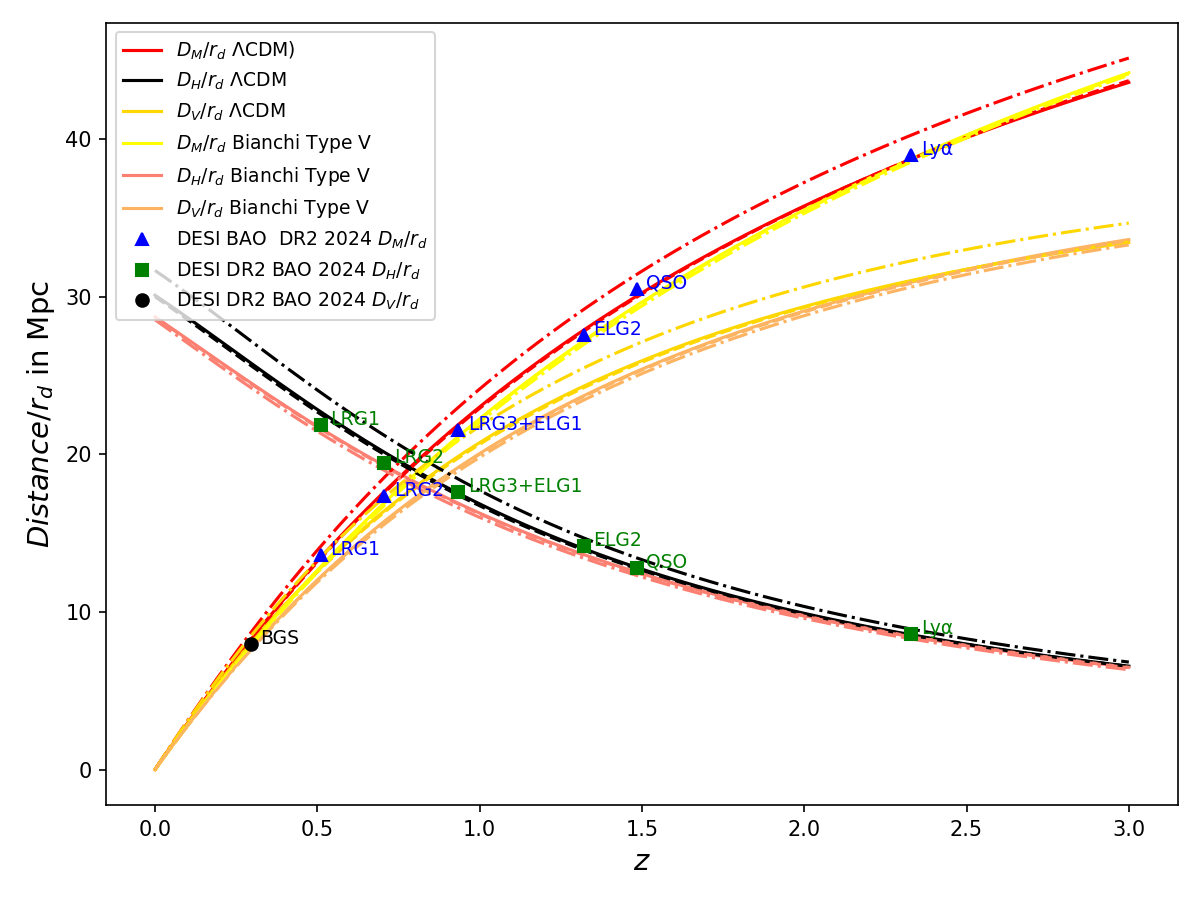}
     \includegraphics[width=1\linewidth]{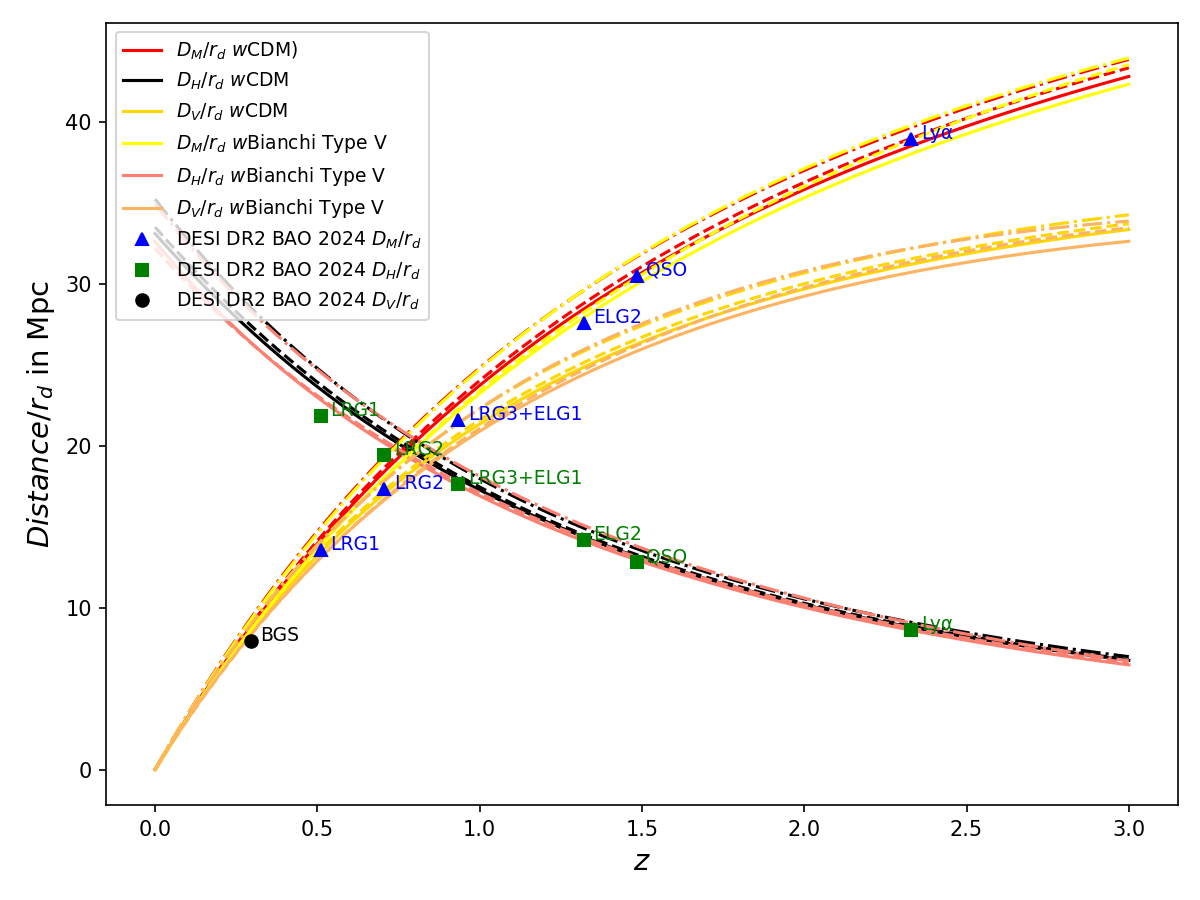}
      \includegraphics[width=1\linewidth]{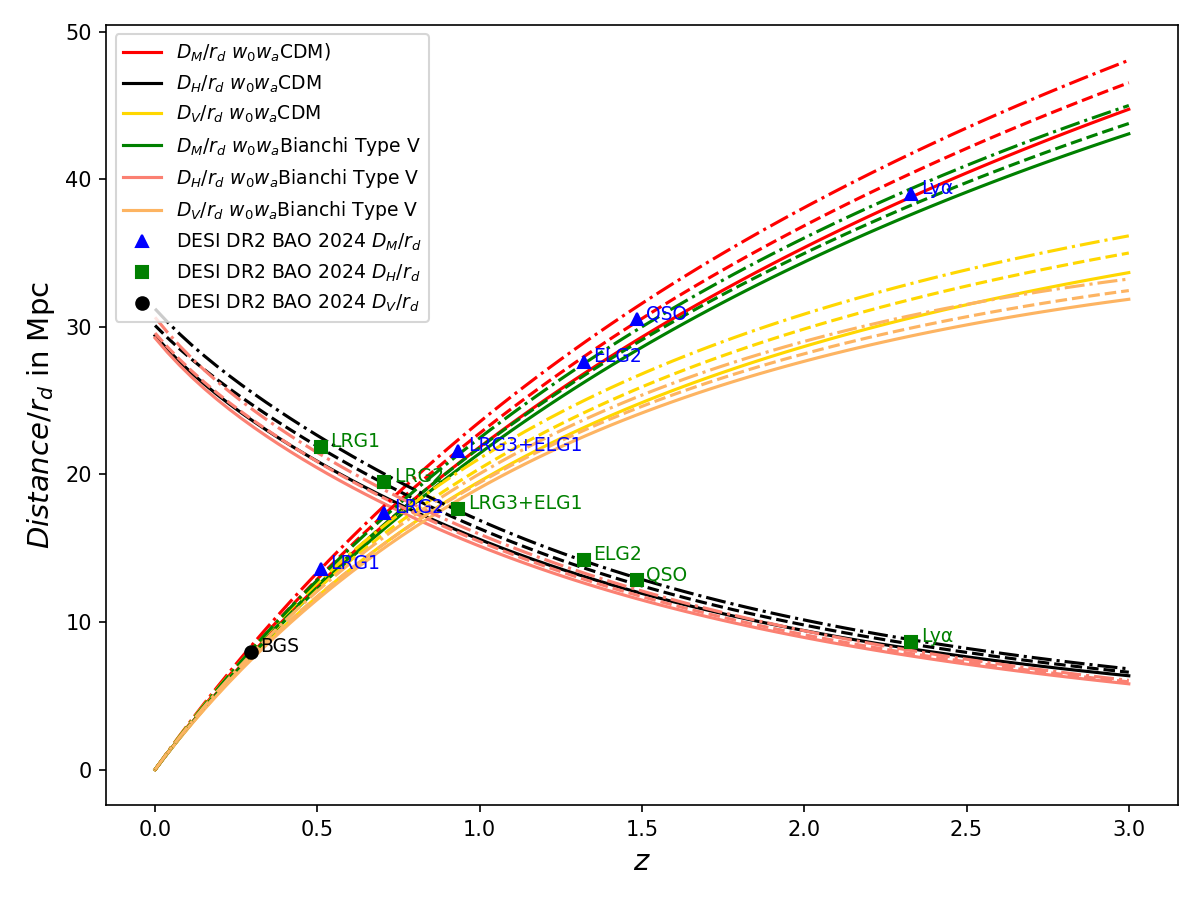}
    \caption{$BAO$ measurements of $D_H/r_d$, $D_M/r_d$ and $D_V/r_d$ with different cosmological models. The \textit{ top panel} presents results for the  $\Lambda$CDM and Bianchi Type V models, the \textit{middle panel}  for the  $w$CDM and $w$Bianchi Type V models, and the \textit{bottom panel} for the  $w_0w_a$CDM and $w_0w_a$Bianchi Type V models. All plots use the best-fit values of the constrained parameters listed in Tables \ref{tab1:cosmo_constraints}, \ref{tab2:cosmo_constraints}, and \ref{tab3:cosmo_constraints} for the combined datasets, respectively.}
    \label{fig:distance}
\end{figure}
\subsection{Numerical results of distance measurements}
In the same manner, using the best-fit values of the constrained parameters as taken from  Tables \ref{tab1:cosmo_constraints} - \ref{tab3:cosmo_constraints}, we also presented the cosmological distance statistically-preferred evolutions together with the recent measurements of $DESI$ $2024$ $BAO$ measurements in Fig. \ref{fig:distance}, where the \textit{top panel} shows the results for the  $\Lambda$CDM and Bianchi Type V models, the \textit{middle panel} does for the  $w$CDM and $w$Bianchi Type V models, and the \textit{bottom panel} doees for the  $w_0w_a$CDM and $w_0w_a$Bianchi Type V models. A similar result is obtained in \citep{de2019baryon}, although within the $\Lambda$CDM approach. The plot in Fig. \ref{fig:distance} shows the transverse comoving distance  $D_M/rd$, which represents the geometry of the expansion universe, the Hubble distance $D_H/rd$ represents the expansion scale along the line of sight,  and the volume-averaged $BAO$ distance measurements $D_V/rd$, which provide the joint measures of distance that have transverse and radial information,  using Eqs. \eqref{DH}, and \eqref{DV}. From this plot, we noticed minimal deviations between: $\Lambda$CDM vs Bianchi-V; $w$CDM vs $w$Bianchi-V; and $w_0w_a$CDM vs $w_0w_a$Bianchi-V  of $D_m/r_d$ and $D_V/r_d$ at lower redshifts and the deviation is slightly higher at higher redshifts, and the opposite conclusion is true for $D_H/r_d$ across all models. 
  \begin{figure}
%     \centering
    \includegraphics[width=1.0\linewidth]{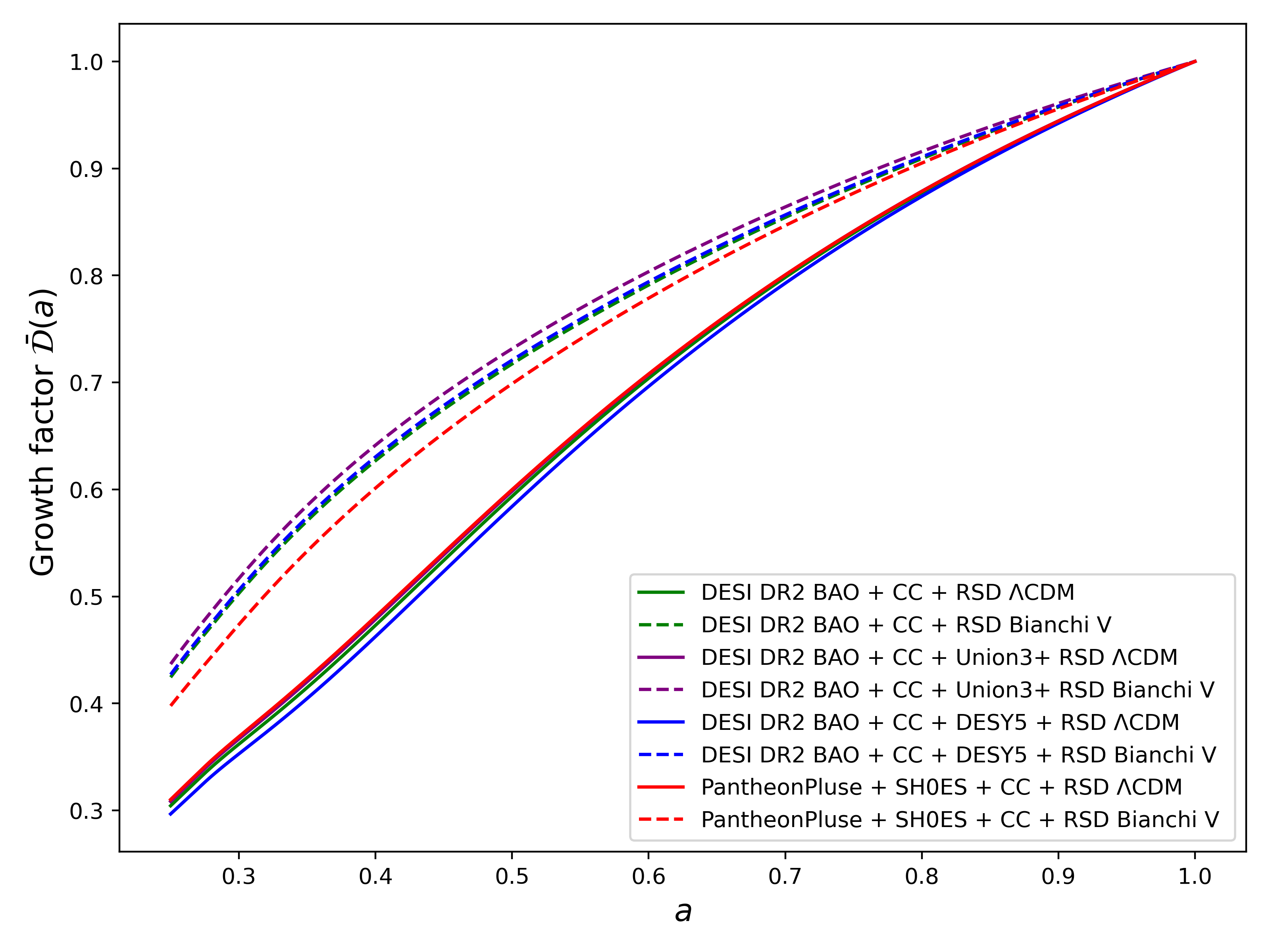}
    \includegraphics[width=1.0\linewidth]{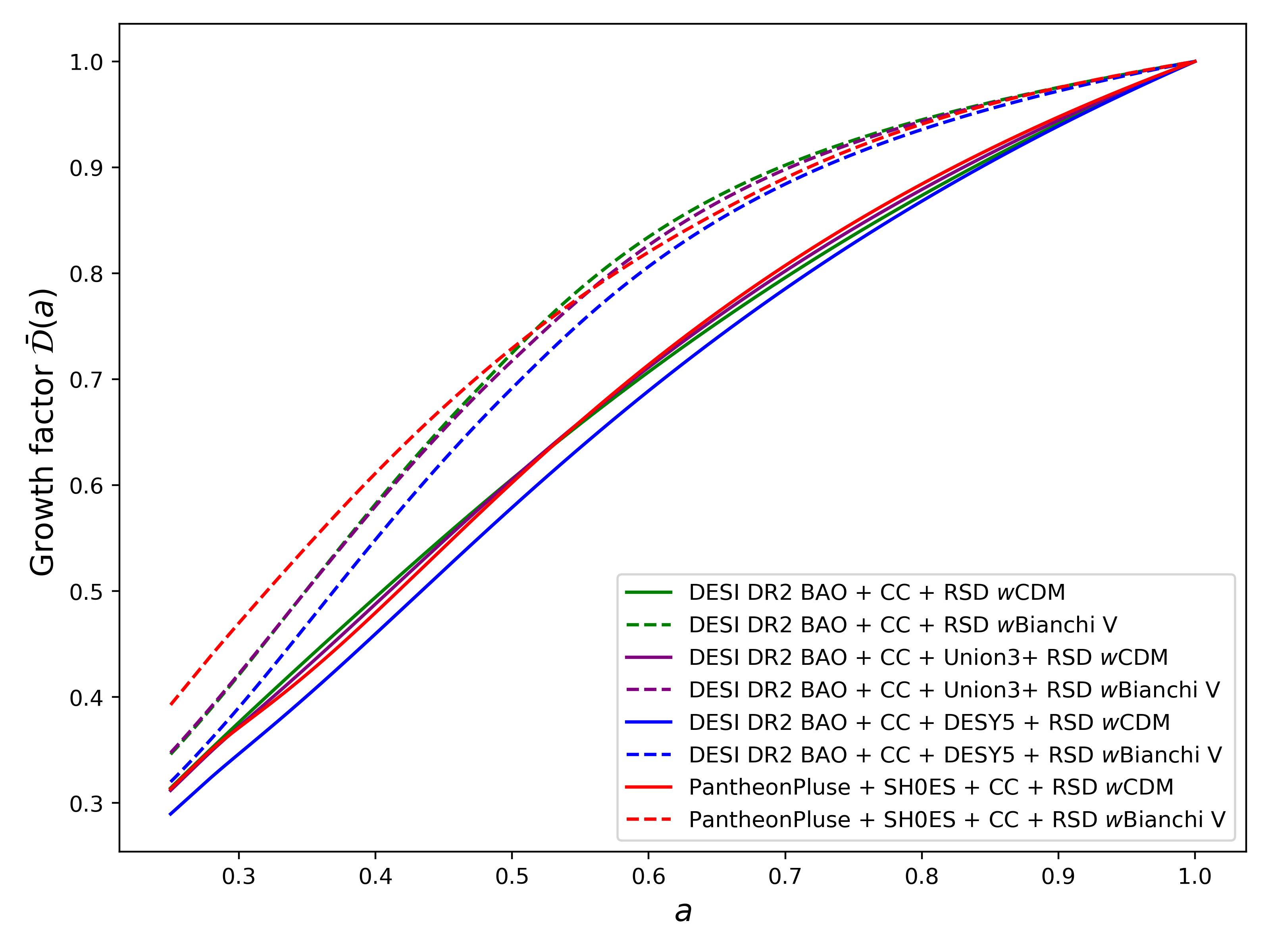}
    \includegraphics[width=1.0\linewidth]{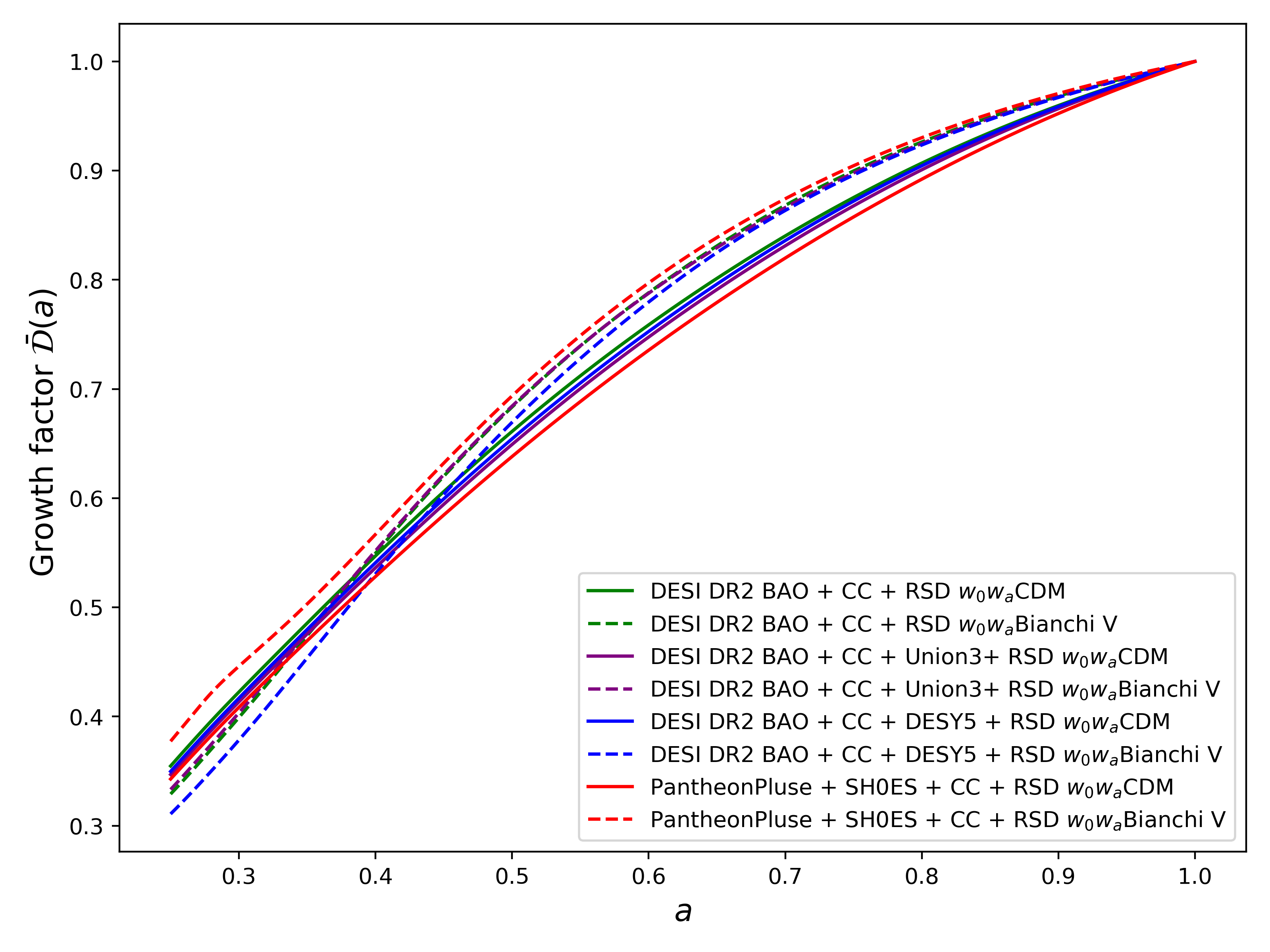}
     \caption{Growth factor $\mathcal{\bar{D}}$ evolution for different cosmological models. The \textit{top panel}  presents results for the  $\Lambda$CDM and Bianchi Type V models, the \textit{ top panel}  for the  $w$CDM and $w$Bianchi Type V models, and the \textit{bottom panel} for the  $w_0w_a$CDM and $w_0w_a$Bianchi Type V models. All plots use the best-fit values of the constrained parameters listed in Tables \ref{tab1:cosmo_constraints}, \ref{tab2:cosmo_constraints}, and \ref{tab3:cosmo_constraints} for the combined datasets, respectively.}
     \label{fig:growthfactor}
\end{figure}
\subsection{Numerical results of growth factor $D(a)$}
The growth factor  \(\mathcal{\bar{D}}(a)\) is numerically obtained using the expression in Eq. \eqref{densitycontarstistopic1} and shown in Fig. \ref{fig:growthfactor} using the best-fit values of the constrained parameters listed in Tables \ref{tab1:cosmo_constraints}, \ref{tab2:cosmo_constraints}, and \ref{tab3:cosmo_constraints} for the combined datasets accordingly. The figure shows the impact of an anisotropic/shear fluid on structure formation in all considered models. The top panel represents the numerical results of \(\mathcal{\bar{D}}(a)\) for the  \(\Lambda\)CDM and Bianchi Type V models and the middle panel shows the \(w\)CDM and \(w\)Bianchi Type V models. Similarly, the bottom panel presents the numerical results of \(\mathcal{\bar{D}}(a)\) for the \(w_0 w_a\)CDM and \(w_0 w_a\)Bianchi Type V models. 

As a result, a significant deviation of the amplitude of $\mathcal{\bar{D}}(a)$ has been shown in the current work between  \(\Lambda\)CDM versus Bianchi Type V models and the \(w\)CDM versus \(w\)Bianchi Type V models, especially when \(a \ll 0\) (see upper and middle panels of Fig. \ref{fig:growthfactor}): indeed the \(w_0w_a\)CDM and \(w_0w_a\)Bianchi Type V models (bottom panel of Fig. \ref{fig:growthfactor}) show relatively smaller variation. These results are consistent with findings in \citep{huterer2015growth}, which indicated that dark energy suppresses structure growth in the \(\Lambda\)CDM model. Even though dark energy suppresses structure growth in the Bianchi Type V model, the shear fluid enhances it. Once the growth factor is obtained, the theoretical redshift-space distortion $f\sigma_8(z)$ curves can be displayed as we did in Fig. \ref{fig:RSD} for all considered models in the current work. 
% From this plot, all models have a better fit \adlcd{How can you quantify that {\it better fit}?} with the data around the redshift $z\le 0.5$, indicating the shear fluid $\Omega_\sigma \propto a^{-6}$ was a stronger influence for the structure growth than today. \adlcd{The last sentence does not make sense. Do you mean {\it was a stronger influence for the structure growth in the past than today}? Please revise what you want to say.} However, the $w_0w_a$Bianchi Type V  has a better fit \adlcd{How do you quantify that better fit? By {\it eye} - this is not scientific - or do you calculate a $\chi^2$?} across all ranges of the redshift with the data than the other mentioned models except the \textit{PantheonP + SH0ES +CC + RSD} case, see the bottom panel of Fig. \ref{fig:RSD}. 
 \subsection{Statistical analysis} \label{stastical}
 In order to perform an adequate statistical comparison of models presented above with $\Lambda$CDM, 
 we employ both the Bayesian/Schwarz Information Criterion (BIC) and Akaike Information Criterion (AIC). As widely known, see for instance \citep{liddle2009statistical,szydlowski2015aic,rezaei2021comparison}, the broad discussion of statistical performance of  $\rm{AIC} = \chi ^{2} +2K, ~\text{and}~ \rm{BIC} = \chi ^{2} +K\log(N_i),$ where $\chi^{2}$ is estimated using the model's Gaussian likelihood function $\mathcal{L}(\hat{\Theta} |\text{data})$, the number of free parameters for that specific model is $K$, and  $N_i$ represents the number of data points for the $i^{th}$ dataset. To get the statistical validation of the considered model with the reference mode (i.e in our case $\Lambda$CDM is a reference model), the following expression: $$\Delta \rm{\rm{AIC}} = \big|\rm{AIC}_{\rm{\rm{Model}}} - \rm{\rm{AIC}}_{\it \rm{}\Lambda\rm{CDM}}\big|\;,$$  is applied to determine whether our theoretical model is accepted or rejected observationally.  If the value of $\rm \Delta AIC \leq 2$, the model has a \textrm{substantial observational support} for the fitted data; if  $ 4 \leq {\rm \Delta AIC} \leq 7$, the model has \textrm{less observational support}, and finally if  $\Delta \rm{AIC} \geq 10$, the model has \textrm{no observational support}, see the work \cite{szydlowski2015aic,sahlu2025constraining,sahlu2024cosmology,sahlu2025structure,sahlu2026observational}.  In the same manner, we also consider the relative difference of the Bayesian of the models as:  $$\Delta\rm{BIC} = - (\rm{BIC}_{i} - \rm{BIC}_{j})\;,
 $$
where model (i) is compared to model (j) in the BIC Bayes factor instance. Then (i) stands for the \(\Lambda\)CDM model, and (j) for the other models we are considering in this paper. The following is a ranking of the evidence against \(\Lambda\)CDM, i.e., in favor of the Bianchi V models, based on the categorization:  negligible if \(0 \leq \Delta \rm{BIC} \leq 2\),  positive if \(2 \leq \Delta\rm{BIC} \leq 6\), strong if \(6 \leq \Delta\rm{BIC} \leq 10\), and extremely strong if \(\Delta\rm{BIC} > 10\).

Table \ref{tab:model_comparison} showed the values of $\Delta$AIC $\ge4$  for all the considered models, indicating the less observational support for the case of \textit{DESI DR2 BAO + CC + RSD} data combinations.   Using the same datasets, the $w$CDM and Bianchi-V models fall within the strongly penalized range of $6\le \Delta\rm{BIC} \ge 10.0$; and extremely strong penalization for   $w_0w_a$CDM and $w$Bianchi-V models. Only the $w_0w_a$Bianchi-V model has a positive survival since its relatively small values of $\mathcal{L}(\hat{\Theta} |\text{data})$.

Similarly, when we include other combined datasets, namely \textit{DESI DR2 BAO   + CC + Union3 + RSD}, \textit{DESI DR2 BAO + CC + DESY5 + RSD}, and \textit{PantheonPlus+SH0ES + CC + RSD} the $w$CDM, $w_0w_a$CDM, Bianchi V, and $w_0w_a$Bianchi-V models are found to have substantial observational support through the $\Delta\text{AIC} \le 2$ threshold.  Using the same datasets  $w$CDM and $w$Bianchi-V models display statistically positive ranges of $\Delta$BIC.  However, for \textit{DESI DR2 BAO + CC + DESY5 + RSD} and \textit{PantheonPlus+SH0ES + CC + RSD}, the rigorous penalization metric predominates because the Bayesian criteria grow logarithmically with the sample volume via $K\log N_i$. With $\Delta\text{BIC} > 10$, the models with more free parameters ($w_0w_a$CDM, $w$Bianchi V, and $w_0w_a$Bianchi V) produce remarkably high deviations $\Lambda$CDM see \citep{szydlowski2015aic,sahlu2025structure} for more.  
 
 % \subsection{Parameter results for  $w$CDM and $w$Bianchi Type V models }
 % The general form of the Friedmannn equation \eqref{Friedmann} is reduced to the Bianchi Type V model for the case of $w_a =0$ and $w_0 \neq-1$. When the density parameter of the shear fluid $\Omega_{\sigma_0} = 0$ $w_0 \neq -1$ and $w_a = 0$, the  $w$CDM is also recovered. 
  % \begin{figure}
  %     \includegraphics[width=1.2\linewidth]{wCDM_wBV_H.png}
  %      \caption{Caption}
  %     \label{fig:placeholder}
  %  \end{figure}
  % \begin{figure}
  %     \includegraphics[width=1.2\linewidth]{wCDM_wBV_S8.png}
  %      \caption{Caption}
  %     \label{fig:placeholder}
  %  \end{figure}
% \subsection{Parameter results for  $w_0w_a$CDM and $w_0w_a$Bianchi Type V models }
% Brodaly discusses Table 3

%  \begin{figure}
%     \includegraphics[width=1.2\linewidth]{cosmoEoS.png}
%     \caption{Caption}
%     \label{fig:placeholder}
% \end{figure}
% \begin{itemize}
%     \item Plot the relation of $\Omega_k$ with $\Omega_{m,0}$ and draw differently, \textbf{Done, will present}
%     \item Plot the distance cosmology \textbf{(1. done)} see \url{https://arxiv.org/pdf/2407.14934v1}
%     \item Plot the Hubbel parameter, discuss with it \textbf{(done)}
%     \item Plot the RSD plot and discuss the effect \textbf{Done}
%     \item Plot the density contrast $\delta(z)$
%     \item plot the Growth factor D(a) and discuss broadly
%     \item Plot the Hubbel results, the Gaussian one, and discuss it
% \end{itemize}
%\newpage
  \begin{figure}
     \includegraphics[width=1.1\linewidth]{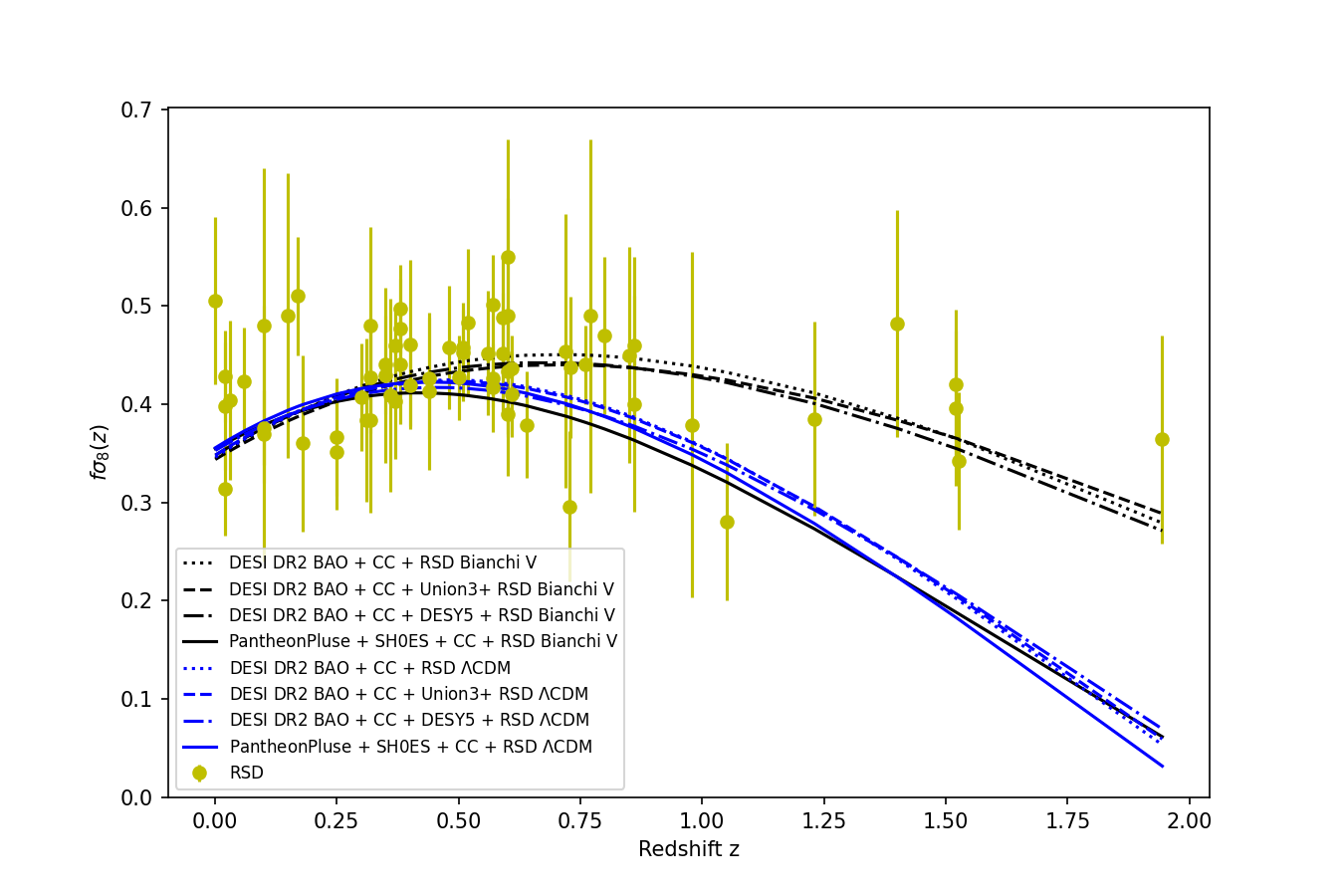}
      \includegraphics[width=1.1\linewidth]{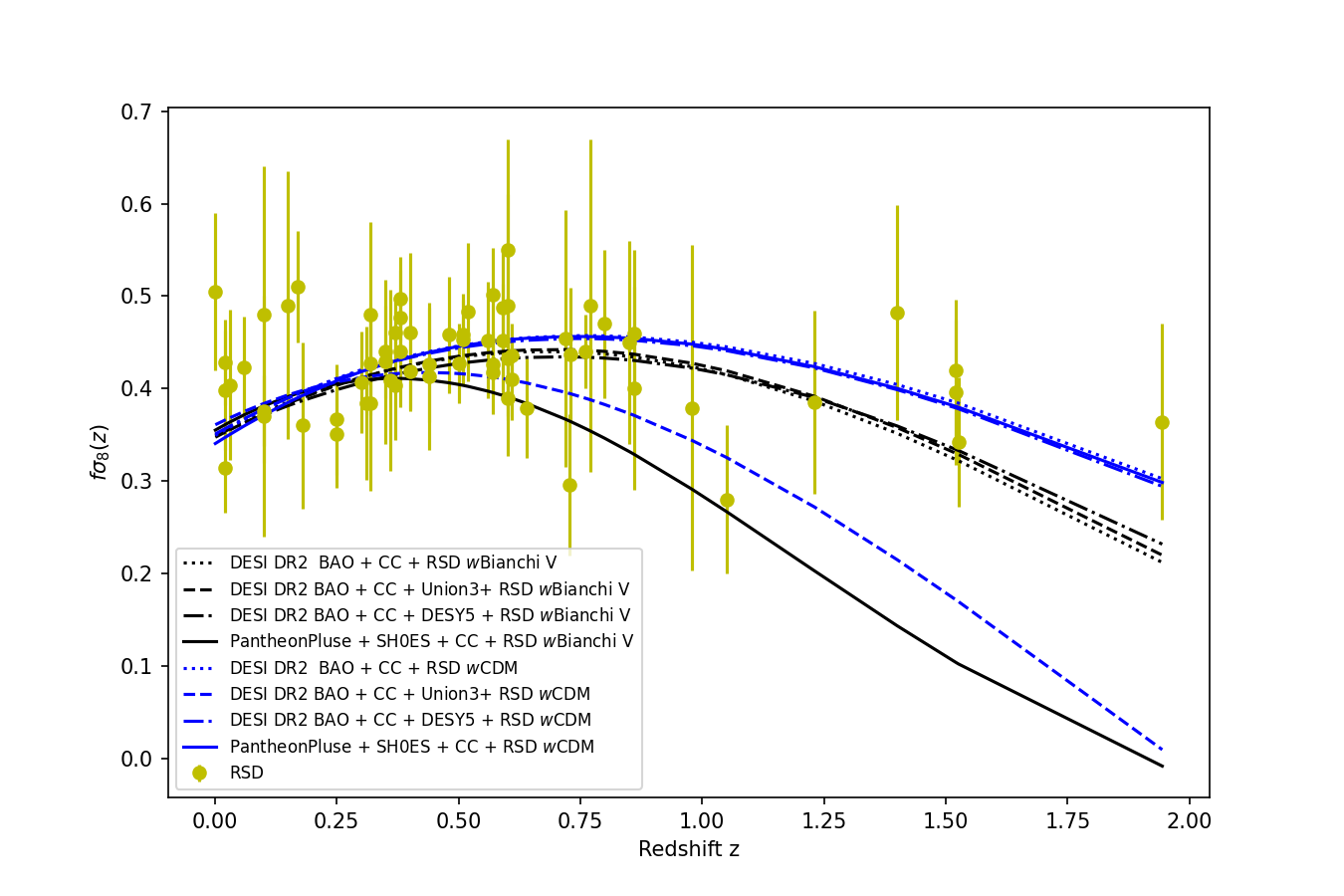}
      \includegraphics[width=1.1\linewidth]{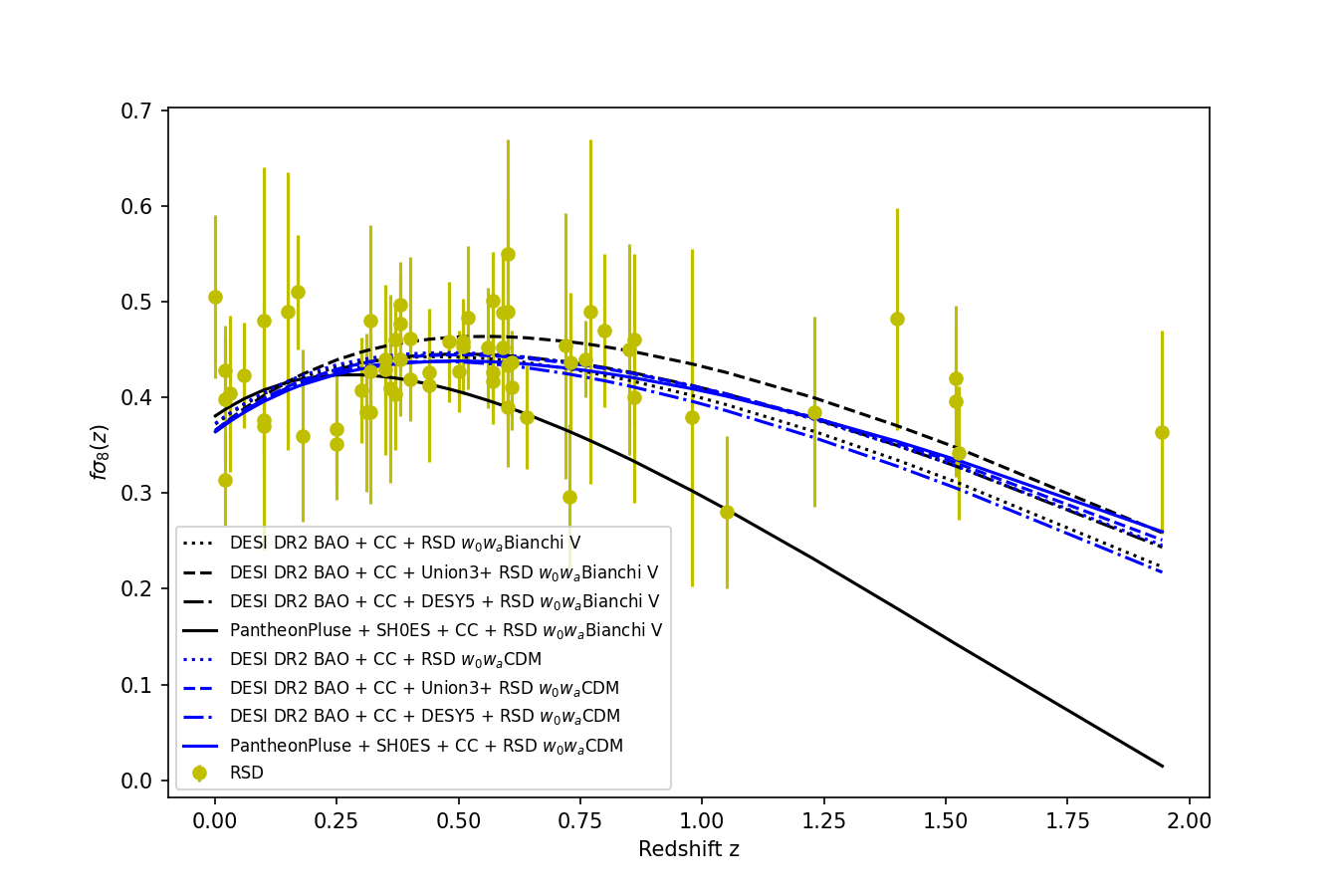}
       \caption{Redshift-space distortion ${{f\sigma_8(z)}}$ for different cosmological models. The \textit{top panel}  presents results for the  $\Lambda$CDM and Bianchi Type V models, the \textit{middle panel} for the  $w$CDM and $w$Bianchi Type V models, and the \textit{bottom panel} for the  $w_0w_a$CDM and $w_0w_a$Bianchi Type V models. All plots use the best-fit values of the constrained parameters listed in Tables \ref{tab1:cosmo_constraints}, \ref{tab2:cosmo_constraints}, and \ref{tab3:cosmo_constraints} for the combined datasets, respectively.}
      \label{fig:RSD}
  \end{figure}
\begin{table*}
\caption{Statistical comparison of the considered models against $\Lambda$CDM.}
\label{tab:model_comparison}
\begin{tabular}{lccccccc}
\hline\hline
\textbf{Dataset} & \textbf{Model} &  $\mathcal{L}(\hat{\Theta} |\text{data})$ & $\chi^2$ & \text{AIC} & $\Delta\text{AIC}$ & \text{BIC} & $\Delta\text{BIC}$ 
\\ \hline
 \textit{DESI DR2 BAO +CC + RSD}         & $\Lambda$CDM       &   -39.294 &    78.588 &    84.588 &     --- &    92.662 &     --- \\
                                    & $w$CDM             &   -40.492 &    80.984 &    88.984 &    4.396 &    99.749 &    7.087 \\
                                    & $w_0 w_a$CDM       &   -39.645 &    79.290 &    89.290 &    4.702 &   102.747 &   10.085 \\
                                    & Bianchi V          &   -40.952 &    81.904 &    89.904 &    5.316 &   100.669 &    8.007 \\
                                    & $w$Bianchi V       &   -39.806 &    79.612 &    89.612 &    5.024 &   103.069 &   10.407 \\
                                    & $w_0w_a$Bianchi V  &   -34.196 &    68.392 &    80.392 &     4.196 &    96.540 &    3.878 \\
\hline
\textit{DESI DR2 BAO +CC + Union3+ RSD} & $\Lambda$CDM       &   -53.709 &   107.418 &   113.418 &     --- &   122.044 &     --- \\
                                    & $w$CDM             &   -53.030 &   106.060 &   114.060 &    0.642 &   125.561 &    3.517 \\
                                    & $w_0w_a$CDM        &   -52.272 &   104.544 &   114.544 &    1.126 &   128.920 &    6.876 \\
                                    & Bianchi V          &   -53.796 &   107.592 &   115.592 &    2.174 &   127.093 &    5.049 \\
                                    & $w$Bianchi V       &   -52.665 &   105.330 &   115.330 &    1.912 &   129.706 &    7.662 \\
                                    & $w_0w_a$Bianchi V  &   -50.469 &   100.938 &   112.938 &     0.480 &   130.189 &    8.146 \\
\hline
\textit{DESI DR2 BAO +CC + DESY5+RSD}   & $\Lambda$CDM       &  -896.591 &  1793.182 &  1799.182 &     --- &  1815.625 &     --- \\
                                    & $w$CDM             &  -895.777 &  1791.554 &  1799.554 &    0.372 &  1821.478 &    5.853 \\
                                    & $w_0w_a$CDM        &  -893.272 &  1786.544 &  1796.544 &     2.638 &  1823.949 &    8.324 \\
                                    & Bianchi V          &  -895.743 &  1791.486 &  1799.486 &    0.304 &  1821.410 &    5.785 \\
                                    & $w$Bianchi V       &  -896.548 &  1793.096 &  1803.096 &    3.914 &  1830.501 &   14.876 \\
                                    & $w_0w_a$Bianchi V  &  -893.893 &  1787.786 &  1799.786 &    0.604 &  1832.672 &   17.047 \\
\hline
\textit{PantheonP + SH0ES +CC +RSD} & $\Lambda$CDM       &  -797.595 &  1595.190 &  1601.190 &     --- &  1617.671 &     --- \\
                                    & $w$CDM             &  -796.073 &  1592.146 &  1600.146 &     1.044 &  1622.121 &    4.450 \\
                                    & $w_0w_a$CDM        &  -797.376 &  1594.752 &  1604.752 &    3.562 &  1632.221 &   14.550 \\
                                    & Bianchi V          &  -795.517 &  1591.034 &  1599.034 &     2.156 &  1621.009 &    3.338 \\
                                    & $w$Bianchi V       &  -795.478 &  1590.956 &  1600.956 &     0.234 &  1628.425 &   10.754 \\
                                    & $w_0w_a$Bianchi V  &  -796.885 &  1593.770 &  1605.770 &    4.580 &  1638.733 &   21.062 \\
                                    \hline\hline
\end{tabular}
\end{table*}
\section{Conclusions}\label{disc}

We have investigated the statistical plausibility of DDE in an anisotropic Bianchi Type V universe using recent late-time cosmological observations, including $DESI$ $DR2$ $BAO$, cosmic chronometers, Type Ia supernovae, and redshift-space distortion measurements. The analysis considered six cosmological scenarios:  $\Lambda$CDM,  $w$CDM,  $w_0w_a$CDM, Bianchi Type V, $w$Bianchi Type V, and $w_0w_a$Bianchi Type V models. Using the constrained parameter values obtained from the combined datasets, we examined both the cosmic expansion history and the evolution of large-scale structures.

Our parameter constraints indicate that the DDE models generally favor a quintessence-like regime, characterized by $w_0>-1$ and $w_a<0$ for most dataset combinations. The departures from the standard $\Lambda$CDM cosmology remain modest, with tensions typically lying in the range $1.4\sigma - 2.7\sigma$. Consequently, our results do not provide conclusive evidence for evolving dark energy, but they do indicate a mild preference for these scenarios, a fact which is broadly consistent with recent $DESI$ analyses.

We also explored the implications of these models for the $H_0$ and $S_8$ tensions. The inferred values of $H_0$ generally lie between the Planck-2018 and SH0ES determinations, reducing the discrepancy relative to the standard flat $\Lambda$CDM scenario. In contrast, the $w_0w_a$CDM model tends to predict lower values of $S_8$, reflecting weaker late-time structure growth. Thus, the corresponding $w_0w_a$Bianchi Type V model remains compatible with current observational constraints and illustrates how anisotropic effects can modify the growth history while remaining consistent with existing late-time observations. 

The numerical results for the growth factor and redshift-space distortion are reported in the present work in Figs. \ref{fig:growthfactor}–\ref{fig:RSD}. These plots consistently show that the deviations of \(\tilde{\mathcal{D}}(a)\) and \(f\sigma_{8}(z)\) between \(\Lambda\)CDM and Bianchi Type V; $w$CDM and $w$Bianchi Type V; and $w_0w_a$CDM and $w_0w_a$Bianchi Type V indicate that the effect of structure formation in the presence of the shear fluid was larger in the past than at present.

Finally, we performed statistical analyses using the Bayesian Information Criterion (BIC) and the Akaike Information Criterion (AIC). The values of $\mathcal{L}(\hat{\Theta} |\text{data})$, $\chi^2$, AIC, $\Delta\text{AIC}$, BIC, and $\Delta\text{BIC}$ were calculated and are presented in Table \ref{tab:model_comparison}. The table indicates that the values of $\Delta\text{AIC}$ are $\le 2$ for all mo\-dels using the combined \textit{DESI DR2 BAO + CC + Union3 + RSD}, \textit{DESI DR2 BAO + CC + DESY5 + RSD}, and \textit{PantheonPlus+SH0ES + CC + RSD} datasets, which have substantial observational su\-pport. On the other hand, the values of the Bayesian criterion $\Delta\text{BIC} \ge 6$ are heavily penalized for the cases of \textit{DESI DR2 BAO + CC + DESY5 + RSD} and \textit{PantheonPlus+SH0ES + CC + RSD} because the criterion scales logarithmically with the sample vo\-lume via $K\log N_i$. Consequently, only the  $w$CDM and Bianchi-V model statistically survive, where their Bayes factor values fall in the range $2\le \Delta\rm{BIC } \leq 6$.  On the other hand, for the \textit{DESI DR2 BAO + CC + RSD} datasets, the $ w_0w_a$ Bianchi-V model is statistically preferred.

Overall, our results show that dynamical dark energy Bianchi Type V cosmologies, when tested against the current late-time observational framework for exploring possible departures from the Concordance $\Lambda$CDM model. However, since the present analysis is restricted to late-time probes, a more conclusive assessment of these models will require the inclusion of early-Universe observables, particularly CMB temperature and polarization measurements, together with weak-lensing and full large-scale-structure datasets. Such analyses - beyond the scope of this study - will determine whether the parameter regions favored by late-time observations remain viable when confronted with the full range of cosmological data.

\section*{Acknowledgments}

This work is supported by the Spanish Grants PID2024-158938NB-I00, PID2023-149560NB-C21, and the Severo Ochoa Excellence Grant CEX2023-001292-S, funded by MICIU/AEI/10.13039/501100011033 (“ERDF A way of making Europe”, “PGC Generacion de Conocimiento”) and FEDER, UE. The authors also acknowledge financial support from the project i-COOPB23096 (funded by CSIC)  and  AdlCD acknowledges support from NRF (South Africa) Grant CSUR23042798041, CNS2024-154286 (Spain) funded by MICIU/AEI/ 10.13039/\-501100011033  {\it ERDF A way of making Europe} and the Project SA097P24 funded by Junta de Castilla y Le\'on (Spain).
This work is also supported by CosmoVerse CA21136 COST action, European Cooperation in Science and Technology. AA, AdlCD, and SS thank Instituto de Física Corpuscular (IFIC), CSIC‐Universitat de Val\`{e}ncia for the hospitality extended during the development of this work in Valencia, Spain. GJO thanks the Cosmology and Gravity Group (UCT), the Centre for Space Research (NWU), and the Astrophysics Research Centre of UKZN (Westville) for their hospitality. GJO extends his gratitude to the Comrades M. Association for their inspiring role in the development of this work.  

%%%%%%%%%%%%%%%%%%%%%%%%%%%%%%%%%%%%%%%%%%%%%%%%%%
\section*{Data Availability}
During the manuscript development, we have used the publicly available datasets from: {\url{https://github.com/CobayaSampler/sn_data}}, {\url{https://github.com/CobayaSampler/bao_data}}, and \url{https://github.com/Ahmadmehrabi/Cosmic_chronometer_data}  for constraining the cosmological parameters, which are listed in Section \ref{resultanddiscussion}.



%\bibliographystyle{mnras}
%\bibliography{referencesbv} % if your bibtex file is called example.bib

\end{document}